\documentclass[longauth,traditabstract]{aa}

\usepackage[nonamebreak]{natbib}
\usepackage[stable]{footmisc}
\usepackage{amsmath}
\usepackage{txfonts}
\usepackage{placeins}
\usepackage{natbib}
\bibpunct{(}{)}{;}{a}{}{,} %
\usepackage{graphicx}
\usepackage{epstopdf}
\usepackage{ifthen}
\usepackage[breaklinks, colorlinks, citecolor=blue]{hyperref}
\usepackage{overpic}
\usepackage{fixltx2e}
\usepackage{rotating}

\usepackage[table,usenames,dvipsnames]{xcolor}

\def\setsymbol#1#2{\expandafter\def\csname #1\endcsname{#2}}
\def\getsymbol#1{\csname #1\endcsname}

\def\Planck{\textit{Planck}}

\def\allearlypapers{\nocite{planck2011-1.1, planck2011-1.3, planck2011-1.4, planck2011-1.5, planck2011-1.6, planck2011-1.7, planck2011-1.10, planck2011-1.10sup, planck2011-5.1a, planck2011-5.1b, planck2011-5.2a, planck2011-5.2b, planck2011-5.2c, planck2011-6.1, planck2011-6.2, planck2011-6.3a, planck2011-6.4a, planck2011-6.4b, planck2011-6.6, planck2011-7.0, planck2011-7.2, planck2011-7.3, planck2011-7.7a, planck2011-7.7b, planck2011-7.12, planck2011-7.13}}

\def\alltwentythirteenresultspapers{\nocite{planck2013-p01, planck2013-p02, planck2013-p02a, planck2013-p02d, planck2013-p02b, planck2013-p03, planck2013-p03c, planck2013-p03f, planck2013-p03d, planck2013-p03e, planck2013-p01a, planck2013-p06, planck2013-p03a, planck2013-pip88, planck2013-p08, planck2013-p11, planck2013-p12, planck2013-p13, planck2013-p14, planck2013-p15, planck2013-p05b, planck2013-p17, planck2013-p09, planck2013-p09a, planck2013-p20, planck2013-p19, planck2013-pipaberration, planck2013-p05, planck2013-p05a, planck2013-pip56, planck2013-p06b, planck2013-p01a}}

\def\alltwentyfifteenresultspapers{\nocite{planck2014-a01, planck2014-a03, planck2014-a04, planck2014-a05, planck2014-a06, planck2014-a07, planck2014-a08, planck2014-a09, planck2014-a11, planck2014-a12, planck2014-a13, planck2014-a14, planck2014-a15, planck2014-a16, planck2014-a17, planck2014-a18, planck2014-a19, planck2014-a20, planck2014-a22, planck2014-a24, planck2014-a26, planck2014-a28, planck2014-a29, planck2014-a30, planck2014-a31, planck2014-a35, planck2014-a36, planck2014-a37, planck2014-ES}}

\newbox\tablebox    \newdimen\tablewidth
\def\leaderfil{\leaders\hbox to 5pt{\hss.\hss}\hfil}
\def\endPlancktable{\tablewidth=\columnwidth 
    $$\hss\copy\tablebox\hss$$
    \vskip-\lastskip\vskip -2pt}
\def\endPlancktablewide{\tablewidth=\textwidth 
    $$\hss\copy\tablebox\hss$$
    \vskip-\lastskip\vskip -2pt}
\def\tablenote#1 #2\par{\begingroup \parindent=0.8em
    \abovedisplayshortskip=0pt\belowdisplayshortskip=0pt
    \noindent
    $$\hss\vbox{\hsize\tablewidth \hangindent=\parindent \hangafter=1 \noindent
    \hbox to \parindent{$^#1$\hss}\strut#2\strut\par}\hss$$
    \endgroup}
\def\doubleline{\vskip 3pt\hrule \vskip 1.5pt \hrule \vskip 5pt}

\def\L2{\ifmmode L_2\else $L_2$\fi}

\def\DeltaT{\ifmmode \Delta T\else $\Delta T$\fi}
\def\deltat{\ifmmode \Delta t\else $\Delta t$\fi}
\def\fknee{\ifmmode f_{\rm knee}\else $f_{\rm knee}$\fi}
\def\Fmax{\ifmmode F_{\rm max}\else $F_{\rm max}$\fi}
\def\solar{\ifmmode{\rm M}_{\mathord\odot}\else${\rm M}_{\mathord\odot}$\fi}
\def\Msolar{\ifmmode{\rm M}_{\mathord\odot}\else${\rm M}_{\mathord\odot}$\fi}
\def\Lsolar{\ifmmode{\rm L}_{\mathord\odot}\else${\rm L}_{\mathord\odot}$\fi}
\def\inv{\ifmmode^{-1}\else$^{-1}$\fi}
\def\mo{\ifmmode^{-1}\else$^{-1}$\fi}
\def\sup#1{\ifmmode ^{\rm #1}\else $^{\rm #1}$\fi}
\def\expo#1{\ifmmode \times 10^{#1}\else $\times 10^{#1}$\fi}
\def\,{\thinspace}
\def\lsim{\mathrel{\raise .4ex\hbox{\rlap{$<$}\lower 1.2ex\hbox{$\sim$}}}}
\def\gsim{\mathrel{\raise .4ex\hbox{\rlap{$>$}\lower 1.2ex\hbox{$\sim$}}}}

\def\simprop{\mathrel{\raise .4ex\hbox{\rlap{$\propto$}\lower 1.2ex\hbox{$\sim$}}}}
\def\deg{\ifmmode^\circ\else$^\circ$\fi}
\def\pdeg{\ifmmode $\setbox0=\hbox{$^{\circ}$}\rlap{\hskip.11\wd0 .}$^{\circ}
          \else \setbox0=\hbox{$^{\circ}$}\rlap{\hskip.11\wd0 .}$^{\circ}$\fi}
\def\arcs{\ifmmode {^{\scriptstyle\prime\prime}}
          \else $^{\scriptstyle\prime\prime}$\fi}
\def\arcm{\ifmmode {^{\scriptstyle\prime}}
          \else $^{\scriptstyle\prime}$\fi}
\newdimen\sa  \newdimen\sb
\def\parcs{\sa=.07em \sb=.03em
     \ifmmode \hbox{\rlap{.}}^{\scriptstyle\prime\kern -\sb\prime}\hbox{\kern -\sa}
     \else \rlap{.}$^{\scriptstyle\prime\kern -\sb\prime}$\kern -\sa\fi}
\def\parcm{\sa=.08em \sb=.03em
     \ifmmode \hbox{\rlap{.}\kern\sa}^{\scriptstyle\prime}\hbox{\kern-\sb}
     \else \rlap{.}\kern\sa$^{\scriptstyle\prime}$\kern-\sb\fi}
\def\ra[#1 #2 #3.#4]{#1\sup{h}#2\sup{m}#3\sup{s}\llap.#4}
\def\dec[#1 #2 #3.#4]{#1\deg#2\arcm#3\arcs\llap.#4}
\def\deco[#1 #2 #3]{#1\deg#2\arcm#3\arcs}
\def\rra[#1 #2]{#1\sup{h}#2\sup{m}}

\def\dots{\relax\ifmmode \ldots\else $\ldots$\fi}
\def\WHzsr{\ifmmode $W\,Hz\mo\,sr\mo$\else W\,Hz\mo\,sr\mo\fi}
\def\mHz{\ifmmode $\,mHz$\else \,mHz\fi}
\def\GHz{\ifmmode $\,GHz$\else \,GHz\fi}
\def\mKs{\ifmmode $\,mK\,s$^{1/2}\else \,mK\,s$^{1/2}$\fi}
\def\muKs{\ifmmode \,\mu$K\,s$^{1/2}\else \,$\mu$K\,s$^{1/2}$\fi}
\def\muKRJs{\ifmmode \,\mu$K$_{\rm RJ}$\,s$^{1/2}\else \,$\mu$K$_{\rm RJ}$\,s$^{1/2}$\fi}
\def\muKHz{\ifmmode \,\mu$K\,Hz$^{-1/2}\else \,$\mu$K\,Hz$^{-1/2}$\fi}
\def\MJysr{\ifmmode \,$MJy\,sr\mo$\else \,MJy\,sr\mo\fi}
\def\MJysrmK{\ifmmode \,$MJy\,sr\mo$\,mK$_{\rm CMB}\mo\else \,MJy\,sr\mo\,mK$_{\rm CMB}\mo$\fi}
\def\microns{\ifmmode \,\mu$m$\else \,$\mu$m\fi}

\def\muK{\ifmmode \,\mu$K$\else \,$\mu$\hbox{K}\fi}
\def\microK{\ifmmode \,\mu$K$\else \,$\mu$\hbox{K}\fi}
\def\muW{\ifmmode \,\mu$W$\else \,$\mu$\hbox{W}\fi}
\def\kms{\ifmmode $\,km\,s$^{-1}\else \,km\,s$^{-1}$\fi}
\def\kmsMpc{\ifmmode $\,\kms\,Mpc\mo$\else \,\kms\,Mpc\mo\fi}
\providecommand{\sorthelp}[1]{}

\def\LCDM{$\Lambda$CDM}
\def\NHUNIT{\ifmmode {\rm \,cm^{-2}} \else $\rm \,cm^{-2}$ \fi} 
\def\clee{$C_\ell^{EE}$}
\def\clbb{$C_\ell^{BB}$}

\def\muKcmb{\ifmmode \,\mu$K$_{\rm CMB}$\else \,$\mu$K$_{\rm CMB}$\fi}
\newcommand{\NH}{N_{\rm H}} 
\newcommand{\planck}{\Planck}
\newcommand{\WMAP}{WMAP}

\newcommand{\lcdm}{$\Lambda$CDM}

\newcommand{\OmegaM}{\ifmmode\Omega_{\rm M}\else $\Omega_{\rm M}$\fi}

\newcommand{\typeone}{{\sc{Type~1}}}
\newcommand{\typetwo}{{\sc{Type~2}}}
\newcommand{\typethree}{{\sc{Type~3}}}

\newcommand{\cooz}{CO(1$\rightarrow$0)}
\newcommand{\coto}{CO(2$\rightarrow$1)}
\newcommand{\cott}{CO(3$\rightarrow$2)}

\begin{document}

\title{\vglue -10mm\Planck\ 2015 results. I. Overview of products and scientific results}

\author{\small
Planck Collaboration: R.~Adam\inst{97}
\and
P.~A.~R.~Ade\inst{114}
\and
N.~Aghanim\inst{79}
\and
Y.~Akrami\inst{83, 134}
\and
M.~I.~R.~Alves\inst{79}
\and
M.~Arnaud\inst{95}
\and
F.~Arroja\inst{87, 101}
\and
J.~Aumont\inst{79}
\and
C.~Baccigalupi\inst{113}
\and
M.~Ballardini\inst{66, 68, 40}
% ...
}
\institute{\small
APC, AstroParticule et Cosmologie, Universit\'{e} Paris Diderot, CNRS/IN2P3, CEA/lrfu, Observatoire de Paris, Sorbonne Paris Cit\'{e}, 10, rue Alice Domon et L\'{e}onie Duquet, 75205 Paris Cedex 13, France\goodbreak
\and
Aalto University Mets\"{a}hovi Radio Observatory and Dept of Radio Science and Engineering, P.O. Box 13000, FI-00076 AALTO, Finland\goodbreak
\and
Aalto University Mets\"{a}hovi Radio Observatory, P.O. Box 13000, FI-00076 AALTO, Finland\goodbreak
\and
African Institute for Mathematical Sciences, 6-8 Melrose Road, Muizenberg, Cape Town, South Africa\goodbreak
\and
Agenzia Spaziale Italiana Science Data Center, Via del Politecnico snc, 00133, Roma, Italy\goodbreak
\and
Agenzia Spaziale Italiana, Viale Liegi 26, Roma, Italy\goodbreak
\and
Aix Marseille Universit\'{e}, CNRS, LAM (Laboratoire d'Astrophysique de Marseille) UMR 7326, 13388, Marseille, France\goodbreak
\and
Astrophysics Group, Cavendish Laboratory, University of Cambridge, J J Thomson Avenue, Cambridge CB3 0HE, U.K.\goodbreak
\and
Astrophysics \& Cosmology Research Unit, School of Mathematics, Statistics \& Computer Science, University of KwaZulu-Natal, Westville Campus, Private Bag X54001, Durban 4000, South Africa\goodbreak
\and
Atacama Large Millimeter/submillimeter Array, ALMA Santiago Central Offices, Alonso de Cordova 3107, Vitacura, Casilla 763 0355, Santiago, Chile\goodbreak
\and
CITA, University of Toronto, 60 St. George St., Toronto, ON M5S 3H8, Canada\goodbreak
\and
CNRS, IRAP, 9 Av. colonel Roche, BP 44346, F-31028 Toulouse cedex 4, France\goodbreak
\and
CRANN, Trinity College, Dublin, Ireland\goodbreak
\and
California Institute of Technology, Pasadena, California, U.S.A.\goodbreak
\and
Centre for Theoretical Cosmology, DAMTP, University of Cambridge, Wilberforce Road, Cambridge CB3 0WA, U.K.\goodbreak
\and
Centro de Estudios de F\'{i}sica del Cosmos de Arag\'{o}n (CEFCA), Plaza San Juan, 1, planta 2, E-44001, Teruel, Spain\goodbreak
\and
Computational Cosmology Center, Lawrence Berkeley National Laboratory, Berkeley, California, U.S.A.\goodbreak
\and
Consejo Superior de Investigaciones Cient\'{\i}ficas (CSIC), Madrid, Spain\goodbreak
\and
DSM/Irfu/SPP, CEA-Saclay, F-91191 Gif-sur-Yvette Cedex, France\goodbreak
\and
DTU Space, National Space Institute, Technical University of Denmark, Elektrovej 327, DK-2800 Kgs. Lyngby, Denmark\goodbreak
\and
D\'{e}partement de Physique Th\'{e}orique, Universit\'{e} de Gen\`{e}ve, 24, Quai E. Ansermet,1211 Gen\`{e}ve 4, Switzerland\goodbreak
\and
Dark Cosmology Centre, Niels Bohr Institute, University of Copenhagen, Juliane Maries Vej 30, 2100 Copenhagen, Denmark\goodbreak
\and
Departamento de F\'{\i}sica, Universidad de Oviedo, Avda. Calvo Sotelo s/n, Oviedo, Spain\goodbreak
\and
Departamento de Matem\'{a}ticas, Estad\'{\i}stica y Computaci\'{o}n, Universidad de Cantabria, Avda. de los Castros s/n, Santander, Spain\goodbreak
\and
Department of Astronomy and Astrophysics, University of Toronto, 50 Saint George Street, Toronto, Ontario, Canada\goodbreak
\and
Department of Astrophysics/IMAPP, Radboud University Nijmegen, P.O. Box 9010, 6500 GL Nijmegen, The Netherlands\goodbreak
\and
Department of Physics \& Astronomy, University of British Columbia, 6224 Agricultural Road, Vancouver, British Columbia, Canada\goodbreak
\and
Department of Physics and Astronomy, Dana and David Dornsife College of Letter, Arts and Sciences, University of Southern California, Los Angeles, CA 90089, U.S.A.\goodbreak
\and
Department of Physics and Astronomy, Johns Hopkins University, Bloomberg Center 435, 3400 N. Charles St., Baltimore, MD 21218, U.S.A.\goodbreak
\and
Department of Physics and Astronomy, University College London, London WC1E 6BT, U.K.\goodbreak
\and
Department of Physics and Astronomy, University of Sussex, Brighton BN1 9QH, U.K.\goodbreak
\and
Department of Physics, Florida State University, Keen Physics Building, 77 Chieftan Way, Tallahassee, Florida, U.S.A.\goodbreak
\and
Department of Physics, Gustaf H\"{a}llstr\"{o}min katu 2a, University of Helsinki, Helsinki, Finland\goodbreak
\and
Department of Physics, Princeton University, Princeton, New Jersey, U.S.A.\goodbreak
\and
Department of Physics, University of Alberta, 11322-89 Avenue, Edmonton, Alberta, T6G 2G7, Canada\goodbreak
\and
Department of Physics, University of California, Berkeley, California, U.S.A.\goodbreak
\and
Department of Physics, University of California, One Shields Avenue, Davis, California, U.S.A.\goodbreak
\and
Department of Physics, University of California, Santa Barbara, California, U.S.A.\goodbreak
\and
Department of Physics, University of Illinois at Urbana-Champaign, 1110 West Green Street, Urbana, Illinois, U.S.A.\goodbreak
\and
Dipartimento di Fisica e Astronomia A. Righi, Universit\`{a} degli Studi di Bologna, Viale Berti Pichat 6/2, I-40127, Bologna, Italy\goodbreak
\and
Dipartimento di Fisica e Astronomia G. Galilei, Universit\`{a} degli Studi di Padova, via Marzolo 8, 35131 Padova, Italy\goodbreak
\and
Dipartimento di Fisica e Scienze della Terra, Universit\`{a} di Ferrara, Via Saragat 1, 44122 Ferrara, Italy\goodbreak
\and
Dipartimento di Fisica, Universit\`{a} La Sapienza, P. le A. Moro 2, Roma, Italy\goodbreak
\and
Dipartimento di Fisica, Universit\`{a} degli Studi di Milano, Via Celoria, 16, Milano, Italy\goodbreak
\and
Dipartimento di Fisica, Universit\`{a} degli Studi di Trieste, via A. Valerio 2, Trieste, Italy\goodbreak
\and
Dipartimento di Fisica, Universit\`{a} di Roma Tor Vergata, Via della Ricerca Scientifica, 1, Roma, Italy\goodbreak
\and
Dipartimento di Matematica, Universit\`{a} di Roma Tor Vergata, Via della Ricerca Scientifica, 1, Roma, Italy\goodbreak
\and
Discovery Center, Niels Bohr Institute, Blegdamsvej 17, Copenhagen, Denmark\goodbreak
\and
Dpto. Astrof\'{i}sica, Universidad de La Laguna (ULL), E-38206 La Laguna, Tenerife, Spain\goodbreak
\and
European Southern Observatory, ESO Vitacura, Alonso de Cordova 3107, Vitacura, Casilla 19001, Santiago, Chile\goodbreak
\and
European Space Agency, ESAC, Camino bajo del Castillo, s/n, Urbanizaci\'{o}n Villafranca del Castillo, Villanueva de la Ca\~{n}ada, Madrid, Spain\goodbreak
\and
European Space Agency, ESAC, Planck Science Office, Camino bajo del Castillo, s/n, Urbanizaci\'{o}n Villafranca del Castillo, Villanueva de la Ca\~{n}ada, Madrid, Spain\goodbreak
\and
European Space Agency, ESTEC, Keplerlaan 1, 2201 AZ Noordwijk, The Netherlands\goodbreak
\and
Facolt\`{a} di Ingegneria, Universit\`{a} degli Studi e-Campus, Via Isimbardi 10, Novedrate (CO), 22060, Italy\goodbreak
\and
Finnish Centre for Astronomy with ESO (FINCA), University of Turku, V\"{a}is\"{a}l\"{a}ntie 20, FIN-21500, Piikki\"{o}, Finland\goodbreak
\and
GEPI, Observatoire de Paris, Section de Meudon, 5 Place J. Janssen, 92195 Meudon Cedex, France\goodbreak
\and
Gran Sasso Science Institute, INFN, viale F. Crispi 7, 67100 L'Aquila, Italy\goodbreak
\and
HGSFP and University of Heidelberg, Theoretical Physics Department, Philosophenweg 16, 69120, Heidelberg, Germany\goodbreak
\and
Haverford College Astronomy Department, 370 Lancaster Avenue, Haverford, Pennsylvania, U.S.A.\goodbreak
\and
Helsinki Institute of Physics, Gustaf H\"{a}llstr\"{o}min katu 2, University of Helsinki, Helsinki, Finland\goodbreak
\and
ICTP South American Institute for Fundamental Research, Instituto de F\'{\i}sica Te\'{o}rica, Universidade Estadual Paulista, S\~{a}o Paulo, Brazil\goodbreak
\and
INAF - Osservatorio Astrofisico di Catania, Via S. Sofia 78, Catania, Italy\goodbreak
\and
INAF - Osservatorio Astronomico di Padova, Vicolo dell'Osservatorio 5, Padova, Italy\goodbreak
\and
INAF - Osservatorio Astronomico di Roma, via di Frascati 33, Monte Porzio Catone, Italy\goodbreak
\and
INAF - Osservatorio Astronomico di Trieste, Via G.B. Tiepolo 11, Trieste, Italy\goodbreak
\and
INAF/IASF Bologna, Via Gobetti 101, Bologna, Italy\goodbreak
\and
INAF/IASF Milano, Via E. Bassini 15, Milano, Italy\goodbreak
\and
INFN, Sezione di Bologna, Via Irnerio 46, I-40126, Bologna, Italy\goodbreak
\and
INFN, Sezione di Roma 1, Universit\`{a} di Roma Sapienza, Piazzale Aldo Moro 2, 00185, Roma, Italy\goodbreak
\and
INFN, Sezione di Roma 2, Universit\`{a} di Roma Tor Vergata, Via della Ricerca Scientifica, 1, Roma, Italy\goodbreak
\and
INFN/National Institute for Nuclear Physics, Via Valerio 2, I-34127 Trieste, Italy\goodbreak
\and
IPAG: Institut de Plan\'{e}tologie et d'Astrophysique de Grenoble, Universit\'{e} Grenoble Alpes, IPAG, F-38000 Grenoble, France, CNRS, IPAG, F-38000 Grenoble, France\goodbreak
\and
ISDC, Department of Astronomy, University of Geneva, ch. d'Ecogia 16, 1290 Versoix, Switzerland\goodbreak
\and
IUCAA, Post Bag 4, Ganeshkhind, Pune University Campus, Pune 411 007, India\goodbreak
\and
Imperial College London, Astrophysics group, Blackett Laboratory, Prince Consort Road, London, SW7 2AZ, U.K.\goodbreak
\and
Infrared Processing and Analysis Center, California Institute of Technology, Pasadena, CA 91125, U.S.A.\goodbreak
\and
Institut N\'{e}el, CNRS, Universit\'{e} Joseph Fourier Grenoble I, 25 rue des Martyrs, Grenoble, France\goodbreak
\and
Institut Universitaire de France, 103, bd Saint-Michel, 75005, Paris, France\goodbreak
\and
Institut d'Astrophysique Spatiale, CNRS (UMR8617) Universit\'{e} Paris-Sud 11, B\^{a}timent 121, Orsay, France\goodbreak
\and
Institut d'Astrophysique de Paris, CNRS (UMR7095), 98 bis Boulevard Arago, F-75014, Paris, France\goodbreak
\and
Institute for Space Sciences, Bucharest-Magurale, Romania\goodbreak
\and
Institute of Astronomy, University of Cambridge, Madingley Road, Cambridge CB3 0HA, U.K.\goodbreak
\and
Institute of Theoretical Astrophysics, University of Oslo, Blindern, Oslo, Norway\goodbreak
\and
Instituto Nacional de Astrof\'{\i}sica, \'{O}ptica y Electr\'{o}nica (INAOE), Apartado Postal 51 y 216, 72000 Puebla, M\'{e}xico\goodbreak
\and
Instituto de Astrof\'{\i}sica de Canarias, C/V\'{\i}a L\'{a}ctea s/n, La Laguna, Tenerife, Spain\goodbreak
\and
Instituto de F\'{\i}sica de Cantabria (CSIC-Universidad de Cantabria), Avda. de los Castros s/n, Santander, Spain\goodbreak
\and
Istituto Nazionale di Fisica Nucleare, Sezione di Padova, via Marzolo 8, I-35131 Padova, Italy\goodbreak
\and
Jet Propulsion Laboratory, California Institute of Technology, 4800 Oak Grove Drive, Pasadena, California, U.S.A.\goodbreak
\and
Jodrell Bank Centre for Astrophysics, Alan Turing Building, School of Physics and Astronomy, The University of Manchester, Oxford Road, Manchester, M13 9PL, U.K.\goodbreak
\and
Kavli Institute for Cosmology Cambridge, Madingley Road, Cambridge, CB3 0HA, U.K.\goodbreak
\and
LAL, Universit\'{e} Paris-Sud, CNRS/IN2P3, Orsay, France\goodbreak
\and
LAPTh, Univ. de Savoie, CNRS, B.P.110, Annecy-le-Vieux F-74941, France\goodbreak
\and
LERMA, CNRS, Observatoire de Paris, 61 Avenue de l'Observatoire, Paris, France\goodbreak
\and
LESIA, Observatoire de Paris, CNRS, UPMC, Universit\'{e} Paris-Diderot, 5 Place J. Janssen, 92195 Meudon, France\goodbreak
\and
Laboratoire AIM, IRFU/Service d'Astrophysique - CEA/DSM - CNRS - Universit\'{e} Paris Diderot, B\^{a}t. 709, CEA-Saclay, F-91191 Gif-sur-Yvette Cedex, France\goodbreak
\and
Laboratoire Traitement et Communication de l'Information, CNRS (UMR 5141) and T\'{e}l\'{e}com ParisTech, 46 rue Barrault F-75634 Paris Cedex 13, France\goodbreak
\and
Laboratoire de Physique Subatomique et Cosmologie, Universit\'{e} Grenoble-Alpes, CNRS/IN2P3, 53, rue des Martyrs, 38026 Grenoble Cedex, France\goodbreak
\and
Laboratoire de Physique Th\'{e}orique, Universit\'{e} Paris-Sud 11 \& CNRS, B\^{a}timent 210, 91405 Orsay, France\goodbreak
\and
Lawrence Berkeley National Laboratory, Berkeley, California, U.S.A.\goodbreak
\and
Lebedev Physical Institute of the Russian Academy of Sciences, Astro Space Centre, 84/32 Profsoyuznaya st., Moscow, GSP-7, 117997, Russia\goodbreak
\and
Leung Center for Cosmology and Particle Astrophysics, National Taiwan University, Taipei 10617, Taiwan\goodbreak
\and
Max-Planck-Institut f\"{u}r Astrophysik, Karl-Schwarzschild-Str. 1, 85741 Garching, Germany\goodbreak
\and
Max-Planck-Institut f\"{u}r Extraterrestrische Physik, Giessenbachstra{\ss}e, 85748 Garching, Germany\goodbreak
\and
McGill Physics, Ernest Rutherford Physics Building, McGill University, 3600 rue University, Montr\'{e}al, QC, H3A 2T8, Canada\goodbreak
\and
Mullard Space Science Laboratory, University College London, Surrey RH5 6NT, U.K.\goodbreak
\and
National University of Ireland, Department of Experimental Physics, Maynooth, Co. Kildare, Ireland\goodbreak
\and
Niels Bohr Institute, Blegdamsvej 17, Copenhagen, Denmark\goodbreak
\and
Observational Cosmology, Mail Stop 367-17, California Institute of Technology, Pasadena, CA, 91125, U.S.A.\goodbreak
\and
Optical Science Laboratory, University College London, Gower Street, London, U.K.\goodbreak
\and
Paris, France\goodbreak
\and
Physics Department, Shahid Beheshti University, Tehran, Iran\goodbreak
\and
SB-ITP-LPPC, EPFL, CH-1015, Lausanne, Switzerland\goodbreak
\and
SISSA, Astrophysics Sector, via Bonomea 265, 34136, Trieste, Italy\goodbreak
\and
School of Physics and Astronomy, Cardiff University, Queens Buildings, The Parade, Cardiff, CF24 3AA, U.K.\goodbreak
\and
School of Physics and Astronomy, University of Nottingham, Nottingham NG7 2RD, U.K.\goodbreak
\and
Simon Fraser University, Department of Physics, 8888 University Drive, Burnaby BC, Canada\goodbreak
\and
Sorbonne Universit\'{e}-UPMC, UMR7095, Institut d'Astrophysique de Paris, 98 bis Boulevard Arago, F-75014, Paris, France\goodbreak
\and
Space Research Institute (IKI), Russian Academy of Sciences, Profsoyuznaya Str, 84/32, Moscow, 117997, Russia\goodbreak
\and
Space Sciences Laboratory, University of California, Berkeley, California, U.S.A.\goodbreak
\and
Special Astrophysical Observatory, Russian Academy of Sciences, Nizhnij Arkhyz, Zelenchukskiy region, Karachai-Cherkessian Republic, 369167, Russia\goodbreak
\and
Stanford University, Dept of Physics, Varian Physics Bldg, 382 Via Pueblo Mall, Stanford, California, U.S.A.\goodbreak
\and
Sub-Department of Astrophysics, University of Oxford, Keble Road, Oxford OX1 3RH, U.K.\goodbreak
\and
Sydney Institute of Astronomy, School of Physics A28, University of Sydney, NSW 2006, Australia\goodbreak
\and
Thales Alenia Space Italia S.p.A., S.S. Padana Superiore 290, 20090 Vimodrone (MI), Italy\goodbreak
\and
Theory Division, PH-TH, CERN, CH-1211, Geneva 23, Switzerland\goodbreak
\and
UPMC Univ Paris 06, UMR7095, 98 bis Boulevard Arago, F-75014, Paris, France\goodbreak
\and
Universit\"{a}t Heidelberg, Institut f\"{u}r Theoretische Astrophysik, Philosophenweg 12, 69120 Heidelberg, Germany\goodbreak
\and
Universit\'{e} Denis Diderot (Paris 7), 75205 Paris Cedex 13, France\goodbreak
\and
Universit\'{e} de Toulouse, UPS-OMP, IRAP, F-31028 Toulouse cedex 4, France\goodbreak
\and
Universities Space Research Association, Stratospheric Observatory for Infrared Astronomy, MS 232-11, Moffett Field, CA 94035, U.S.A.\goodbreak
\and
University Observatory, Ludwig Maximilian University of Munich, Scheinerstrasse 1, 81679 Munich, Germany\goodbreak
\and
University of Granada, Departamento de F\'{\i}sica Te\'{o}rica y del Cosmos, Facultad de Ciencias, Granada, Spain\goodbreak
\and
University of Granada, Instituto Carlos I de F\'{\i}sica Te\'{o}rica y Computacional, Granada, Spain\goodbreak
\and
University of Heidelberg, Institute for Theoretical Physics, Philosophenweg 16, 69120, Heidelberg, Germany\goodbreak
% ...
}

\date{\vglue -1.5mm 1 August 2015\vglue -5mm}
\abstract{\vglue -3mm 
The European Space Agency's \Planck\ satellite, dedicated to studying the early Universe and its subsequent evolution, was launched 14~May 2009 and scanned the microwave and submillimetre sky continuously between 12~August 2009 and 23~October 2013.  In February~2015, ESA and the \Planck\ Collaboration released the second set of cosmology products based on data from the entire
\Planck\ mission, including both temperature and polarization, along with a set of scientific and technical papers and a web-based explanatory supplement.  This paper gives an overview of the main characteristics of the data and the data products in the release, as well as the associated cosmological and astrophysical science results and papers.  The science products include maps of the cosmic microwave background (CMB), the thermal Sunyaev-Zeldovich effect, and diffuse foregrounds in temperature and
polarization, catalogues of compact Galactic and extragalactic sources (including separate catalogues of Sunyaev-Zeldovich clusters and Galactic cold clumps), and extensive simulations of signals and noise used in assessing the performance of the analysis methods and assessment of uncertainties.  The likelihood code used to assess cosmological models against the \Planck\ data are described, as well as a CMB lensing likelihood.  Scientific results include cosmological parameters deriving from CMB power spectra, gravitational lensing, and cluster counts, as well as constraints on inflation, non-Gaussianity, primordial magnetic fields, dark energy, and modified gravity.}

\authorrunning{Planck Collaboration}
\titlerunning{The \Planck\ mission}
   \maketitle

\allearlypapers

\alltwentythirteenresultspapers

\alltwentyfifteenresultspapers

\section{Introduction $\tau$}

The \Planck\ satellite\footnote{\Planck\ (\url{http://www.esa.int/Planck}) is a project of the European Space Agency (ESA) with  instruments provided by two scientific consortia funded by ESA member states and led by Principal Investigators from France and Italy, telescope reflectors provided through a collaboration between ESA and a scientific consortium led and funded by Denmark, and additional contributions from NASA (USA).}  \citep{tauber2010a, planck2011-1.1} was launched on 14 May 2009 and observed the sky stably and continuously from 12~August 2009 to 23~October 2013.  \Planck's scientific payload contained an array of 74~detectors in nine frequency bands sensitive to frequencies between 25 and 1000\,GHz, which scanned the sky with angular resolution between 33\arcm\ and 5\arcm.  The
detectors of the Low Frequency Instrument \citep[LFI;][]{Bersanelli2010, planck2011-1.4} were pseudo-correlation radiometers, covering bands centred at 30, 44, and 70\,GHz.  The detectors of the High Frequency Instrument \citep[HFI;][]{Lamarre2010, planck2011-1.5} were
bolometers, covering bands centred at 100, 143, 217, 353, 545, and $857\,$GHz.  \Planck\ imaged the whole sky twice in one year, with a combination of sensitivity, angular resolution, and frequency coverage never before achieved.  \Planck, its payload, and its performance as predicted at the time of launch are described in 13~papers included in a special issue of Astronomy \& Astrophysics (Volume 520).

The main objective of \Planck, defined in 1995, was to measure the spatial anisotropies in the temperature of the cosmic microwave background (CMB), with an accuracy set by fundamental astrophysical limits, thereby extracting essentially all the cosmological information embedded in the temperature anisotropies of the CMB. \Planck\ was not initially designed to measure to high accuracy the CMB polarization anisotropies, which encode not only a wealth of cosmological information, but also provide a unique probe of the early history of the Universe, during the time when the first stars and galaxies formed.  However, during its development it was significantly
enhanced in this respect, and its polarization measurement capabilities have exceeded all original expectations.  \Planck\ was also designed to produce a wealth of information on the properties of extragalactic sources and of clusters of galaxies (via the Sunyaev-Zeldovich effect), and on the dust and gas in the Milky Way.  The scientific objectives of \Planck\ were described in detail in
\cite{planck2005-bluebook}.  With the results presented here and in a series of accompanying papers, 
\Planck\ has already achieved all of its planned science goals.

An overview of the scientific operations of the \Planck\ mission has been presented in \citet{planck2013-p01}. Further operational details---extending this description to the end of the mission---are presented in the 2015 Explanatory Supplement \citep{planck2014-ES}.
This paper presents an overview of the main data products and scientific results of \Planck's third release,\footnote{In January
of 2011, ESA and the \Planck\ Collaboration released to the public a first set of scientific data, the Early Release Compact Source Catalogue (ERCSC), a list of unresolved and compact sources extracted from the first complete all-sky survey carried out by \Planck\ \citep{planck2011-1.10}.  At the same time, initial scientific results related to astrophysical foregrounds were published in a special issue of Astronomy and Astrophysics (Vol.\ 520, 2011).  Since then, 34 {\bf UPDATE} ``Intermediate'' papers have been submitted for publication to A\&A, containing further astrophysical investigations by the Collaboration.  In March of 2013, the second release of scientific data took place, consisting mainly of temperature maps of the whole sky; these products and associated scientific
results are described in a special issue of A\&A (Vol.~571, 2014).} based on data acquired in the period 12~August 2009 to 23~October 2013, and hereafter referred to as the ``2015 products.''

\section{Data products in the 2015 release}
\label{DataProds}

The 2015 distribution of released products, which can be freely accessed via the \Planck\ Legacy Archive interface (PLA),\footnote{\url{http://pla.esac.esa.int}} is based on all the data acquired by \Planck\ during routine operations, starting on 12~August 2009 and ending on 23~October 2014.  The distribution contains the following.

\begin{itemize}

\item Cleaned and calibrated timelines of the data for each detector.  

\vskip 4pt

\item Maps of the sky at nine frequencies (Sect.~\ref{sec:FreqMaps}) in temperature, and at seven frequencies (30--353\,GHz) in polarization.  Additional products serve to quantify the characteristics of the maps to a level adequate for the science results being presented, such as noise maps, masks, and instrument characteristics. 

\vskip 4pt

\item Four high-resolution maps of the CMB sky in temperature and polarization, and accompanying characterization products (Sect.~\ref{subsec:CMBmapNG})\footnote{It has become the norm in CMB studies to use the COSMO (\url{http://healpix.sourceforge.net/html/intronode6.htm}) convention for polarization angles, rather than the IAU \citep{1996Hamaker} convention, and the \Planck\ data products have followed this trend. The net effect of using the COSMO convention is a sign inversion on Stokes U with respect to the IAU convention. All \Planck\ fits files containing polarization data include a keyword that emphasizes the convention used. Nonetheless, users should be keenly aware of this fact.}.

\vskip 4pt

\item Four high-pass-filtered maps of the CMB sky in polarization, and accompanying characterization products (Sect.~\ref{subsec:CMBmapNG}). The rationale for providing these maps is explained in the following section.

\vskip 4pt

\item A low-resolution CMB temperature map (Sect.~\ref{subsec:CMBmapNG}) used in the low-$\ell$ likelihood code, with an associated set of foreground temperature maps produced in the process of separating the low-resolution CMB from foregrounds, with accompanying characterization products.

\vskip 4pt

\item Maps of thermal dust and residual cosmic infrared background (CIB), carbon monoxide (CO), synchrotron, free-free, and spinning dust temperature emission, plus maps of dust temperature and opacity (Sect.~\ref{sec:AstroProds}). 

\vskip 4pt

\item Maps of synchrotron and dust polarized emission.

\vskip 4pt

\item A map of the estimated CMB lensing potential over 70\,\%\ of the sky. 

\vskip 4pt

\item A map of the Sunyaev-Zeldovich effect Compton parameter. 

\vskip 4pt

\item Monte Carlo chains used in determining cosmological parameters from
the \Planck\ data.

\vskip 4pt

\item The Second \Planck\ Catalogue of Compact Sources (PCCS2, Sect.~\ref{sec:PCCS2}), comprising lists of compact sources over the entire sky at the nine \Planck\ frequencies. The PCCS2 includes polarization information, and supersedes the previous Early Release Compact Source Catalogue \citep{planck2011-6.2} and the PCCS1 \citep{planck2013-p05}.

\vskip 4pt

\item The Second \Planck\ Catalogue of Sunyaev-Zeldovich Sources (PSZ2, Sect.~\ref{sec:PSZ2}), comprising a list of sources detected by their SZ distortion of the CMB spectrum.  The PSZ2 supersedes the previous Early Sunyaev-Zeldovich Catalogue \citep{planck2013-p05a} and the PSZ1 \citep{planck2013-p05a}.

\vskip 4pt

\item The \Planck\ catalogue of Galactic Cold Clumps \citep[PGCC,][]{planck2014-a37}, providing a list of Galactic cold sources
over the whole sky (see Sect.~\ref{sec:PCCC}).  The PGCC supersedes the previous Early Cold Core Catalogue (ECC), part of the Early Release Compact Source Catalogue \citep[ERCSC, ][]{planck2011-1.10}.

\vskip 4pt

\item A full set of simulations, including Monte Carlo realizations.

\vskip 4pt

\item A likelihood code and data package used for testing cosmological models against the \Planck\ data, including both the CMB (Sect.~\ref{sec:CMBLike}) and CMB lensing (Sect.~\ref{sec:LensLike}).

\end{itemize}

The first 2015 products were released in February 2015, and the full release will be complete by July 2015.  In parallel, the \Planck\ Collaboration is developing the next generation of data products, which will be delivered in the early part of 2016.

\subsection{The state of polarization in the \Planck\ 2015 data}
\label{subsec:stateofpolarization}

\noindent{\bf LFI---}The 2015 \Planck\ release includes polarization data at the LFI frequencies 30, 44, and 70\,GHz. The 70\,GHz polarization data are used for the 2015 \Planck\ likelihood at low-multipoles ($\ell <30$). The 70\,GHz map is cleaned with the 30\,GHz and the 353\,GHz channels for synchrotron and dust emission, respectively \citep{planck2014-a15}.

Control of systematic effects is a challenging task for polarization measurements, especially at large angular scales. We carry out extensive analyses of systematic effects impacting the 2015 LFI polarization data \citep{planck2014-a04}. Our approach follows two complementary paths. First, we use the redundancy in the \Planck\ scanning strategy to produce difference maps that, in principle, contain the same sky signal (``null tests''). Any residuals in these maps blindly probe all non-common-mode systematics present in the data. Second, we use our knowledge of the instrument to build physical models of all the known relevant systematic effects. We then simulate timelines and project them into sky maps following the LFI map-making process. We quantify the results in terms of power spectra and compare them to the FFP8 LFI noise model.

Our analysis shows no evidence of systematic errors significantly affecting the 2015 LFI polarization results. On the other hand, our model indicates that at low multipoles the dominant LFI systematics (gain errors and ADC non-linearity) are only marginally dominated by noise and the expected signal. Therefore, further independent tests are being carried out and will be discussed in a forthcoming paper, as well as in the final 2016 \Planck\ release. These include polarization cross-spectra between the LFI 70\,GHz and the HFI\,100 and 143\,GHz maps (that are not part of this 2015 release -- see below). Because systematic effects between the two \Planck\
instruments are expected to be largely uncorrelated, such cross-instrument approach may prove particularly effective.

\noindent{\bf HFI---}The February 2015 data release included polarization data at 30, 44, 70, and 353\,GHz. 
The release of the remaining three HFI channels -- 100, 143, and 217\,GHz -- was delayed because of residual systematic errors in the polarization data, particularly but not exclusively at $\ell < 10$.  The sources of these systematic errors were identified, but insufficiently characterized to support reliable scientific analyses of such things as the optical depth to ionization $\tau$ and the isotropy and statistics of the polarization fluctuations.  Due to an internal mixup, however, the unfiltered polarized sky maps ended up in PLA instead of the high-pass-filtered ones. This was discovered in July 2015, and the high-pass-filtered maps at 100, 143, and 217\,GHz were added to the PLA.  The unfiltered maps have been left in place to avoid confusion, but warnings about their unsuitability for science have been added.  
Since February our knowledge of the causes of residual systematic errors and our characterization of the polarization maps have improved.  Problems that will be encountered in the released 100--353\,GHz maps include the following:

\begin{itemize}

\item Null tests on data splits indicate inconsistency of polarization measurements on large angular scales at a level much larger than our instrument noise model (see Fig.~10 of \citealt{planck2014-a09}).  The reasons for this are numerous and will be described in detail in a future paper. 

\item While analogue-to-digital converter (ADC) nonlinearity is corrected to a much better level than in previous releases, some residual effects remain, particularly in the distortion of the dipole that leaks dipole power to higher signal frequencies.

\item Bandpass mismatches leak dust temperature to polarization, particularly on large angular scales.

\item While the measured beam models are improved, main beam mismatches cause temperature-to-polarization leakage in the maps (see Fig.~17 of \citealt{planck2014-a08}).  In producing the results given in the \Planck\ 2015 release, we correct for this at the spectrum level \citep{planck2014-a13}, but the public maps contain this effect.

\end{itemize}

The component separation work described in Sect.~9, \citealt{planck2014-a11}, and \citealt{planck2014-a12} was performed on all available data, and produced unprecedented full-sky polarization maps of foreground emission (Figs.~\ref{fig:Psynch} and \ref{fig:Pdust}), as well as maps of polarized CMB emission. The polarized CMB maps, derived using four independent component separation methods, were the basis for quantitative statements about the level of residual polarization systematics and the conclusion that reliable science results could not be obtained from them on the largest angular scales.

Recent improvements in mapmaking methodology that reduce the level of residual systematic errors in the maps, especially at low multipoles, will be described in a future paper.  A more fundamental ongoing effort aimed at correcting systematic polarization effects in the time-ordered data will produce the final legacy \Planck\ data, to be released in 2016.

\section{Papers accompanying the 2015 release }

The characteristics, processing, and analysis of the \Planck\ data, as well as
a number of scientific results, are described in a series of papers released
with the data.  The titles of the papers begin with ``\Planck\ 2015 results.'',
followed by the specific titles below.  

\def\CPPtitle#1{\vskip 2pt\noindent\vbox{\raggedright\hsize=\columnwidth\hangafter=1\hangindent=3em\noindent\strut#1\strut\par}}
\bigskip

\CPPtitle{I. Overview of products and scientific results (\textit{this paper})}
\CPPtitle{II. Low Frequency Instrument data processing}
\CPPtitle{III. LFI systematic uncertainties}
\CPPtitle{IV.  LFI beams and window functions}
\CPPtitle{V. LFI calibration}
\CPPtitle{VI. LFI maps}
\CPPtitle{VII. High Frequency Instrument data processing: Time-ordered information and beam processing}
\CPPtitle{VIII. High Frequency Instrument data processing: Calibration and maps}
\CPPtitle{IX. Diffuse component separation: CMB maps}
\CPPtitle{X. Diffuse component separation: Foreground maps}
\CPPtitle{XI. CMB power spectra, likelihoods, and robustness of parameters}
\CPPtitle{XII. Simulations}
\CPPtitle{XIII. Cosmological parameters}
\CPPtitle{XIV. Dark energy and modified gravity}
\CPPtitle{XV. Gravitational lensing}
\CPPtitle{XVI. Isotropy and statistics of the CMB}
\CPPtitle{XVII. Constraints on primordial non-Gaussianity}
\CPPtitle{XVIII. Background geometry and topology of the Universe}
\CPPtitle{XIX. Constraints on primordial magnetic fields}
\CPPtitle{XX. Constraints on inflation}
\CPPtitle{XXI. The integrated Sachs-Wolfe effect}
\CPPtitle{XXII. A map of the thermal Sunyaev-Zeldovich effect}
\CPPtitle{XXIII. The thermal Sunyaev-Zeldovich effect--cosmic infrared background correlation}
\CPPtitle{XXIV. Cosmology from Sunyaev-Zeldovich cluster counts}
\CPPtitle{XXV. Diffuse, low-frequency Galactic foregrounds}
\CPPtitle{XXVI. The Second Planck Catalogue of Compact Sources}
\CPPtitle{XXVII. The Second Planck Catalogue of Sunyaev-Zeldovich Sources}
\CPPtitle{XXVIII. The Planck Catalogue of Galactic Cold Clumps}

\bigskip

This paper contains an overview of the main aspects of the \Planck\ project that have contributed to the 2015 release, and points to the papers that contain full descriptions.  It proceeds as follows.  Section~\ref{sec:FFP} describes the simulations that have been generated to support the analysis of \Planck\ data.  Section~\ref{sec:DataProcessing} describes the basic processing steps leading to the generation of the \Planck\ timelines.  Section~\ref{sec:Timelines} describes the timelines themselves.  Section~\ref{sec:FreqMaps} describes the generation of the nine \Planck\ frequency maps and their characteristics.  Section~\ref{sec:CMBProds} describes the \Planck\ 2015 products related to the cosmic microwave background, namely the CMB maps, the lensing products, and the likelihood code.  Section~\ref{sec:AstroProds} describes the \Planck\ 2015 astrophysical products, including catalogues of compact sources and maps of diffuse
foreground emission.  Section~\ref{sec:CMBcosmology} describes the main cosmological science results based on the 2015 CMB products.
Section~\ref{sec:astroresults} describes some of the astrophysical results based on the 2015 data.  Section~\ref{sec:summary} concludes with a summary and a look towards the next generation of \Planck\ products.

\section{Simulations}
\label{sec:FFP}

We simulate time-ordered information (TOI) for the full focal plane (FFP) for the nominal mission.  The first five FFP realizations were less comprehensive and were primarily used for validation and verification of the \Planck\ analysis codes and for cross-validation of the data processing centres' (DPCs) and FFP simulation pipelines. The first \Planck\ cosmology results \citep{planck2013-p01} were supported primarily by the sixth FFP simulation-set, hereafter FFP6. The current results were supported by the next generation of simulations, FFP8, which is described in detail in \cite{planck2014-a14}.

Each FFP simulation comprises a single ``fiducial'' realization (CMB, astrophysical foregrounds, and noise), together with separate Monte Carlo (MC) realizations of the CMB and noise. The CMB component contains the effect of our motion with respect to the CMB rest frame. This induces an additive dipolar aberration, a frequency-dependent dipole modulation, and a frequency-dependent quadrupole in the CMB data. Of these effects, the additive dipole and frequency-independent component of the quadrupole are removed (see \citealt{planck2014-a14} for details), the residual quadrupole, aberration, and modulation effects are left in the simulations and are also left in the LFI and HFI data.  To mimic the \Planck\ data as closely as possible, the simulations use the actual pointing, data flags, detector bandpasses, beams, and noise properties of the nominal mission.  For the fiducial realization, maps were made of the total observation (CMB, foregrounds, and noise) at each frequency for the nominal mission period, using the Planck Sky Model \citep{delabrouille2012}.  In addition, maps were made of each component separately, of subsets of detectors at each frequency, and of half-ring and single Survey subsets of the data.  The noise and CMB Monte Carlo realization-sets also included both all and subsets of detectors (so-called ``DetSets'') at each frequency, and full and half-ring data sets for each detector combination. 

To check that the PR2-2015 results are not sensitive to the exact cosmological parameters used in FFP8, we subsequently generated FFP8.1, exactly matching the PR2-2015 cosmology.

All of the FFP8 and FFP8.1 simulations are available to be used at NERSC (\url{http://crd.lbl.gov/cmb-data}); in addition, a limited subset of the simulations are available for download from the PLA.

\section{Data Processing}
\label{sec:DataProcessing}

\subsection{Timeline processing}

\subsubsection{LFI}
\label{sec:LFITOI}

The main changes in the LFI data processing compared to the earlier release \citep{planck2013-p02} are related to the way in which we take into account the beam information in the pipeline processing, as well as an entire overhaul of the iterative algorithm used to calibrate the raw data.  The process starts at Level~1, which retrieves all the necessary information from data packets and auxiliary data received from the Mission Operation Centre, and transforms the scientific packets and housekeeping data into a form manageable by
Level~2.  Level~2 uses scientific and housekeeping information to:

\begin{itemize}

\item build the LFI reduced instrument model (RIMO), which contains the main characteristics of the instrument;

\item remove ADC non-linearities and 1\,Hz spikes diode by diode;

\item compute and apply the gain modulation factor to minimize $1/f$ noise;

\item combine signals from the diodes with associated weights;

\item compute the appropriate detector pointing for each sample, based on auxiliary data and beam information, corrected by a model (PTCOR) built using solar distance and radiometer electronics box assembly (REBA) temperature information;

\item calibrate the scientific timelines to physical units ($\mathrm{K}_\mathrm{CMB}$), fitting the total CMB dipole convolved with the
$4\pi$ beam representation, without taking into account the signature due to Galactic straylight;

\item remove the solar and orbital dipole (convolved with the $4\pi$ beam) representation and the Galactic emission (convolved with the beam sidelobes) from the scientific calibrated timeline;

\item combine the calibrated time-ordered information (TOI) into aggregate products, such as maps at each frequency.

\end{itemize}

Level~3 collects Level~2 outputs from both HFI \citep{planck2014-a09} and LFI and derives various products, such as component-separated maps of astrophysical foregrounds, catalogues of different classes of sources, and the likelihood of cosmological and astrophysical models given in the maps.

\subsubsection{HFI}
\label{sec:HFITOI}

The HFI data processing for this release is very similar to that used for the
2013 release~\citep{planck2013-p03}. The main improvement is carried out in
the very first step in to the pipeline, namely the correction for ADC
non-linearities.

The HFI bolometer electronic readout, described in the Planck Explanatory
Supplement \citep{planck2014-ES}, ends with a 16-bit Analogue-to-Digital Converter.  Its tolerance on
the differential non-linearity (the maximum deviation from one least
significant bit, LSB, between two consecutive levels, on the whole range)
is specified to be better than $\pm$1.6\,LSB.  The consequences of this
feature on HFI performances had not been anticipated, nor did it produce any
detected effect on ground-test data, but it proved to be a major systematic
effect impacting the flight data.  A method that reduces the ADC effect by
more than an order of magnitude for most channels has been implemented. 

No changes were made to any software module involved in the TOI processing,
from ADC-corrected TOI to clean TOI that are ready for qualification,
calibration and mapmaking.  However, several input parameters of the modules
have been fine-tuned for better control of some residual systematic errors
that were noticed in the 2013 data. 

\begin{figure*}[htpb]
\begin{center}
  \includegraphics[width=\textwidth]{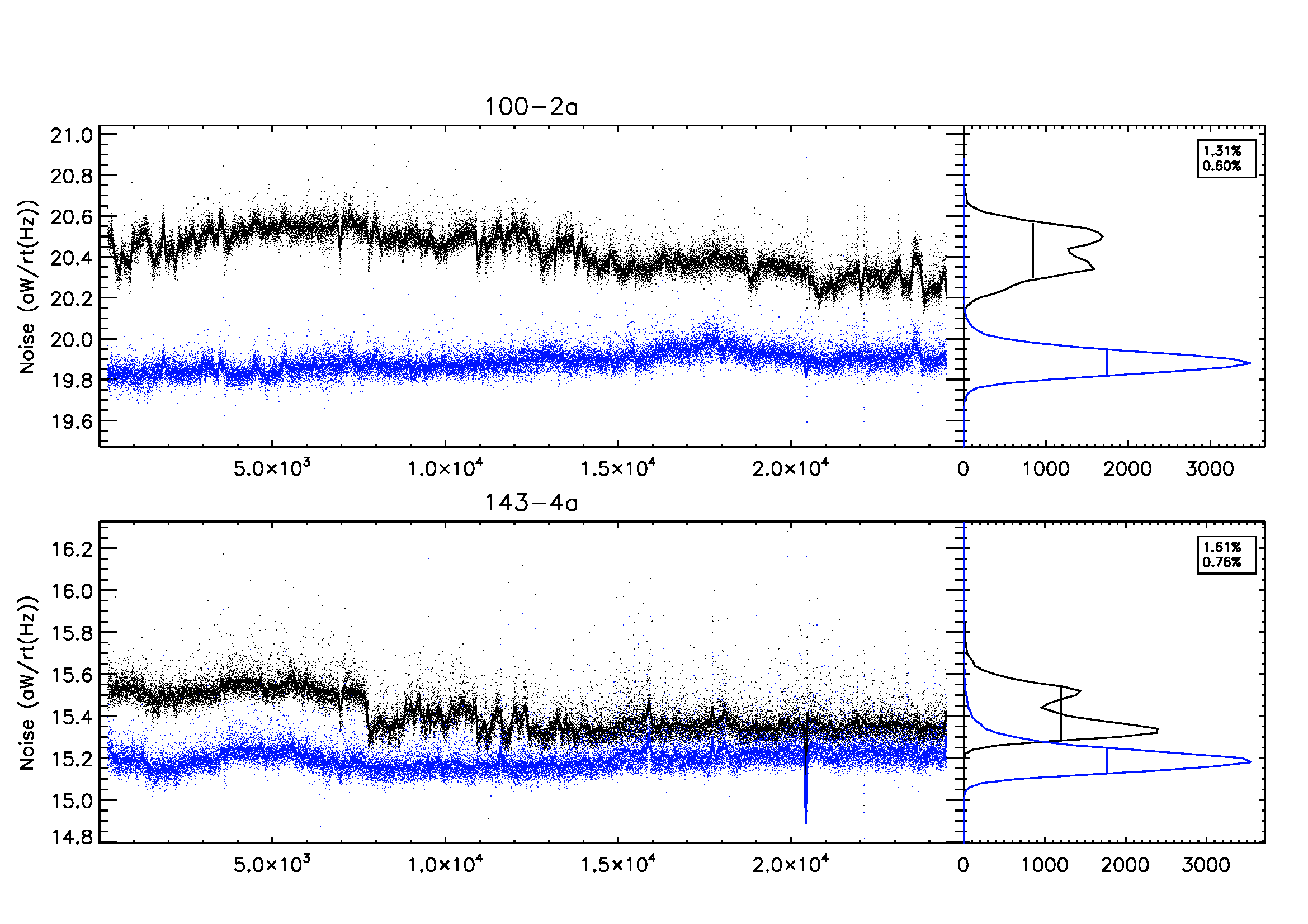}
\caption{\label{fig:noise_statio} Noise stationarity for a selection of two
bolometers. The left panels show the total noise trends for each bolometer
(dots). The solid line shows a running box average. The black dots are from
the 2013 data release and the blue dots concern this release. The right panels
show a histogram of the trends on the left. The box gives the width of the
distribution at half maximum, as measured on the histogram, normalized to the
mean noise level. The time response deconvolution has changed between the two
data release and hence the absolute noise level is different.}
\end{center}
\end{figure*}

Improvements can be assessed by comparing the noise stationarity in the 2013
and 2015 data. Trends of the so-called total noise versus ring number before
(black dots, 2013 release) and after the ADC correction (blue dots, this
release) are shown in Fig.~\ref{fig:noise_statio}. There is a significant
decrease in the relative width of the distribution when the ADC correction is
included.  For most bolometers, the noise stationarity is ascertained to be
within the percent level \citep{planck2014-a09}. 

For strong signals, the threshold for cosmic ray removal (``deglitching'') is
auto-adjusted to cope with source noise, due to the small pointing drift
during a ring.  Thus, more glitches are left in data in the vicinity of bright
sources, such as the Galactic centre, than are left elsewhere. To mitigate this
effect near bright planets, the signal at the planet location is flagged and
interpolated prior to the TOI processing. For the 2015 release, this is done
for Jupiter at all HFI frequency bands, for Saturn at
$\nu\ge 217\,\mathrm{GHz}$ and for Mars at $\nu\ge 353\,\mathrm{GHz}$.

Nevertheless, for beam and calibration studies (see Sect.~\ref{sec:HFIbeams},
\citealt{planck2014-a08} and \citealt{planck2014-a09}),
the TOI of all planet crossings, including the planet signals,
are needed at all frequencies. Hence, a dedicated production is done in
parallel for those pointing periods and bolometers. In that case, in order to
preserve the quality of the deglitching, an iterative 3-level deglitcher
has been added in the 2015 data analysis.

As noted in \cite{planck2013-p01}, \Planck\ scans a given ring on the sky for
roughly 45 minutes before moving on to the next ring. The data between these
rings, taken while the spacecraft spin-axis is moving, are discarded as
``unstable.'' The data taken during the intervening ``stable'' periods are
subjected to a number of statistical tests to decide whether they should be
flagged as usable or not \citep{planck2013-p03}; this procedure continues to
be used for the present data release. An additional selection process has been
introduced to mitigate the effect of the 4-K lines (i.e., periodic cooler
variations) on the data, especially the 30\,Hz line signal, which is correlated
across bolometers. It is therefore likely that the 4-K line-removal procedure
leaves correlated residuals on the 30\,Hz line. The consequence of this
correlation is that the angular cross-power spectra between different
detectors can show excess power at multipoles around $\ell\approx1800$.
To mitigate this effect, we discard all 30\,Hz resonant rings for the 16
bolometers between 100 and 353\,GHz for which the median average of the
30\,Hz line amplitude is above $10\,\mathrm{aW}$.  As a result, the
$\ell\approx1800$ feature has now disappeared.

\begin{figure*}[htpb]
\begin{center}
\includegraphics[width=18cm]{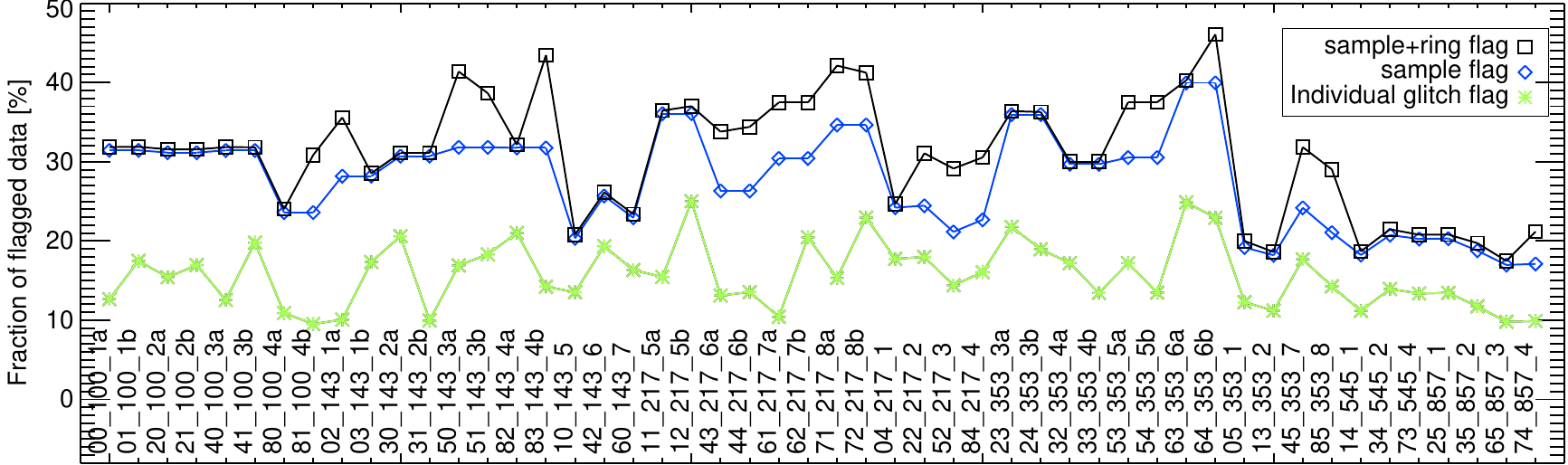}
\caption{Fraction of discarded data per bolometer (squares with thick black
line). The fraction of data discarded from glitch-flagging alone is shown with
stars and the thin green line. The blue line with diamonds indicates the
average fraction of discarded samples in valid rings. The two RTS bolometers
(143\_8 and 545\_3) are not shown, since they are not used in the data
processing.}
\label{fig:DiscardedData}
\end{center}
\end{figure*}

Figure~\ref{fig:DiscardedData} summarizes the situation, showing the fraction
of discarded samples for each detector over the full mission. It gathers the
flags at the sample level (blue line), which are mainly due to glitches
(green line) plus the pointing maneuvers between rings (about 8\,\%) and the glitch
flag combination for the polarization-sensitive bolometers (PSBs)
and secondly, at the ring level (black line),
which are mostly due to the 4-K lines, but also due to Solar flares, big
manoeuvres, and end-of-life calibration sequences, which are common to all
detectors. With respect to the nominal mission, presented in the 2013 papers,
the main difference appears in Survey~5, which is somewhat disjointed, due to
Solar flares arising from the increased Solar activity, and to special
calibration sequences. The full cold \Planck\ HFI mission lasted 885~days,
excluding the Calibration and Performance Verification (CPV) period of 1.5
months.  Globally, for this duration, the total amount of HFI data discarded amounted to 31\,\%, the
majority of which came from glitch flagging.

Details of the TOI processing are given in the \cite{planck2014-a08}.

\subsection{Beams}
\label{sec:Opt}

\subsubsection{LFI beams}
\label{sec:LFIbeams}

As described in \cite{planck2014-a05}, the in-flight assessment of the LFI main
beams relied on the measurements performed during seven Jupiter crossings: the
first four transits occurred in nominal scan mode (spin shift 2\arcm,
1\deg\,day\mo); and the last three scans in deep mode (spin shift 0\parcm5,
15\arcm\,day\mo).  By stacking data from the seven scans, the main beam
profiles are measured down to $-$25\,dB at 30 and 44\,GHz, and down to
$-$30\,dB at 70\,GHz.  Fitting the main beam shapes with an elliptical Gaussian
profile, we have expressed the uncertainties of the measured scanning beams in
terms of statistical errors for the Gaussian parameters: ellipticity;
orientation; and FWHM.  In this release, the error on the reconstructed beam
parameters is lower with respect to that in the 2013 release.  Consequently,
the error envelope on the window functions is lower as well.  For example, the
beam FWHM is determined with a typical uncertainty of 0.2\,\% at 30 and
44\,GHz, and 0.1\,\% at 70\,GHz, i.e., a factor of two better than the value
achieved in 2013.

The scanning beams\footnote{The term ``scanning beam" refers to the angular response of a single detector to a compact source, including the optical beam and (for HFI) the effects of time domain filtering.  In the case of HFI, a Fourier filter deconvolves the bolometer/electronics time response and lowpass-filters the data.  In the case of LFI, the sampling tends to smear signal in the time domain.  The term ``effective beam'' refers to a beam defined in the map domain, obtained by averaging the scanning beams pointing at a given pixel of the sky map taking into account the scanning strategy and the orientation of the beams themselves when they point along the direction to that pixel.  See \citep{planck2013-p02d}.} used in the LFI pipeline (affecting calibration, effective
beams, and beam window functions) are based on {\tt GRASP} simulations,
properly smeared to take into account the satellite motion, and are similar to
those presented in \cite{planck2013-p02d}.  They come from a tuned optical
model, and represent the most realistic fit to the available measurements of
the LFI main beams.  In \cite{planck2013-p02d}, calibration was performed
assuming a pencil beam, the main beams were full-power main beams, and the
resulting beam window functions were normalized to unity.  For the 2015
release, a different beam normalization has been used to properly take into
account the power entering the main beam (typically about 99\,\% of the total
power).  Indeed, as described in \citet{planck2014-a06}, the current LFI
calibration takes into account the full 4$\pi$ beam (i.e., the main beam, as
well as near and far sidelobes).  Consequently, in the calculation of the
window function, the beams are not normalized to unity; instead, their
normalization uses the value of the efficiency calculated taking into account
the variation across the band of the optical response (coupling between feed
horn pattern and telescope) and the radiometric response (band shape). 

Although the {\tt GRASP} beams are computed as the far-field angular
transmission function of a linearly polarized radiating element in the focal
plane, the far-field pattern is in general not perfectly linearly polarized,
because there is a spurious component induced by the optical system, named
``beam cross-polarization.''  The Jupiter scans allowed us to measure only the
total field, that is, the co- and cross-polar components combined in
quadrature.  The adopted beam model has the added value of defining the co- and
cross-polar pattern separately, and it permits us to properly consider the beam
cross-polarization in every step of the LFI pipeline.  The {\tt GRASP} model,
together with the pointing information derived from the focal plane geometry
reconstruction, gives the most advanced and precise noise-free representation
of the LFI beams.  

The polarized main beam models were used to calculate the
effective beams$^5$, which take into account the specific scanning strategy in
order to include any smearing and orientation effects on the beams themselves.
Moreover, the sidelobes were used in the calibration pipeline to correctly
evaluate the gains and to subtract Galactic straylight from the calibrated
timelines \citep{planck2014-a03}.

To evaluate the beam window functions, we adopted two independent approaches,
both based on Monte Carlo simulations. In one case, we convolved a fiducial
CMB signal with realistic scanning beams in harmonic space to generate the
corresponding timelines and maps.  In the other case, we convolved the fiducial
CMB map with effective beams in pixel space with the {\tt FEBeCoP}
\citep{mitra2010} method. Using the first approach, we have also evaluated the
contribution of the near and far sidelobes on the window functions.  The
impact of sidelobes on low multipoles is about 0.1\,\% (for details see
\citealt{planck2014-a05}).

The error budget was evaluated as in the 2013 release and it comes from two
contributions: the propagation of the main beam uncertainties throughout the
analysis; and the contribution of near and far sidelobes in the Monte Carlo
simulation chain.  The two error sources have different relevance, depending
on the angular scale. Ignoring the near and far sidelobes is the dominant
error at low multipoles, while the main beam uncertainties dominate the total
error budget at $\ell\geq 600$.  The total uncertainties in the effective beam
window functions are 0.4\,\% and 1\,\% at 30 and 44\,GHz, respectively
(at $\ell \approx 600$), and 0.3\,\% at 70\,GHz at $\ell \approx 1000$.

\subsubsection{HFI beams}
\label{sec:HFIbeams}

The HFI main beam measurement is described in detail in \cite{planck2014-a08}
and is similar to that of \cite{planck2013-p03c}, although with several
important changes.  The HFI scanning beam model is a ``Bspline'' decomposition
of the time-ordered data from planetary observations.  The domain of
reconstruction of the main beam is extended from a 40\arcm\ square to a
100\arcm\ square and is no longer apodized in order to preserve near sidelobe
structure in the main beam model \cite{planck2013-p01a}, as well as to
incorporate residual time-response effects into the beam model.  A combination
of Saturn and Jupiter data (rather than Mars data) is used for an improved
signal-to-noise ratio, and a simple model of diffraction (consistent with
physical optics predictions) is used to extend the beam model below the noise
floor from planetary data.  A second stage of cosmic ray glitch removal is
added to reduce bias from unflagged cosmic ray hits.

The effective beams and effective beam window functions are computed using the
{\tt FEBeCoP} and {\tt Quickbeam} codes, as in \cite{planck2013-p03c}.  While
the scanning beam measurement produces a total intensity map only, effective
beam window functions appropriate for both temperature and polarixed angular
power spectra are produced by averaging the individual detector window
functions weighed by temperature sensitivity and polarization sensitivity.
Temperature-to-polarization leakage due to main beam mismatch is subdominant
to noise in the polarization measurement, and is corrected as an additional
nuisance parameter in the likelihood.

The uncertainty in the beam measurement is derived from an ensemble of 100
Monte Carlo simulations of the planet observations that include random
realizations of detector noise, cosmic ray hits, and pointing uncertainty
propagated through the same pipeline as the data to simulated scanning beam
products and simulated effective beam window functions. The error is expressed
in multipole space as a set of error eigenmodes, which capture the correlation
structure of the errors.  Additional consistency checks are performed to
validate the error model, splitting the planet data to construct Year~1 and
Year~2 beams and to create Mars-based beams.  With improved control of
systematics and higher signal-to-noise ratio, the uncertainties in the HFI
beam window function have decreased by more than a factor of 10 relative to
the 2013 release.

Several differences between the beams in 2013 and 2015 may be listed.

\begin{itemize}

\item {\em Finer polar grid.} Instead of the cartesian grid 40\arcm\ on each
side used previously, the beam maps were produced on both a cartesian grid of
200\arcm\ on each side and 2\arcs\ resolution, and a polar grid with a radius
of 100\arcm\ and a resolution of 2\arcs\ in radius and 30\arcm\ in azimuth.
The latter grid has the advantage of not requiring any extra interpolation to
compute the beam spherical harmonic coefficients $b_{\ell m}$ required by
{\tt quickbeam}, and therefore improves the accuracy of the resulting
$B(\ell)$. 

\vskip 3pt

\item {\em Scanning beam elongation.} To account for the elongation of the
scanning beam induced by the time response deconvolution residuals, the
{\tt quickbeam} computations are conducted with the $b_{\ell m}$ for
$-6 \le m \le 6$. We checked that the missing terms account for less than
$10^{-4}$ of the effective $B^2(\ell)$ at $\ell=2000$. Moreover, spotcheck
comparisons with the effective $B(\ell)$ obtained by {\tt FEBeCoP} show very
good agreement.

\vskip 3pt

\item {\em Finite size of Saturn .} Even though its rings seem invisible at
\Planck\ frequencies (and unlike Mars), Saturn has an angular size that must be
accounted for in the beam window function.  The planet was assumed to be a
top-hat disc of radius 9.5\arcs\ at all HFI frequencies, whose window function
is well approximated by that of a 2D Gaussian profile of FWHM 11\parcs185; the
effective $B(\ell)$ were therefore divided by that window function.

\vskip 3pt

\item {\em Cut sky and pixel shape variability.} The effective beam window
functions do not include the (nominal) pixel window function, which must be
accounted for separately when analysing \Planck\ maps. However, the shape and
individual window function of the {\tt HEALPix} \cite{gorski2005}
pixels have large-scale variations around their nominal values across the sky.
These variations impact the effective beam window functions applicable to
\Planck\ maps, in which the Galactic plane has been masked more or less
conservatively, and are included in the effective $B(\ell)$ that are provided.

\vskip 3pt

\item {\em Polarization and detector weights.} Each 143, 217 and 353\,GHz
frequency map is a combination of measurement by polarization-sensitive and
polarization-insensitive detectors, each having a different optical response.
As a consequence, at each of these frequencies, the $Q$ and $U$ maps will have
a different beam window function than the $I$ map.  When cross-correlating
the 143 and 217\,GHz maps for example, the $TT$, $EE$, $TE$, and $ET$ spectra
will each have a different beam window function.

\vskip 3pt

\item {\em Polarization and beam mismatch.} Since polarization measurements are differential by nature, any mismatch in the effective beams of the detectors involved will couple with temperature anisotropies to create spurious polarization signals \citep{hhz2003}.  In the likelihood pipeline \citep{planck2014-a13} this additive leakage is modelled as a polynomial whose parameters are fit on the power spectra.

\vskip 3pt

\item {\em Beam error model.} See above. The improved S/N compared to 2013
leads to smaller uncertainties.  At $\ell=1000$ the uncertainties on
$B^2_{\ell}$ are $2.2\times10^{-4}$, $0.84\times10^{-4}$, and
$0.81\times10^{-4}$ for 100, 143, and 217\,GHz, respectively.  At $\ell=2000$,
they are $11\times10^{-4}$, $1.9\times10^{-4}$, and $1.3\times10^{-4}$.

\end{itemize}

A reduced instrument model (RIMO) containing the effective $B(\ell)$ for
temperature and polarization detector assemblies will be provided, for both
auto- and cross-spectra.  The RIMO will also contain the beam error eigenmodes
and their covariance matrices.

\subsection{Focal plane geometry and pointing}
\label{sec:FocPlaGeo}

The focal plane geometry of LFI was determined independently for each Jupiter
crossing \citep{planck2014-a05}, using the same procedure adopted in the 2013
release. The solutions for the seven crossings agree within 4\arcs\ at 70\,GHz
(and 7\arcs\ at 30 and 44\,GHz).  The uncertainty in the determination of the
main beam pointing directions evaluated from the single scans is about 4\arcs\
for the nominal scans, and 2\parcs5 for the deep scans at 70\,GHz (27\arcs\
for the nominal scan and 19\arcs\ for the deep scan, at 30 and 44\,GHz).
Stacking the seven Jupiter transits, the uncertainty in the reconstructed main
beam pointing directions becomes 0\parcs6 at 70\,GHz and 2\arcs\ at 30 and
44\,GHz.  With respect to the 2013 release, we have found a difference in the
main beam pointing directions of about 5\arcs\ in the cross-scan direction and
0\parcs6 in the in-scan direction.

Throughout the extended mission, \Planck\ continued to operate star camera
STR1, with the redundant unit, STR2, only briefly swapped on for testing. No
changes were made to the basic attitude reconstruction. We explored the
possibility of updating the satellite dynamic model and using the fibre-optic
gyro for additional high frequency attitude information. Neither provided
significant improvements to the pointing and were actually detrimental to
overall pointing performance; however, they may become useful in future attempts to 
recover accurate pointing during the ``unstable" periods.

\begin{figure}[htbp]
\centering
\includegraphics[width=\columnwidth]{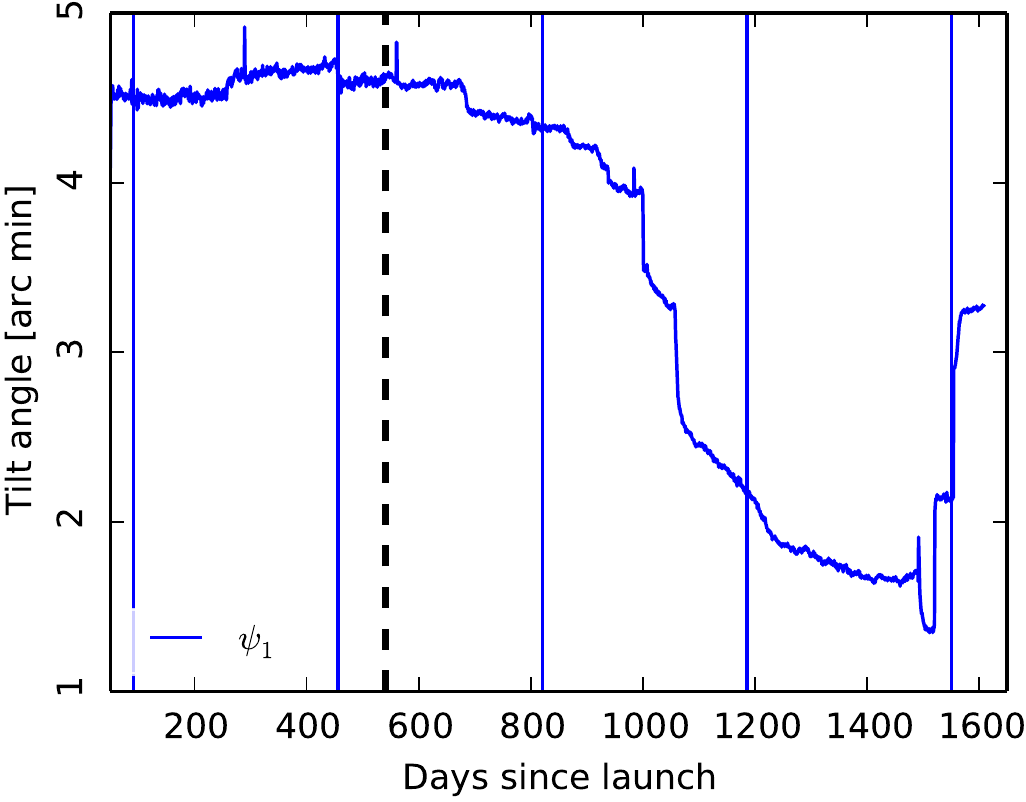}
\includegraphics[width=\columnwidth]{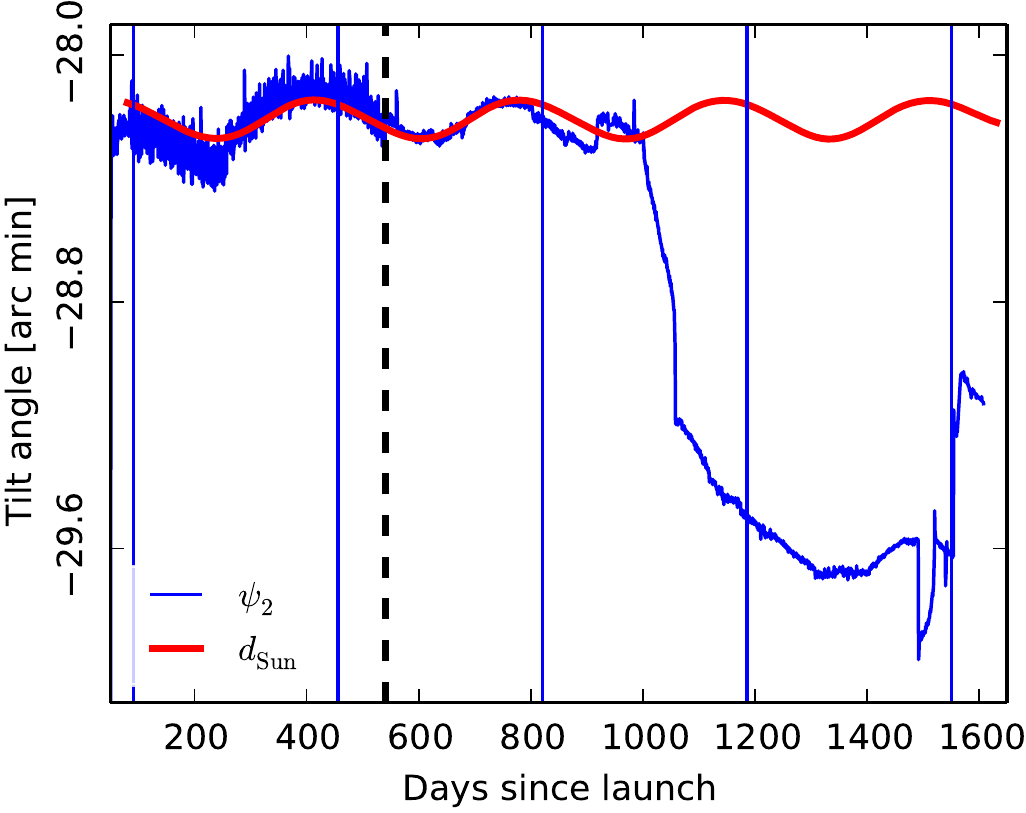}
\caption{Reconstructed tilt (wobble) angles between the satellite body frame
 and the principal axis frame. Vertical blue lines mark the ends of operation
 years and the dashed black line indicates day 540 after launch, when the
 thermal control on the LFI radiometer electronics box assembly (REBA) was
 adjusted.
  \emph{Top:}
  First angle, $\psi_1$, corresponds to a rotation about the satellite axis
  just 5\deg\ off the focal plane centre. Observed changes in $\psi_1$ only
  have a small impact on focal plane line-of-sight.
  \emph{Bottom:}
  Second angle, $\psi_2$, is perpendicular to a plane defined by the nominal
  spin axis and the telescope line of sight. Rotation in $\psi_2$ immediately
  impacts the opening angle and thus the cross-scan position of the focal
  plane. We also plot a scaled and translated version of the Solar distance
  that correlates well with $\psi_2$ until the reconstructed angles became
  compromised around day 1000 after launch.
} \label{fig:wobble_angles}
\end{figure}

Attitude reconstruction delivers two quantities: the satellite body reference
system attitude; and the angles between it and the principal axis reference
system (so-called ``tilt'' or ``wobble'' angles). The tilt angles are needed
to reconstruct the focal plane line-of-sight from the raw body reference frame
attitude. For unknown reasons, the reconstructed tilt angles became irregular
at the start of the LFI-only extension (cf.~Fig.~\ref{fig:wobble_angles}).
Starting near day 1000 after launch, the tilt angles began to indicate a drift
that covered $1\parcm5$ over about a month of operations. We found that the
drift was not present in observed planet positions and we were therefore forced
to abandon the reconstructed tilt angles and include the tilt correction into
our ad hoc pointing correction, PTCOR.

\begin{figure}[htbp]
\centering
\includegraphics[width=\columnwidth]{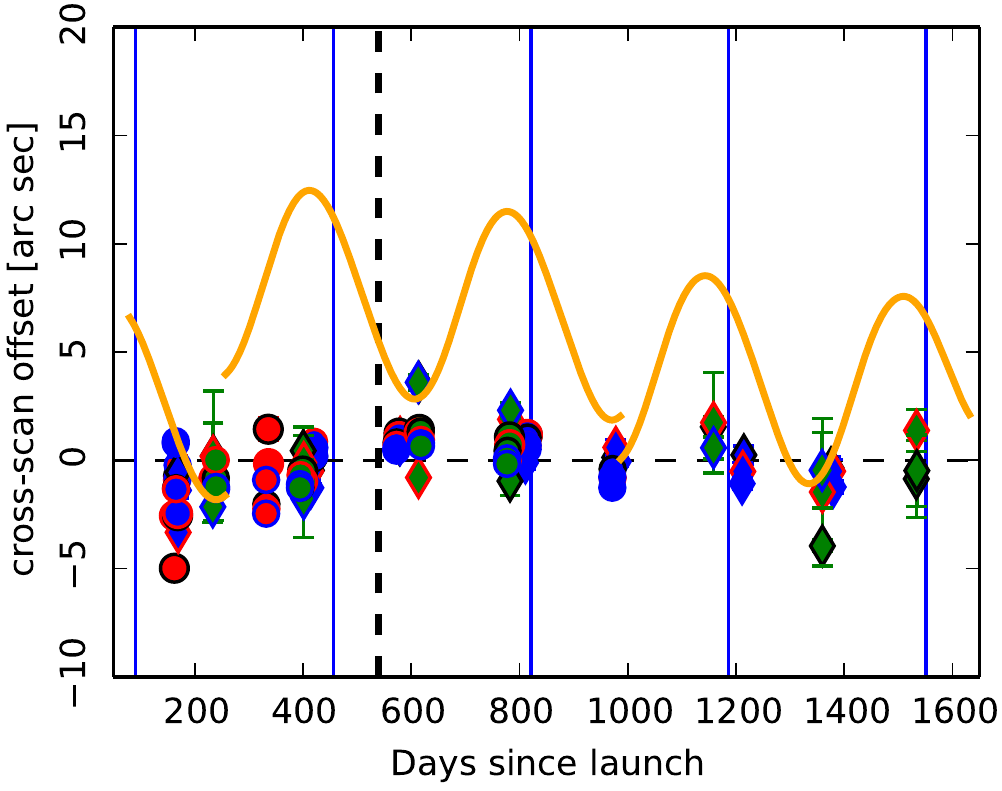}
\includegraphics[width=\columnwidth]{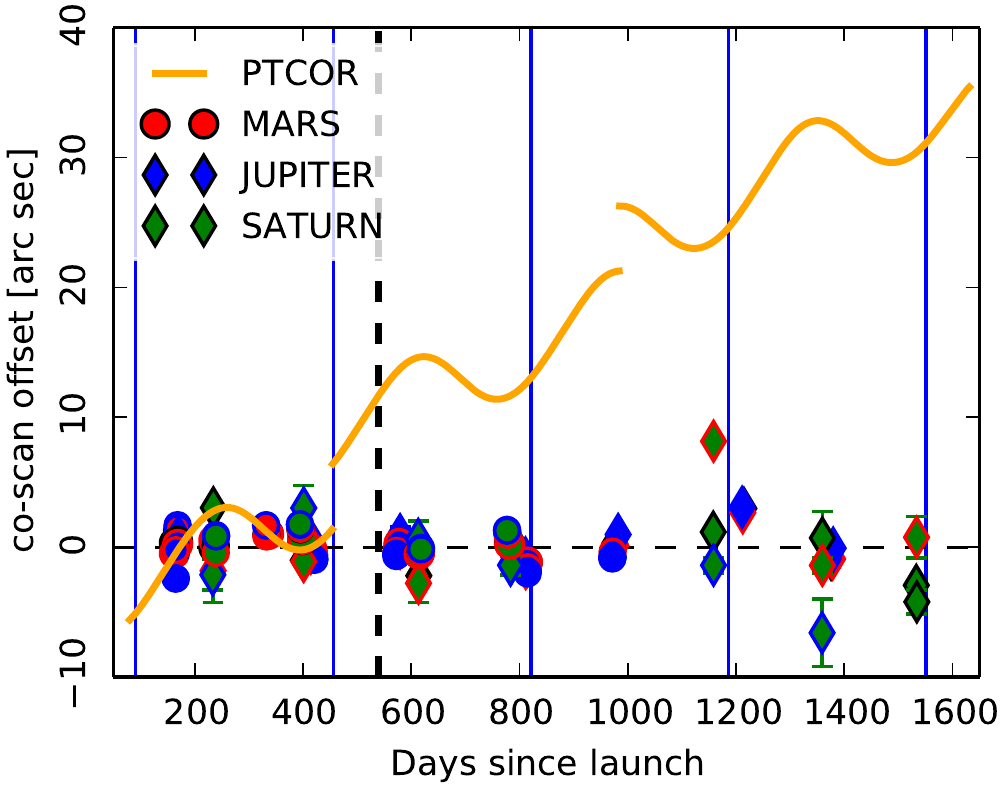}
\caption{
  Our ad hoc pointing correction, PTCOR, and a selection of observed planet
  position offsets after applying the correction.
  \emph{Top:}
  Cross-scan pointing offset. This angle is directly affected by the second
  tilt angle, $\psi_2$, in Fig.~\ref{fig:wobble_angles}.
  \emph{Bottom:}
  In-scan pointing offset. This angle corresponds to the spin phase and
  matches the third satellite tilt angle, $\psi_3$. Since $\psi_3$ is poorly
  resolved by standard attitude reconstruction, the in-scan pointing was
  already driven by PTCOR in the 2013 release.
} \label{fig:PTCOR}
\end{figure}

We noticed that the most significant tilt angle corrections prior to the LFI
extension tracked well the distance, $d_{\rm Sun}$, between the Sun and \Planck\
(see Fig.~\ref{fig:wobble_angles}, bottom panel), so we decided to replace the
spline fitting from 2013 with the use of the Solar distance as a fitting
template. The fit was improved by adding a linear drift component and
inserting breaks at events known to disturb the spacecraft thermal enviroment.
In Fig.~\ref{fig:PTCOR} we show the co- and cross-scan pointing corrections
and a selection of planet position offsets after the correction was applied.
The template-based pointing correction differs only marginally from the 2013
PTCOR, but an update was absolutely necessary to provide consistent, high
fidelity pointing for the entire \Planck\ mission.

Finally we addressed the LFI radiometer electronics box assembly (REBA)
interference that was observed in the 2013 release by constructing, fitting,
and subtracting another template from the REBA thermometry. This greatly
reduced short timescale pointing errors prior to REBA thermal tuning on day~540
after launch. The REBA template removal reduced the pointing period timescale
errors from $2\parcs7$ to $0\parcs8$ (in-scan) and $1\parcs9$ (cross-scan).

\subsection{Calibration}
\label{sec:GainCal}

In this section we compare the relative photometric calibration of the all-sky
CMB maps between LFI and HFI, as well as between \Planck\ and WMAP. The two
\Planck\ instruments use different technologies and are subject to different
foregrounds and systematic effects. The \Planck\ and WMAP measurements overlap
in frequency range, but have independent spacecraft, telescopes, and scanning
strategies. Consistency tests between these three data sets are very demanding
tests of the control of calibration, transfer functions, systematic effects,
and foreground contamination.

\subsubsection{The orbital dipole}
\label{sec:orbitaldipole}

In the 2013 data release, photometric calibration from 30 to 353\,GHz was based on the ``Solar dipole'', that is, the dipole induced in the CMB by the motion of the Solar System barycentre with respect to the CMB.  We used the value of the dipole measured by WMAP5 (\citealt{hinshaw2009}; \citealt{jarosik2011}).  

In the 2015 data release, photometric calibration of both LFI and HFI is based on the ``orbital dipole,'' i.e., the modulation of the Solar dipole induced by the orbital motion of the satellite around the Solar System barycentre.  By using this primary calibrator, we can derive for each \Planck\ detector (or combination of detectors) an independent measurement of the Solar dipole, which is then used in the \Planck\ calibration pipeline.  The orbital motion is known with exquisite accuracy, making the orbital dipole potentially the most accurate calibration source in all of astrophysics, limited ultimately by the accuracy of the temperature of the CMB.  The amplitude of this modulation, however, is only about 250\muK, varying with the details of the satellite motion, an order of magnitude smaller than the Solar dipole.  Realizing its advantages as a fundamental calibration source requires low noise and good control of foregrounds, sidelobes, and large-angular-scale systematics.  For the 2015 release, improvements in the control of systematic effects and foregrounds on both LFI and HFI, including the availability of 2.5 and 4 orbital cycles for HFI and LFI, respectively (compared to 1.25 cycles in the 2013 release), have allowed accurate calibration of both instruments on the orbital dipole, summarized in the following subsections and described in detail in \citet{planck2014-a03} and \citet{planck2014-a09}. The dipole component of the CMB and the frequency-independent part of the quadrupole (induced by the Solar dipole) are removed from both the LFI and HFI data; however, higher order effects of the Solar dipole \citep[see][]{planck2013-pipaberration} are left in the data, which is matched by what is contained in the
simulations \citet{planck2014-a14}.

With the 2015 data calibrated on the orbital dipole, \Planck\ has made independent measurements of the Solar dipole (Table~\ref{tab:dipole}), which can be compared to the WMAP5 measurement \citep{hinshaw2009}.  Amplitudes agree within 0.28\,\%; directions agree to better than 2\arcm.  Although the difference in amplitude between the \Planck\ and the WMAP5 measurements of the Solar dipole is small and within uncertainties, it had non-negligible consequences in 2013.  WMAP was calibrated on the orbital dipole, so errors in its Solar dipole measurement did not contribute to its overall calibration errors.  \Planck\ in 2013, however, was calibrated on the WMAP5 Solar dipole, which is 0.28\,\% lower than the orbital-dipole-calibrated 2015 \Planck\ measurement.  Calibrating LFI and HFI against WMAP5 in the 2013 results, therefore, resulted in 2013 gains that were 0.28\,\% {\it too low\/} for both LFI and HFI.  This factor is included in Tables~\ref{tab:LFIcal} and \ref{tab:HFIcal}.

\begin{table*}[tb]
\newdimen\tblskip \tblskip=5pt
\caption{LFI, HFI, and WMAP measurements of the Solar dipole.}
\label{tab:dipole}
\vskip -6mm
\footnotesize
\setbox\tablebox=\vbox{
 \newdimen\digitwidth
 \setbox0=\hbox{\rm 0}
 \digitwidth=\wd0
 \catcode`*=\active
 \def*{\kern\digitwidth}
  \newdimen\dpwidth
  \setbox0=\hbox{.}
  \dpwidth=\wd0
  \catcode`!=\active
  \def!{\kern\dpwidth}
\halign{\hbox to 4.5cm{#\leaderfil}\tabskip 2em&
    \hfil$#$\hfil \tabskip 2em&
    \hfil$#$\hfil \tabskip 2em&
    \hfil$#$\hfil \tabskip 0em\cr
\noalign{\doubleline}
\omit&&\multispan2\hfil\sc Galactic coordinates\hfil\cr
\noalign{\vskip -3pt}
\omit&\omit&\multispan2\hrulefill\cr
\noalign{\vskip 3pt} 
\omit&\omit\hfil\sc Amplitude\hfil&l&b\cr
\omit\hfil\sc Experiment\hfil&[\muK_{\rm CMB}]&\omit\hfil[deg]\hfil&\omit\hfil[deg]\hfil\cr
\noalign{\vskip 3pt\hrule\vskip 5pt}
\noalign{\vskip 2pt}
LFI$^{\rm a}$&                               3365.5*\pm3.0&                           264.01*\pm0.05*&          48.26*\pm0.02*\cr
HFI$^{\rm a}$&                               3364.29\pm1.1&                           263.914\pm0.013&          48.265\pm0.002\cr
\noalign{\vskip 3pt}
\bf Planck 2015 nominal\rlap{$^{\rm a}$}&\bf 3364.5*\pm2.0\rlap{$^{\rm b}$}&\bf       264.\bf00*\pm0.\bf03*&\bf 48.\bf24*\pm0.\bf02*\cr
\noalign{\vskip 3pt}
WMAP$^{\rm c}$&                              3355\rlap{$^{\rm d}$}*\phantom{.5}\pm8!*&263.99*\pm0.14*&          48.26*\pm0.03*\cr
\noalign{\vskip 3pt}
\noalign{\vskip 3pt\hrule\vskip 5pt}}}
\endPlancktablewide
\tablenote {{\rm a}} The ``nominal'' \Planck\ dipole was chosen as a plausible combination of the LFI and HFI measurements early in the analysis, to carry out subtraction of the dipole from the frequency maps (see Sect.~\ref{sec:calibrationconsistency}). The current best determination of the dipole comes from an average of 100 and 143\,GHz results \citep{planck2014-a09}.\par
\tablenote {{\rm b}} Uncertainties include an estimate of systematic errors.\par
\tablenote {{\rm c}} \citet{hinshaw2009}.\par
\tablenote {{\rm d}} See Sect.~\ref{sec:orbitaldipole} for the effect of this amplitude on \Planck\ calibration in 2013.\par
\end{table*}

\subsubsection{Instrument level calibration}
\label{sec:instrumentcalibration}

\noindent{\bf LFI---}%
There were four significant changes related to LFI calibration between the 2013 and the 2015 results. First (as anticipated in the 2013 LFI calibration paper, \citealt{planck2013-p02b}), the convolution of the beam with the overall dipole (Solar and orbital dipoles, including their induced frequency independent quadrupoles) is performed with the full $4\pi$ beam rather than a pencil beam. This dipole model is used to extract the gain calibration parameter. Because the details of the beam pattern are unique for each detector even within the same frequency channel, the reference signal used for the calibration is different for each of the 22~LFI radiometers. This change improves the results of null tests and the quality of the polarization maps. When taking into account the proper window function \citep{planck2014-a05}, the new convolution scheme leads to a shifts of +0.32, +0.03, and +0.30\,\% in gain calibration at 30, 44, and 70\,GHz, respectively (see Table 2).  Second, a new destriping code, {\tt Da Capo} \citep{planck2014-a06}, is used; this supersedes the combination of a dipole fitting routine and the {\tt Mademoiselle} code used in the 2013 data release and offers improved handing of $1/f$ noise and residual Galactic signals. Third, Galactic contamination entering via sidelobes is subtracted from the timelines after calibration. Finally, a new smoothing algorithm is applied to the calibration parameters. It adapts the length of the smoothing window depending on a number of parameters, including the
peak-to-peak amplitude of the dipole seen within each ring and sudden temperature changes in the instrument. These changes improve the results of null tests, and also lead to overall shifts in gain calibration a few per mill, depending on frequency channel. The values reported in the third column of Table 2 are approximate estimates from the combination of improved destriping, Galactic contamination removal, and smoothing. They are calculated under the simplifying assumption that these effects are completely independent of the beam convolution and can therefore be combined linearly with the latter (for more details see \cite{planck2014-a06}).

In total, these four improvements give an overall increase in gain calibration for LFI of +0.17, +0.36, and +0.54\,\% at 30, 44, and 70\,GHz, respectively. Adding the 0.28\,\% error introduced by the WMAP Solar dipole in 2013 (discussed in Sect.~\ref{sec:orbitaldipole}), for the three LFI frequency channels we find overall shifts of about 0.5, 0.6 and 0.8\,\% in gain calibration with respect to our LFI 2013 analysis (see Table 2). 

As shown in \citet{planck2014-a06}, relative calibration between LFI radiometer pairs is consistent within their statistical uncertainties.  At 70\,GHz, using the deviations of the calibration of single channels, we estimate that the relative calibration error is about 0.10\,\%.

\begin{table}[tb]
\newdimen\tblskip \tblskip=5pt
\caption{LFI calibration changes at map level, 2013 $\rightarrow$ 2015.}
\label{tab:LFIcal}
\vskip -6mm
\footnotesize
\setbox\tablebox=\vbox{
  \newdimen\digitwidth
  \setbox0=\hbox{\rm 0}
  \digitwidth=\wd0
  \catcode`*=\active
  \def*{\kern\digitwidth}
   \newdimen\dpwidth
   \setbox0=\hbox{.}
   \dpwidth=\wd0
   \catcode`!=\active
   \def!{\kern\dpwidth}
\halign{\hbox to 2.5cm{#\leaderfil}\tabskip 0.5em&
     \hfil#\hfil \tabskip 0.5em&
     \hfil#\hfil \tabskip 0.8em&
     \hfil#\hfil \tabskip 2em&
     \hfil#\hfil \tabskip 0em\cr
\noalign{\doubleline}
\omit&Beam solid&Pipeline&Orbital\cr
\noalign{\vskip 1pt}
\omit\hfil Frequency\hfil&angle&improvements\rlap{$^{\rm 
a}$}&Dipole\rlap{$^{\rm b}$}&Total\cr
\noalign{\vskip 3pt}
\omit\hfil [GHz]\hfil&[\%]&[\%]&[\%]&[\%]\cr
\noalign{\vskip 3pt\hrule\vskip 5pt}
\noalign{\vskip 2pt}
30&$0.32$&$-0.15$&0.28&0.45\cr
44&$0.03$&$+0.33$&0.28&0.64\cr
70&$+0.30$&$+0.24$&0.28&0.82\cr
\noalign{\vskip 3pt\hrule\vskip 5pt}}}
\endPlancktable
\tablenote {{a}} This term includes the combined effect of the new 
destriping code, subtraction of Galactic contamination from timelines, 
new smoothing algorithm. It has been calculated under the hypothesis 
that it is fully independent of the beam convolution.\par
\tablenote {{b}} Change from not being dependent on the amplitude error 
of the
  WMAP9 Solar dipole (Sect.~\ref{sec:orbitaldipole}).\par
\end{table}

\vskip 2mm

\noindent{\bf HFI---}%
There were three significant changes related to HFI calibration between the
2013 and the 2015 results: improved determination and handling of near and far
sidelobes; improved ADC non-linearity correction; and improved handling of
very long time constants.  The most significant changes arise from the introduction
of near sidelobes of the main beam (0\pdeg5--5\deg) that were not detected in
observations of Mars, and from the introduction of very long time constants.
We consider these in turn.

Observations of Jupiter were not used in 2013 results, because its signal is
so strong that it saturates some stages of the readout electronics. The
overall transfer function for each detector is corrected through the
deconvolution of a time transfer function, leaving a compact effective beam
used together with the maps in the science analysis.  In the subsequent
``consistency paper'' \citet{planck2013-p01a}, it was found that lower-noise
hybrid beams built by using Mars, Saturn, and Jupiter observations reveal near
sidelobes leading to significant corrections of 0.1 to 0.3\,\%.  Far sidelobes
give a very small calibration correction that is almost constant for $\ell>3$.
The zodiacal contribution was removed in the timelines, as it does not project
properly on the sky. It gives an even smaller and negligible correction, except
in the submillimetre (hereafter ``submm'') channels at 545 and 857 GHz. 

The most significant change results from the recognition of the existence of very
long time constants (VLTC) and their inclusion in the analysis.  VLTCs
introduce a significant shift in the apparent position of the dominant
anisotropy in the CMB, the Solar dipole, away from its true position.  This in
effect creates a leakage of the Solar dipole into the orbital dipole.  This disturbance is
the reason why calibration on the orbital dipole did not work as expected from
simulations, and why calibration in 2013 was instead based on the WMAP5 Solar
dipole.  As discussed in Sect.~\ref{sec:orbitaldipole}, the WMAP5 Solar dipole
was underestimated by 0.28\,\% when compared with the \Planck\ best-measured
amplitude, leading to an under-calibration of 0.28\,\% in the \Planck\ 2013
maps. With VLTCs included in the analysis, calibration on the orbital dipole
worked as expected, and gave more accurate results, while at the same time
eliminating the need for the WMAP5 Solar dipole and removing the 0.28\,\%
error that it introduced in 2013.

These HFI calibration changes are reported in Table 3. Together, they give an average shift of gain calibration of typically 1\,\% \citep{planck2014-a09} for the three CMB channels, accounting for the previously unexplained difference in calibration on the first acoustic peak observed between HFI and WMAP.
 
\begin{table}[tb]
\newdimen\tblskip \tblskip=5pt
\caption{HFI calibration changes at map level, 2013 $\rightarrow$ 2015.}
\label{tab:HFIcal}
\vskip -6mm
\footnotesize
\setbox\tablebox=\vbox{
 \newdimen\digitwidth
 \setbox0=\hbox{\rm 0}
 \digitwidth=\wd0
 \catcode`*=\active
 \def*{\kern\digitwidth}
  \newdimen\dpwidth
  \setbox0=\hbox{.}
  \dpwidth=\wd0
  \catcode`!=\active
  \def!{\kern\dpwidth}
\halign{\hbox to 2.5cm{#\leaderfil}\tabskip 2em&
    \hfil#\hfil \tabskip 1em&
    \hfil#\hfil \tabskip 2em&
    \hfil#\hfil \tabskip 1em&
    \hfil#\hfil \tabskip 2em&
    \hfil#\hfil \tabskip 0em\cr
\noalign{\doubleline}
\omit&\multispan2\hfil\sc Sidelobes \hfil&\multispan2\hfil\sc Orbital dipole\hfil\cr
\noalign{\vskip -3pt}
\omit&\multispan2\hrulefill&\multispan2\hrulefill\cr
\noalign{\vskip 2pt} 
\omit\hfil\sc Frequency\hfil&Near&Far&Dipole\rlap{$^{\rm a}$}&\sc VLTC&\sc Total\cr
\noalign{\vskip 3pt}
\omit\hfil[GHz]\hfil&[\%]&[\%]&[\%]&[\%]&[\%]\cr
\noalign{\vskip 3pt\hrule\vskip 5pt}
\noalign{\vskip 2pt}
100&0.2**&0.087&0.28&0.49&1.06\cr
143&0.2**&0.046&0.28&0.47&1.00\cr
217&0.2**&0.043&0.28&0.66&1.17\cr
353&0.275&0.006&0.28&1.5*&2.06\cr
\noalign{\vskip 3pt\hrule\vskip 5pt}}}
\endPlancktable
\tablenote {{a}} Change from not being dependent on the amplitude error of the
WMAP9 Solar dipole (Sect.~\ref{sec:orbitaldipole}).\par
\end{table}

The relative calibration between detectors operating at the same frequency is within
0.05\,\% for 100 and 143\,GHz, 0.1\,\% at 217\,GHz, and 0.4\,\% at 353\,GHz
\citet{planck2014-a09}. These levels for the CMB channels are within a factor
of 3 of the accuracy floor set by noise in low $\ell$ polarization
\citep{Tristram2011}.

The 545 and 857\,GHz channels are calibrated separately using models of planetary
atmospheric emission.  As in 2013, we used both Neptune and Uranus. The main
difference comes from better handling of the systematic errors affecting the
planet flux density measurements. Analysis is now performed on the timelines,
using aperture photometry, and taking into account the inhomogeneous spatial
distribution of the samples.  On frequency maps, we estimate statistical errors
on absolute calibration of 1.1\,\% and 1.4\,\% at 545 and 857\,GHz,
respectively, to which we add the 5\,\% systematic uncertainty arising from
the planet models. Errors on absolute calibration are therefore 6.1 and
6.4\,\% at 545 and 857\,GHz, respectively. Since the reported relative
uncertainty of the models is of the order of 2\,\%, we find the relative
calibration between the two HFI high-end frequencies to be better than 3\,\%.
Relative calibration based on diffuse foreground component separation gives
consistent numbers (see table~6 of \citealt{planck2014-a12}).  Compared to
2013, calibration factors changed by 1.9 and 4.1\,\%, at 545 and 857\,GHz,
respectively. Combined with other pipeline changes (such as the ADC
corrections), the brightness of the released 2015 frequency maps has decreased by
1.8 and 3.3\,\% compared to 2013.

\subsubsection{Relative calibration and consistency}
\label{sec:calibrationconsistency}

The relative calibration of LFI, HFI, and WMAP can be assessed on several
angular scales.  At $\ell=1$, we can compare the amplitude and direction of
the Solar dipole, as measured in the frequency maps of the three instruments.
On smaller scales, we can compare the amplitude of the CMB fluctuations
measured frequency by frequency by the three instruments, during and after
component separation.

\vskip 2mm\noindent{\bf Comparison of independent measurements of the Solar
dipole---}Table~\ref{tab:dipole} gives the LFI and HFI measurements of the
Solar dipole, showing agreement well within the uncertainties.  The amplitudes
agree within 1.5\muK\ (0.05\,\%), and the directions agree within 3\arcm.
Table~\ref{tab:dipole} also gives the ``nominal'' \Planck\ dipole that has
been subtracted from the \Planck\ frequency maps in the 2015 release.  This is
a plausible combination of the LFI and HFI values, which satisfied the need
for a dipole that could be subtracted uniformly across all \Planck\
frequencies early in the data processing, before the final systematic
uncertainties in the dipole measurements were available and a rigorous
combination could be determined.  See \citet{planck2014-a09} Sect.~5.1 for
additional measurements. 

Nearly independent determinations of the Solar dipole can be extracted from
individual frequency maps using component-separation methods relying on
templates from low and high frequencies where foregrounds dominate
(\citealt{planck2014-a06}; \citealt{planck2014-a09}). The amplitude and
direction of these Solar dipole measurements can be compared with each other
and with the statistical errors.  This leads to relative gain calibration
factors for the $\ell=1$ mode of the maps expressed in K$_{\rm CMB}$, as shown
for frequencies from 70 to 545\,GHz in Table~\ref{tab:intercalibration}.  For
components of the signal with spectral distribution different from the CMB, a
colour correction is needed to take into account the broad bands of these
experiments. 

\begin{figure}[htpb]
\includegraphics[width=\columnwidth]{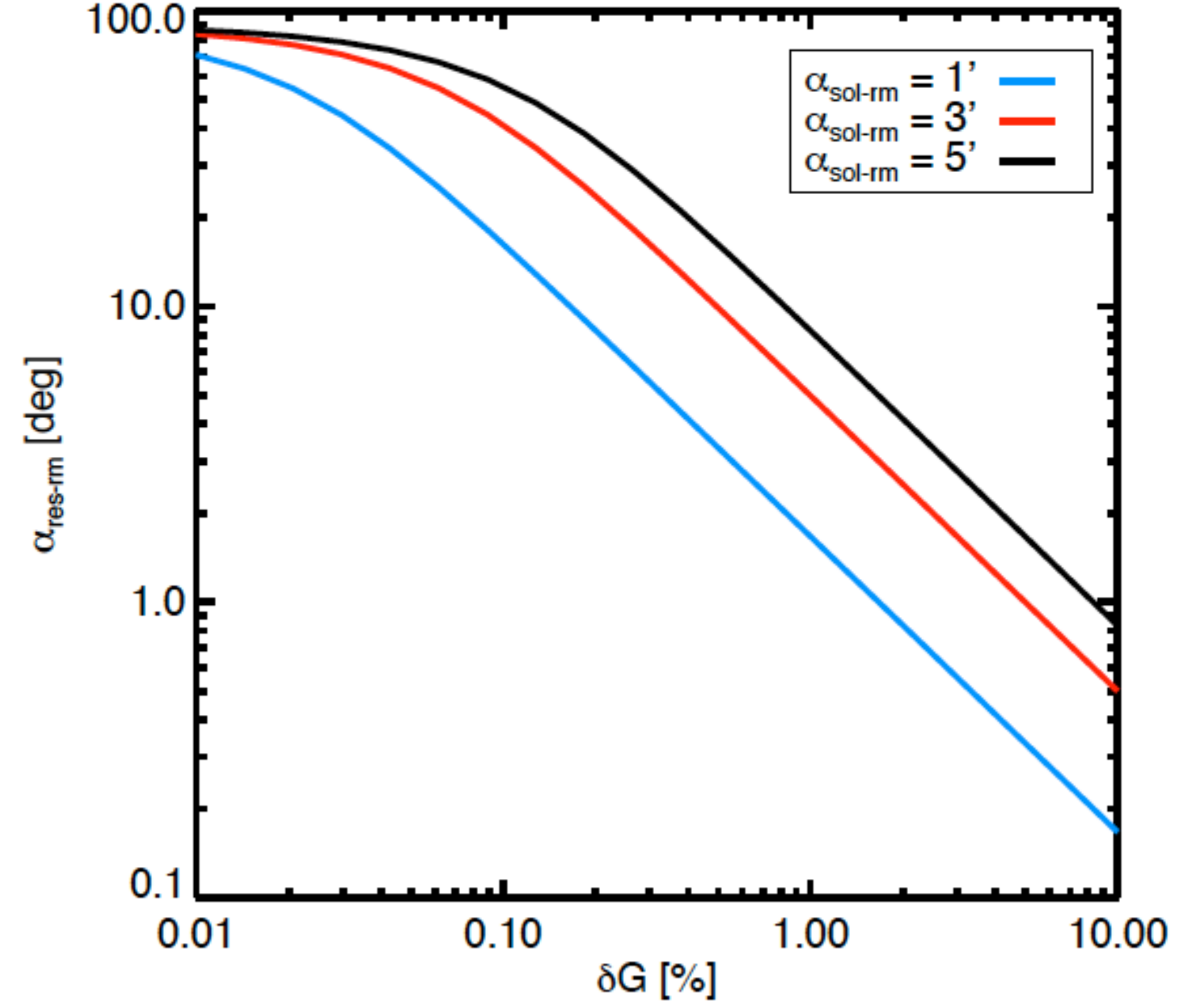}
\caption{Angle difference $\alpha_{\rm res-rm}$ between the removed Solar
dipole and the residual dipole for given errors on the dipole direction
(i.e., the angle difference between the removed dipole and the true Solar
dipole, $\alpha_{\rm sol-rm}$) and on the calibration
($\delta G = 1 - A_{\rm rm}/A_{\rm sol}$, expressed in percent).}
\label{fig:dipoledifferences}
\end{figure}

\vskip 2mm\noindent{\bf Comparison of the residuals of the Solar dipole left
in the CMB maps after removal of the best common estimate---}Another
measurement of relative calibration is given by the residuals of the Solar
dipole left in CMB maps after removing the best common estimate, the nominal
\Planck\ dipole. The residual dipole comes from two terms as illustrated in
Fig.~\ref{fig:dipoledifferences}, one associated with the error in direction,
with an axis nearly orthogonal to the Solar dipole, and one associated with
the error in amplitude aligned with the Solar dipole. Using the 857\,GHz map
as a dust template (extrapolated with optimized coefficients derived per patch
of sky), we find residual dipoles dominated by errors orthogonal to the direction of the
Solar dipole at 100 and 143\,GHz, and residuals associated with calibration
errors for the other frequencies.  The relative residual amplitudes are given
in Table~\ref{tab:intercalibration}.  This shows that a minimization of the
dipole residuals can and will be introduced in the HFI calibration pipeline
for the final 2016 release.

\vskip 2mm\noindent{\bf Comparison of CMB anisotropies frequency by frequency
during and after component separation---}Table~\ref{tab:intercalibration} also
shows the relative calibration between frequencies and detectors determined by
{\tt SMICA} (\citealt{planck2013-p08}; \citealt{planck2014-a11}) and
{\tt Commander} (\citealt{planck2014-a11}; \citealt{planck2014-a12}), two of
the map-based diffuse component separation codes used by \Planck.  The
calculation is over different multipole ranges for the two methods, so
variation between the two could reflect uncertainties in transfer functions;
moreover, {\tt Commander} uses different constraints in order to deal with the
complexities and extra degrees of freedom involved in fitting foregrounds
individually (see \citealt{planck2014-a12} for details), so we do not expect
identical results with the two codes.  Nevertheless, the agreement is
excellent, at the 0.2\,\% level between the first acoustic peak, intermediate
$\ell$, and dipole residuals, and the intercalibration offsets between
frequencies are within 0.3\,\% of zero from 30\,GHz to 217\,GHz.

\begin{table*}
\caption{Intercalibration factors by frequency between LFI, HFI, and WMAP. }
\label{tab:intercalibration}
\vskip -6mm
\footnotesize
\newdimen\tblskip \tblskip=2.5pt
\setbox\tablebox=\vbox{
 \newdimen\digitwidth
 \setbox0=\hbox{\rm 0}
 \digitwidth=\wd0
 \catcode`*=\active
 \def*{\kern\digitwidth}
  \newdimen\signwidth
  \setbox0=\hbox{+}
  \signwidth=\wd0
  \catcode`!=\active
  \def!{\kern\signwidth}
\halign{\hbox to 5.0cm{#\leaderfil}\tabskip 4em&
    \hfil$#$\hfil \tabskip 4em&
    \hfil$#$\hfil \tabskip 4em&
    \hfil$#$\hfil \tabskip 0em\cr
\noalign{\doubleline\vskip 1pt}
\omit&&\multispan2\hfil\sc CMB Anisotropy [\%]\hfil\cr
\noalign{\vskip -3pt}
\omit&&\multispan2\hrulefill\cr
\omit&\omit\hfil\sc Solar dipole [\%]\hfil&\omit\hfil {\tt Commander}\hfil&\omit\hfil {\tt SMICA}\hfil\cr
\omit\hfil\sc Frequency [GHz] (Detector)\hfil&\ell = 1&25\le\ell\le100&50\le\ell\le500\cr
\noalign{\vskip 5pt\hrule\vskip 5pt}
\noalign{\vskip 2pt}
*30&\dots&                 -0.3\rlap{$^{\rm a}$}*\pm0.1*&\dots\cr
\noalign{\vskip\tblskip}
*44&\dots&                 !0.3\rlap{$^{\rm a}$}*\pm0.1*&\dots\cr
\noalign{\vskip\tblskip}
*70&!0.04\rlap{$^{\rm a}$}&!0.0\rlap{$^{\rm a}$}*\pm0.1*&!0.21\rlap{$^{\rm a}$}\pm0.06\cr
\noalign{\vskip\tblskip}
100&!0.03&                   !0.09\pm0.02&               !0.03\pm0.02\cr
\noalign{\vskip\tblskip}
143&0\rlap{$^{\rm b}$}&0; -0.1\pm0.1 \rlap{$^{\rm c}$}&         0\rlap{$^{\rm b}$}\cr
\noalign{\vskip\tblskip}
217&!0.20&             0; 0.02\pm0.03 \rlap{$^{\rm c}$}&               !0.28\pm0.02\cr
\noalign{\vskip\tblskip}
353&!0.53&                   !0.5*\pm0.1*&               !0.73\pm0.11\cr
\noalign{\vskip\tblskip}
545&!1.25&          -1.0\rlap{$^{\rm d}$}&               !1.09\pm1.5*\cr
\noalign{\vskip\tblskip}
WMAP (23)&\dots&       0\rlap{$^{\rm b}$}&                      \dots\cr
\noalign{\vskip\tblskip}
WMAP (33)&\dots&             !0.1*\pm0.1*&                      \dots\cr
\noalign{\vskip\tblskip}
WMAP 41 (Q)&\dots&           !0.1*\pm0.1*&                      \dots\cr
\noalign{\vskip\tblskip}
WMAP 61 (V)&\dots&           !0.2*\pm0.1*&                      \dots\cr
\noalign{\vskip\tblskip}
WMAP 94 (W)&-0.26&           !0.2*\pm0.1*&               !0.28\pm0.15\cr
\noalign{\vskip 6pt\hrule\vskip 10pt}}}
\endPlancktablewide
\tablenote {{\rm a}} LFI map rescaling factors that are incorporated in the beam transfer functions, as described in \citet{planck2014-a03}, have been applied.\par
\tablenote {{\rm b}} Reference frequency; no intercalibration calculated.\par
\tablenote {{\rm c}} For {\tt Commander} at 143\,GHz, detector set `ds1' was used as a reference (intercalibration factor = 0).  The mean intercalibration factor for detectors ds2+5+6+7 was $-0.1\pm0.1$.  Similarly, at 217\,GHz detector `1' was used as a reference (intercalibration factor = 0), and the mean intercalibration factor for detectors 2+3+4 was $0.02\pm0.03$.  See Table~6 in \citet{planck2014-a12} for details.\par 
\tablenote {{\rm d}} For {\tt Commander}, the effective recalibration of the 545\,GHz channel measured in units of $\mu{\rm K}_{\rm cmb}$ is the product of a multiplicative calibration factor and a unit conversion correction due to revised bandpass estimation. See Sect.~5.3 in \citet{planck2014-a12} for details.\par
\end{table*}

The following points highlight the remarkable internal consistency of \Planck\ calibration.

\begin{itemize}
 
\item The small Solar dipole residuals measured for the 100 and 143\,GHz channels ($<4\muK$) are close to 90\deg away from the adopted \Planck\ Solar dipole, reflecting in both cases a small 2\parcm8 shift in the measured direction of the dipole compared to the adopted dipole, but amplitudes (hence calibrations) within 0.1\,\% of the adopted (``mean'') value. The {\tt Commander} and {\tt SMICA} inter-comparisons below and on the first acoustic peak give a calibration difference between 100 and 143\,GHz of $\le0.09$\,\%, confirming the very high calibration accuracy of these two channels.

\vskip 3pt
 
\item  The amplitude of the Solar dipole measured by the 70\,GHz channel shows a difference of 1\muK\ (0.03\,\%) with respect to the best HFI Solar dipole amplitude.

\vskip 3pt
 
\item The 217, 353, and 545\,GHz channels show dipole residuals aligned with the Solar dipole, which thus measure directly calibration errors with respect to 143\,GHz of 0.2, 0.53, and 1.25\,\%. 

\vskip 3pt
 
\item The {\tt SMICA} first peak intercalibration of 217 and 353\,GHz with respect to 143\,GHz, taken again as reference, shows similar intercalibration to the dipole residuals with differences 0.08 and 0.20\,\%. In fact, Table~\ref{tab:intercalibration} suggests that we can now achieve significantly better intercalibration of all CMB channels from 70 to 353\,GHz. 

\vskip 3pt
 
\item Comparison of the Solar dipole and first acoustic peak intercalibration factors for the 545\,GHz channel gives a difference of only 0.16\,\%.  This shows that the 545\,GHz channel could be calibrated using the first acoustic peak of the CMB instead of planets. Use of the planet model could then be limited to intercalibration between 545 and 857\,GHz. The roughly 1\,\% agreement between the planets and CMB absolute calibrations also show that the current uncertainties on the absolute calibration of the high-frequency channels, dominated by the roughly 5\,\% error on the models, are probably overestimated.

\vskip 3pt
 
\item The intercalibration factors derived from {\tt Commander} in all frequency bands from 70\,GHz to 217\,GHz are less than 0.1\,\%. Considering all \Planck\ bands from 30\,GHz to 353\,GHz, they are within 0.5\,\%. 

\end{itemize}

\begin{figure*}[htbp]
\centering
\includegraphics[width=\columnwidth]{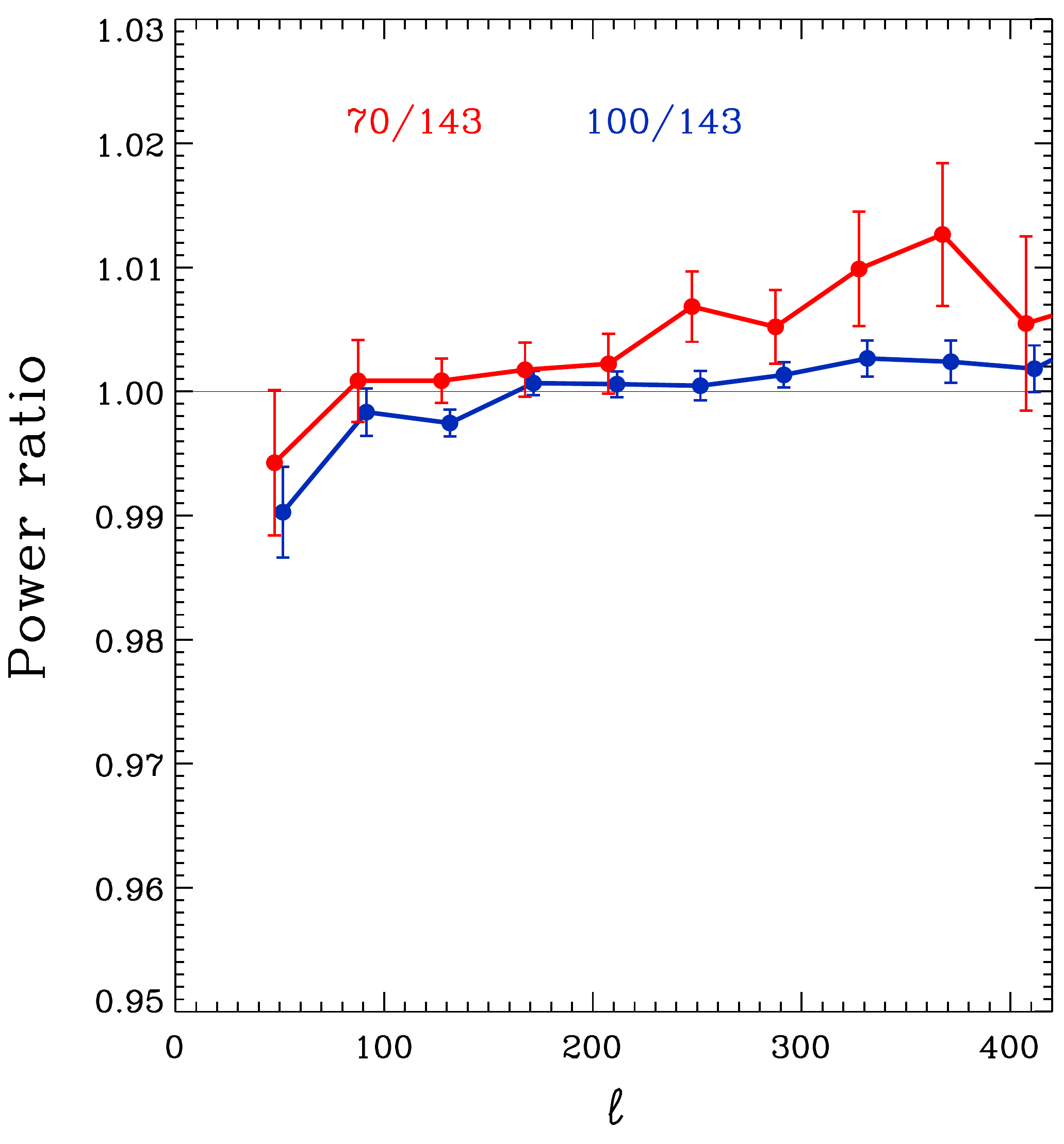}
\includegraphics[width=\columnwidth]{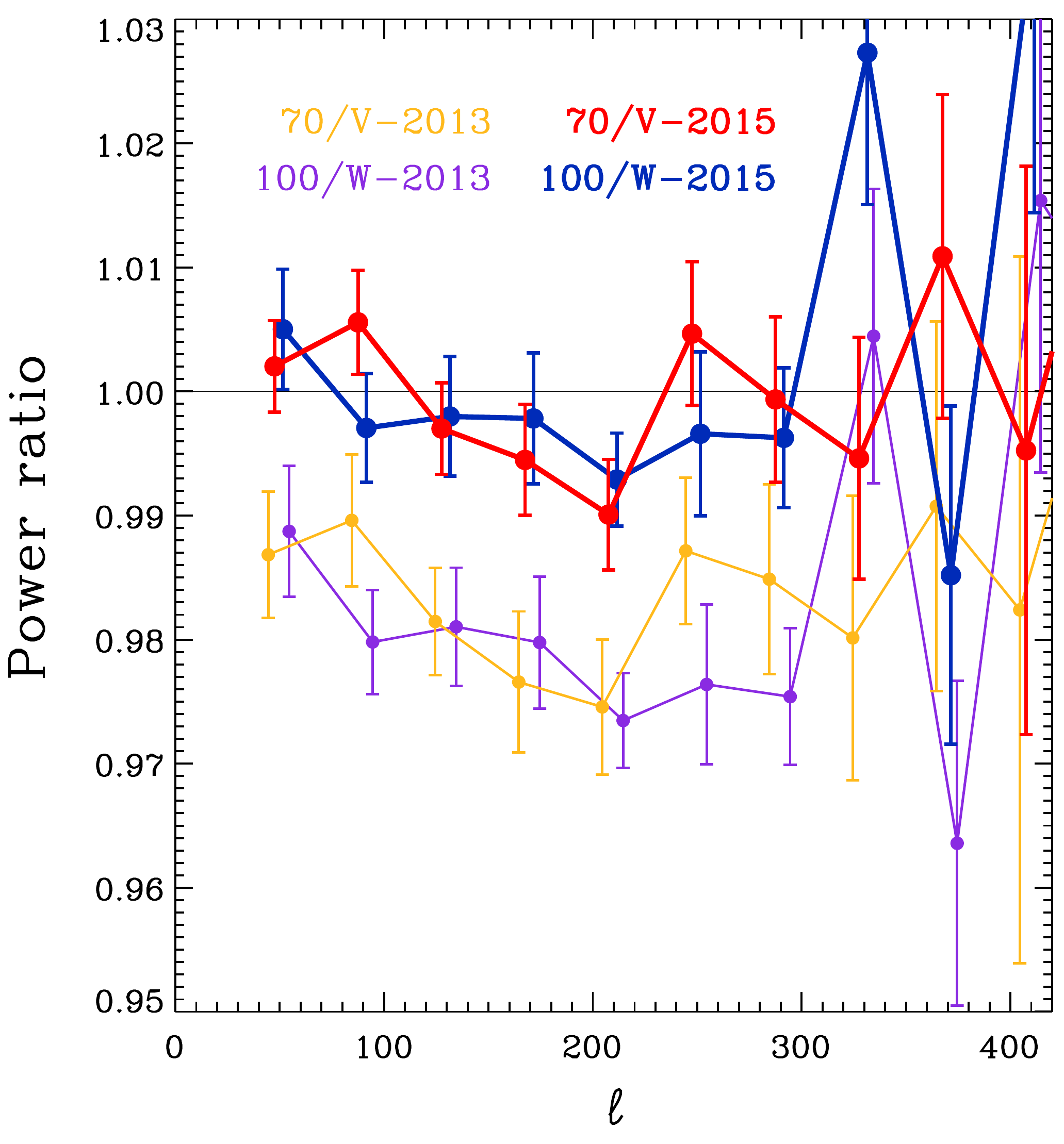}
\caption{Ratios of power spectra over the region of the first acoustic peak, uncorrected for foregrounds (which vary over the three frequencies), over 60\,\% of the sky.  The uncertainties are the errors on the mean within each $\Delta\ell = 40$ bin of the ratios computed $\ell$ by~$\ell$. \quad {\it Left\/}: Ratios of 70 and 100\,GHz $TT$ spectra to 143\,GHz.  The low values at $\ell=50$ are due to diffuse foregrounds at 143\,GHz.  The rise to higher multipoles in the 70/143 ratio is due to discrete foregrounds.  \quad {\it Right\/}: Ratio of $TT$
spectra of \Planck\ 70 and 100\,GHz to WMAP V and W bands, as calculated for \Planck\ 2013 data \citep{planck2013-p01a} and for the 2015 data.  The near-overlap of frequencies between the \Planck\ and WMAP bands means that foregrounds have no appreciable effect on the ratios.  The effect of the calibration changes in \Planck\ between 2013 and 2015 that are discussed in this paper is clear.  There is now excellent agreement within statistical errors between \Planck\ and WMAP in the region of the spectrum where both have high S/N.}
\label{fig:LFIHFIratio}
\end{figure*}

This comparison can also be made at the power spectrum level.  The left-hand panel in Fig.~\ref{fig:LFIHFIratio} compares the 70, 100, and 143\,GHz channels of LFI and HFI in the multipole range of the first acoustic peak, $50< \ell < 500$, uncorrected for foregrounds, over 60\,\% of the sky. The low values a $\ell=50$ show the effect of unremoved diffuse foregrounds at 143\,GHz, and the rise of the 70/143 ratio is at least partly driven by unremoved discrete foregrounds; the uncertainties are larger at 70\,GHz as well.  In the middle region, the agreement is very good, at a level of a few tenths of a percent.  This result is a direct test that all systematic effects in calibration have been corrected on both instruments to better than this value.

The right-hand panel of Fig.~\ref{fig:LFIHFIratio} shows the ratios of
\Planck\ $TT$ spectra at 70 and 100\,GHz to those of WMAP in the V and W bands,
as calculated for \Planck\ 2013 data \citep{planck2013-p01a} and for the 2015
data.  While the scatter is significantly larger than that in the left-hand
panel, due to the higher noise in WMAP, the agreement is very good, and within
the statistical errors.  We can now say that within the uncertainties, LFI,
HFI, and WMAP agree, and the difference seen in the 2013 data
\citep{planck2013-p01a} is gone.

\subsubsection{Summary of calibration}
\label{sec:calibrationsummary}

The \Planck\ 70 and 100\,GHz channels belong to instruments based on different
technologies, with different systematic effects, and close to the minimum of
the diffuse foregrounds.  They thus provide a very good test of the consistency
of calibration and transfer functions.  The internal consistency between LFI
and HFI is remarkable.  Figure~\ref{fig:LFIHFIratio} represents a stringent
test of calibration, systematic effects, beams, and transfer functions, and
demonstrates overall consistency at a level of a few parts per thousand
between independent instruments and spacecraft. 

The \Planck\ CMB-channels from 70 to 217\,GHz show calibration difference below
0.3\,\%, measured from both residual dipoles and first acoustic peak.  Using a
Solar dipole reference established on the 100 and 143\,GHz channels, it is
likely that all detectors could be inter-calibrated to 0.05\,\% in subsequent
data processing versions.  The agreement of the measured calibration factors
from dipole residuals ($\ell = 1$) and first acoustic peak ($\ell = 200$) shows
that the transfer functions are controlled to better than 0.2\,\%  in this
multipole range.  Corrections for systematic effects in HFI cover a dynamic
range from detector to detector larger than 2 at 100 and 143\,GHz, but have
reduced the calibration errors by an order of magnitude. This suggests that
the corrections lead now to an absolute photometric calibration accuracy on
the orbital dipole (limited only by systematics and noise) of 0.1\,\%.
  
As in other instances in the \Planck\ data processing, when very small
systematic effects are detected and measured in a posteriori characterization,
their removal from the data is complicated.  Their determinations are often
degenerate, and complete reprocessing is necessary. The calibration improvement
demonstrated by the minimization of the dipole residuals using the 857\,GHz
dust template will be introduced in a self-consistent way in the HFI
calibration pipeline and overall processing for the final 2016 release.
Furthermore, the use of the Solar dipole parameters from the best \Planck\ CMB
channels (100 and 143\,GHz) will be introduced in the processing of the
channels more affected by foregrounds and noise. The LFI calibration accuracy
is now close to noise-limited, but improvements will be made in 2015 according
to a complete simulation plan to improve our understanding of calibration and
systematics affecting low multipoles, particularly for polarization analysis.

\section{Timelines}
\label{sec:Timelines}

For the first time, the 2015 \Planck\ release includes time series of the
observations acquired by individual detectors in LFI and HFI
\citep[see][for details]{planck2014-a03,planck2014-a08}. These timelines
will be of use for those wishing to construct maps using specific time periods
or mapmaking algorithms.

The delivered timelines have been cleaned of all major instrumental systematic
effects.  For HFI timelines this means that the raw timelines are
ADC-corrected, demodulated, despiked, corrected for rare baseline jumps,
and a dark template has been removed; they are converted to absorbed power
units, and the time transfer function has been deconvolved. 
For LFI timelines this means that the raw timelines are ADC-corrected,
despiked, and demodulated; furthermore, the raw diode outputs (two per
receiver) are combined and gain regularization is applied before calibration. 

The timelines are calibrated to astrophysical units and corrected for a
zero-point level (determined at map level). The Solar and orbital dipole
signals have been removed. In addition, for LFI, an estimation of Galactic
straylight has been removed.

The timelines still contain the low-frequency noise that is later removed by
destriping at the mapmaking stage. However, a set of offsets are provided
(determined during mapmaking), which can be used to convert the calibrated
timelines to maps without destriping. For HFI a single offset per ring is
determined; for LFI the offsets are computed every 0.246, 0.988, and 1.000\,s
for the 30, 44, and 70\,GHz channels, respectively. For HFI, the offsets are
determined during mapmaking using the full mission data set and all valid
detectors per channel, and they are then applied to all the maps produced,
i.e., using any fraction of the mission (year, survey) or any subset of
detectors (single detector, detector set). For LFI the offsets are similarly
determined using the full mission and all valid detectors per channel. These
offsets have been used to produce the full-mission LFI maps; however, for
shorter period maps, different offsets are used, which optimize noise
cross-correlation effects, and these are not delivered.

The timelines are accompanied by flags that determine which data have been
used for mapmaking, as well as pointing timelines, which are sampled at the
same frequency as the data themselves.

\section{Frequency Maps}
\label{sec:FreqMaps}

\begin{sidewaysfigure*}[htbp]
\vspace*{18cm}~\\
\centering
\includegraphics[width=\columnwidth]{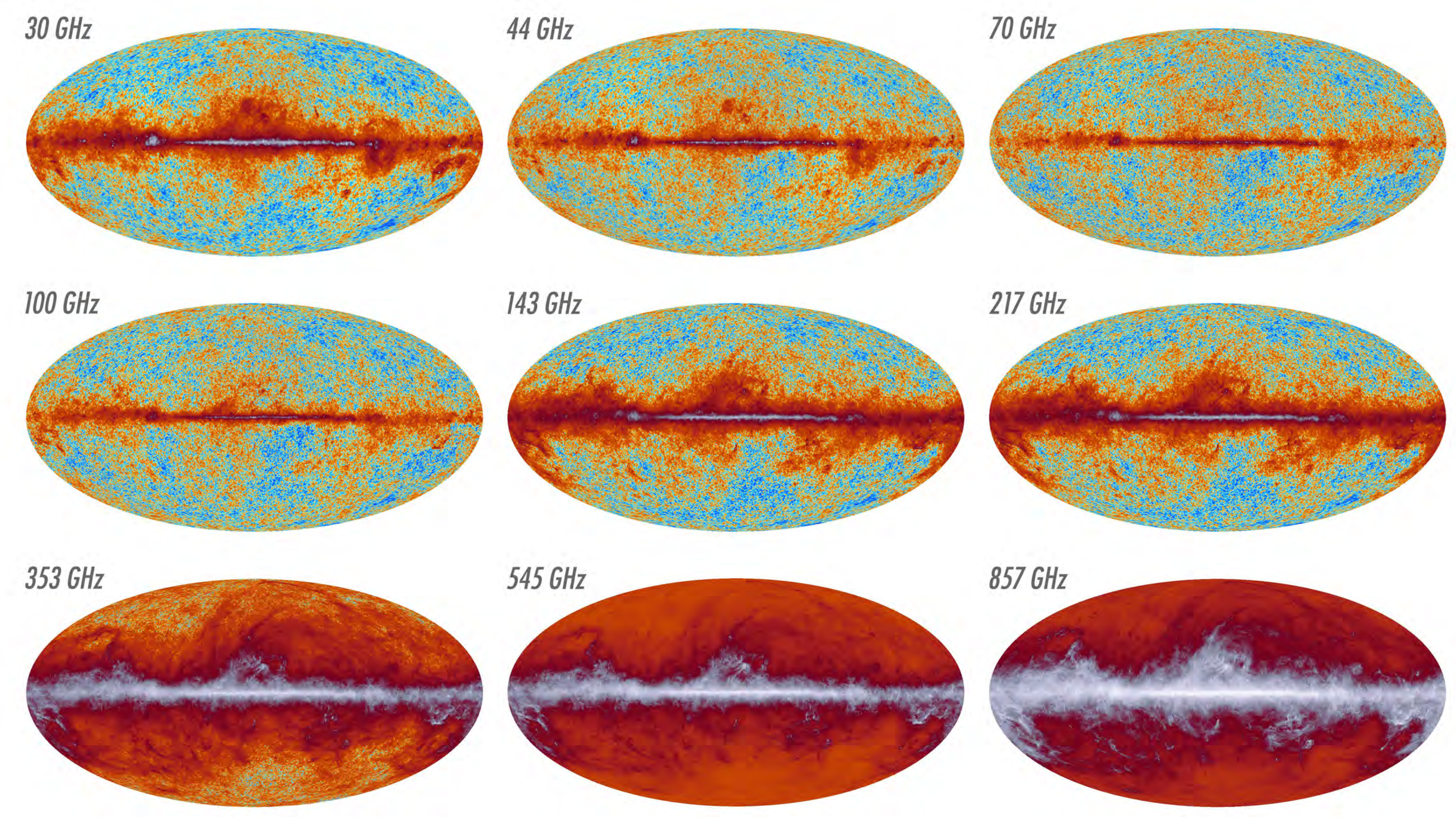}
\includegraphics[width=84mm]{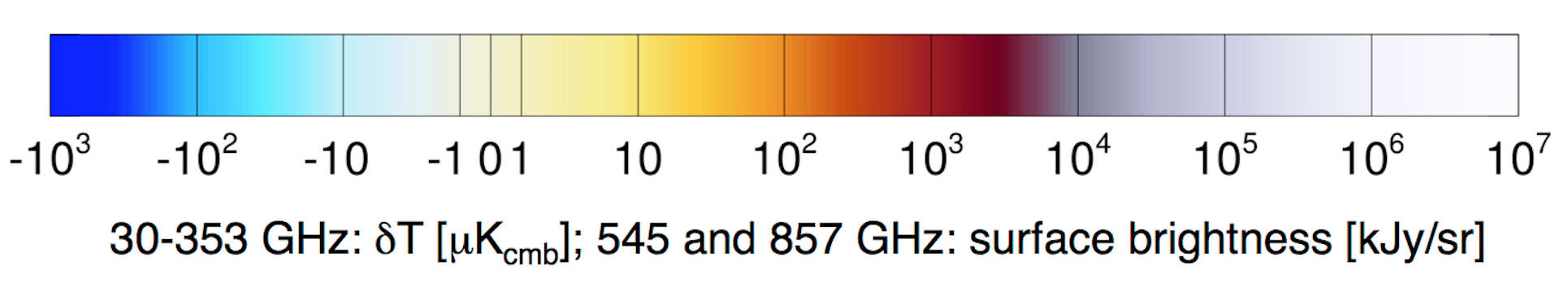}
\caption{The nine \Planck\ frequency maps show the broad frequency response of the individual channels. The color scale (identical to the one used in 2013), based on inversion of the function $y = 10^x - 10^{-x}$, is tailored to show the full dynamic range of the maps. }
\label{fig:FreqMaps}
\end{sidewaysfigure*}

\begin{figure*}[th!]
\centering
\includegraphics[width=18cm]{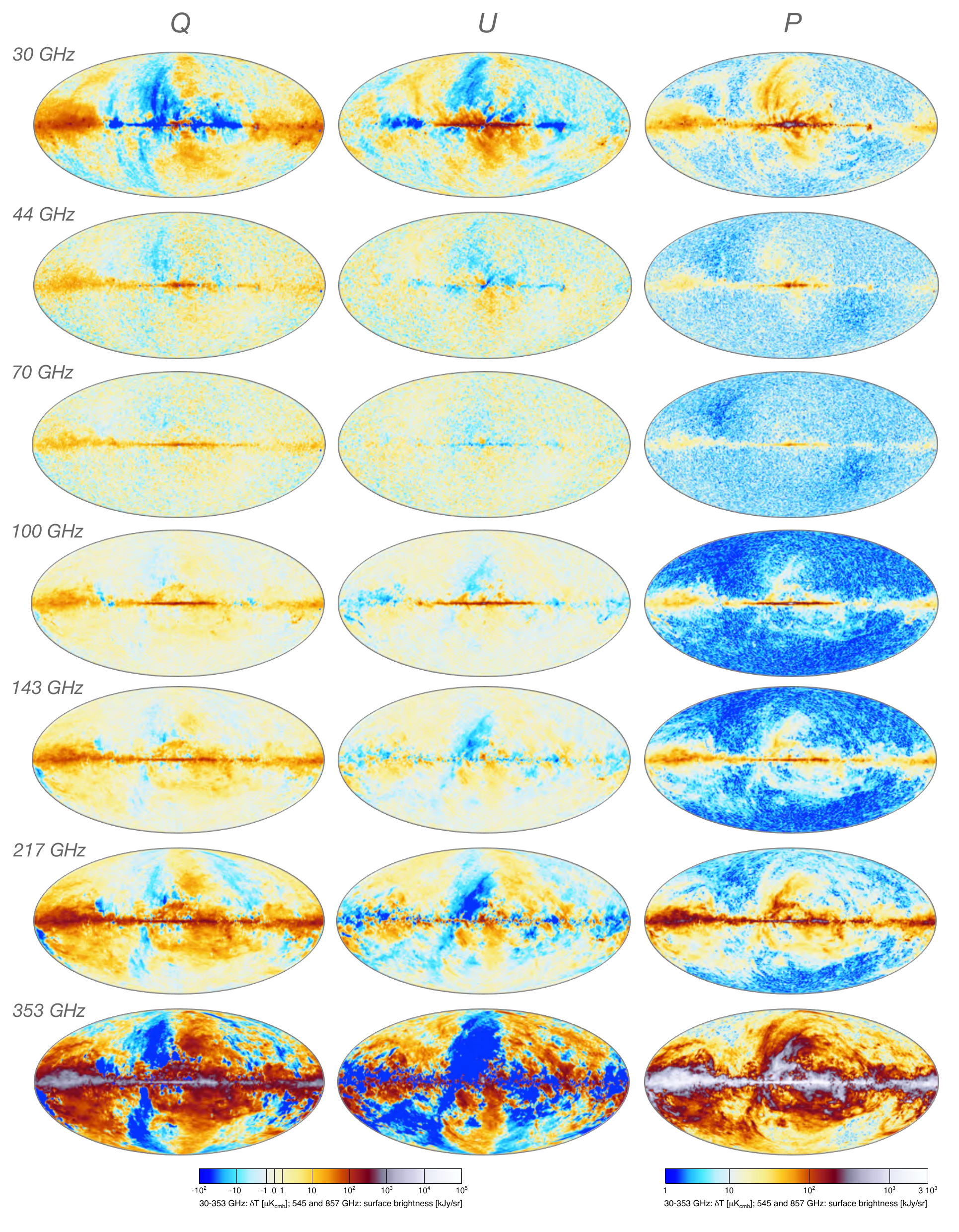}
\caption{The seven \Planck\ polarization maps between 30 and 353\,GHz, shown in Stokes $Q$ and $U$, as well as in total polarized intensity ($P$).  The LFI maps are not bandpass-corrected; the HFI maps are. The color scale uses the same function as in Fig. \ref{fig:FreqMaps}, but the range limits have been adjusted.}
\label{fig:PolMaps}
\end{figure*}

In Fig. \ref{fig:FreqMaps} and Fig. \ref{fig:PolMaps} we show the \Planck\ 2015 maps. 
Note that \Planck\ uses {\tt HEALPix} \citep{gorski2005} as its basic representation
scheme for maps, with resolution labelled by the $N_{\rm side}$ value. 

\subsection{Mapmaking}
\label{sec:mapmaking}

\subsubsection{LFI}
\label{sec:lfi_maps}

Mapmaking takes as its input the calibrated timelines, from which the
cosmological and orbital dipole signals have been removed. An estimate of
Galactic straylight is subtracted from the timelines prior to mapmaking,
since this is difficult to correct for at map level. As for the 2013 release,
the LFI maps are produced using the {\tt Madam} destriping code
\citep{keihanen2010}, enhanced with a noise prior, which enables accurate
removal of correlated $1/f$ noise, while simultaneously minimizing systematic
errors by judicious use of masks. The production of maps and covariance
matrices is validated using the FFP8 simulations. The output of the code
consists of sky maps of temperature and Stokes $Q$ and $U$ polarization,
and a statistical description of residual noise in the maps in the form of
pixel-pixel noise covariance matrices. These matrices are produced at
$N_{\rm side} = 64$.  In addition to full-mission maps at both high and low
resolution ($N_{\rm side}=16$), many other types of maps are produced,
including those from single horns, single radiometers, single surveys, odd and
even surveys, single years, and halves of the mission.  The LFI maps are not
corrected for beam shape, so that point sources in the map have the shape of
the effective beam at that location. Zero-levels are estimated by fitting a
cosecant-law Galactic latitude model to the CMB-subtracted maps, and
subtracting this from the maps. The polarization maps must be corrected for
bandpass leakage through multiplication with leakage template maps, which are
estimated via a process similar to component separation.

A summary of the characteristics of the LFI maps is presented in
Table~\ref{tab_summary_performance}.

\begin{table*}
\begingroup
\newdimen\tblskip \tblskip=5pt
\caption{Main characteristics of LFI full mission maps.}
\label{tab_summary_performance}
\nointerlineskip
\vskip -3mm
\footnotesize
\setbox\tablebox=\vbox{
   \newdimen\digitwidth
   \setbox0=\hbox{\rm 0}
   \digitwidth=\wd0
   \catcode`*=\active
   \def*{\kern\digitwidth}
   \newdimen\signwidth
   \setbox0=\hbox{+}
   \signwidth=\wd0
   \catcode`!=\active
   \def!{\kern\signwidth}
\halign{\hbox to 3.0in{#\leaderfil}\tabskip=3em&
        \hfil#\hfil&
        \hfil#\hfil&
        \hfil#\hfil\tabskip=0pt\cr
\noalign{\doubleline}
\omit&\multispan3\hfil Frequency band\hfil\cr
\noalign{\vskip -3pt}\omit&\multispan3\hrulefill\cr
\noalign{\vskip 2pt}
\omit\hfil Characteristic\hfil&30\,GHz&44\,GHz&70\,GHz\cr
\noalign{\vskip 3pt\hrule\vskip 5pt}
Centre frequency [GHz]&28.4&44.1&70.4\cr
\noalign{\vskip 3pt}
Effective beam FWHM$^{\rm a}$ [arcmin]&32.29&27.00&13.21\cr
\noalign{\vskip 3pt}
Effective beam ellipticity$^{\rm a}$& 1.32& 1.04& 1.22\cr
\noalign{\vskip 3pt}
Temperature noise (1\deg)$^{\rm b}$ [\muKcmb]& 2.5& 2.7& 3.5\cr
\noalign{\vskip 3pt}
Polarization noise (1\deg)$^{\rm b}$ [\muKcmb]& 3.5& 4.0& 5.0\cr
\noalign{\vskip 3pt}
Overall calibration uncertainty$^{\rm c}$ [\%]& 0.35& 0.26& 0.20\cr
\noalign{\vskip 3pt}
Systematic effects uncertainty in Stokes $I^{\rm d}$ [\muKcmb]& 0.19& 0.39&
 0.40\cr
\noalign{\vskip 3pt}
Systematic effects uncertainty in Stokes $Q^{\rm d}$ [\muKcmb]& 0.20& 0.23&
 0.45\cr
\noalign{\vskip 3pt}
Systematic effects uncertainty in Stokes $U^{\rm d}$ [\muKcmb]& 0.40& 0.45&
 0.44\cr
\noalign{\vskip 5pt\hrule\vskip 3pt}}}
\endPlancktablewide
\tablenote a Calculated from the main beam solid angle of the effective beam,
 $\Omega_{\rm eff} = \hbox{mean}(\Omega)$. These values
 are used in the source extraction pipeline
 ~\citep{planck2014-a35}.\par
\tablenote b Noise rms computed after smoothing to 1\deg . \par
\tablenote c Sum of the error determined from the absolute and relative
 calibration, see \citet{planck2014-a05}.\par
\tablenote d Estimated rms values over the full sky and after full mission
 integration. Not included here are gain reconstruction uncertainties,
 estimated to be of order 0.1\,\%\ . \par
\endgroup
\end{table*}

\subsubsection{HFI}
\label{sub:hfi_maps}

As for the \Planck\ 2013 release, the measurements in each {\tt HEALPix}
pixel visited during a stable pointing period (i.e., ``ring'') are
averaged for each detector, keeping track of the bolometer orientation on the
sky. The calibration and mapmaking operations use this intermediate product as
an input. For each detector, the TOIs are only modified by a single offset
value per ring, determined using the destriping method described in
\citet{Tristram2011}.  The offsets are computed simultaneously for all
bolometers at a given frequency, using the full mission data. For a given
bolometer, the same offset per ring is applied whatever the map (e.g.,
full-mission, half-mission, detector-set maps; but for half-ring maps, see
\citealt{planck2014-a08}). Each data sample is calibrated in $K_{\rm CMB}$ for
the 100, 143, 217, and 353\,\GHz\ channels, and MJy\,sr\mo\
(assuming $\nu I_{\nu}=\hbox{constant}$) for the 545 and 857\,\GHz\ channels,
using the calibration scheme presented in
Sect.~\ref{sec:instrumentcalibration}.  Contrary to the 2013 release, the
bolometer gains are assumed to be constant throughout the mission. The final
mapmaking is a simple projection of each unflagged sample to the nearest grid
pixel.  For polarization data, when several detectors are solved
simultaneously, the polarization mapmaking equation is inverted on a
per-pixel basis \citep{planck2014-a08}.

The products of the HFI mapmaking pipelines are pixelized maps of $I$, $Q$,
and $U$, together with their covariances. Map resolution is
$N_{\rm side} = 2048$, and the pixel size is 1\parcm07.  The basic
characteristics of the maps are given in Table~\ref{tab:HFImapsummary}. For
details, see \citep{planck2014-a08}.

Maps are cleaned for the zodiacal light component, which varies in time, based
on templates fitted on the survey difference maps
\citep[see][]{planck2013-pip88}.
These templates are systematically subtracted prior to mapmaking. The \Planck\
total dipole (Solar and orbital) is computed and also subtracted from the data.
Contrary to 2013, the far sidelobes (FSL) are not removed from the maps.

The 2015 HFI maps delivered via the PLA have had zodiacal light removed, include the CIB and the zero
level of the temperature maps has been adjusted for Galactic emission. 
However,  the zero level adjustment was based on  maps which contained zodiacal light, and therefore 
the released maps require an additional frequency-dependent correction which has to be applied manually.
For work requiring all astrophysical sources of emission to be present in the maps, the corrections provided under Note ``e2" of  Table~\ref{tab:HFImapsummary}  must be added to the maps. 
For work requiring Galactic emission only, the ``e2" corrections should be added to the maps, and the CIB levels provided under Note ``e"  should be removed.

\begin{table*}[t]
\caption{Main characteristics of HFI full mission maps.}
\label{tab:HFImapsummary}
\begingroup
\newdimen\tblskip \tblskip=5pt
\nointerlineskip
\vskip -3mm
\footnotesize
\setbox\tablebox=\vbox{
   \newdimen\digitwidth 
   \setbox0=\hbox{\rm 0} 
   \digitwidth=\wd0 
   \catcode`*=\active 
   \def*{\kern\digitwidth}
   \newdimen\signwidth 
   \setbox0=\hbox{+} 
   \signwidth=\wd0 
   \catcode`!=\active 
   \def!{\kern\signwidth}
\halign{\hbox to 2.5in{#\leaderfil}\tabskip 2.2em&
\hfil#\hfil\tabskip 1.2em&
\hfil#\hfil&
\hfil#\hfil&
\hfil#\hfil&
\hfil#\hfil&
\hfil#\hfil\tabskip 2.2em&
\hfil#\hfil\tabskip=0pt\cr
\noalign{\doubleline}
\omit&\multispan6\hfil Reference frequency $\nu$ [\GHz]\hfil\cr
\noalign{\vskip -3pt}
\omit&\multispan6\hrulefill\cr
\noalign{\vskip 3pt}
\omit\hfil Characteristic\hfil&100&143&217&353&545&857&Notes\cr
\noalign{\vskip 5pt\hrule\vskip 5pt}
Number of bolometers&                                        8&    11&    12&    12&     3&     4& a1\cr
\noalign{\vskip 3pt\vskip 3pt}  
Effective beam FWHM$_1$ [arcmin]&                      9.68&  7.30&  5.02&  4.94&  4.83&  4.64& b1\cr
Effective beam FWHM$_2$ [arcmin]&                      9.66&  7.22&  4.90&  4.92&  4.67&  4.22& b2\cr
Effective beam ellipticity $\epsilon$&                1.186& 1.040& 1.169& 1.166& 1.137& 1.336& b3\cr
\noalign{\vskip 3pt\vskip 5pt}                                      
Noise per beam solid angle [$\mu\mathrm{K_{CMB}}$]&     7.5&   4.3&   8.7&  29.7&&&             c1\cr
\phantom{Noise per beam solid angle }[kJy sr$^{-1}$]& \dots& \dots& \dots& \dots&   9.1&   8.8& c1\cr
 Temperature noise [$\mu\mathrm{K_{CMB}}$\,deg]& 1.29& 0.55&   0.78&  2.56&\dots& \dots&        c2\cr
 \phantom{Temperature noise }[kJy sr$^{-1}$\,deg]&    \dots& \dots& \dots& \dots&  0.78&  0.72& c2\cr
 Polarization noise [$\mu\mathrm{K_{CMB}}$\,deg]&      1.96&  1.17&  1.75&  7.31& \dots& \dots& c3\cr
\noalign{\vskip 3pt\vskip 3pt}
Calibration accuracy [\%]&                              0.09& 0.07&  0.16&  0.78&1.1(+5)&1.4(+5)&d\cr
\noalign{\vskip 3pt\vskip 3pt}  
CIB monopole prediction [\MJysr]&                        0.0030&0.0079&0.033&  0.13& 0.35& 0.64& e\cr
Zodiacal light level correction [K$_{\mathrm{CMB}}$] & $4.3 \times 10^{-7}$ & $9.4 \times 10^{-7}$ & $3.8\times 10^{-6}$ & $3.4 \times 10^{-5}$ &  &  & e2\cr
\phantom{Zodiacal light level correction} [\MJysr] &  &  &  &  &                    0.04& 0.12& e2\cr
\noalign{\vskip 5pt\hrule\vskip 3pt}}}
\endPlancktablewide
\tablenote {{\rm a1}} {~Number of bolometers whose data were used in producing
 the channel map.}\par
\tablenote {{\rm b1}} {~FWHM of the Gaussian whose solid angle is equivalent
 to that of the effective beams.}\par
\tablenote {{\rm b2}} {~FWHM of the elliptical Gaussian fit.}\par
\tablenote {{\rm b3}} {~Ratio of the major to minor axis of the best-fit
 Gaussian averaged over the full sky. }\par
\tablenote {{\rm c1}} {~Estimate of the noise per beam solid angle, as given
 in \textit{b1}.}\par
\tablenote {{\rm c2}} {~Estimate of the noise in intensity scaled to $1\deg$
 assuming that the noise is white.}\par
\tablenote {{\rm c3}} {~Estimate of the noise in polarization scaled to
 $1\deg$ assuming that the noise is white.}\par
\tablenote {{\rm d\phantom{1}}} {~Calibration accuracy (at 545 and 857\GHz,
 the 5\,\% accounts for the model uncertainty).}\par
\tablenote {{\rm e\phantom{1}}} {~According to the~\cite{bethermin2012} model,
 whose uncertainty is estimated to be at the 20\,\% level (also for constant
 $\nu I_\nu$).}\par
\tablenote {{\rm e2}} {~Zero-level correction to be applied on zodiacal-light corrected maps.}\par
\endgroup
\end{table*}

\section{CMB Products}
\label{sec:CMBProds}

\subsection{CMB maps}
\label{subsec:CMBmapNG}

\begin{figure*}
\begin{center}
\includegraphics[width=18cm]{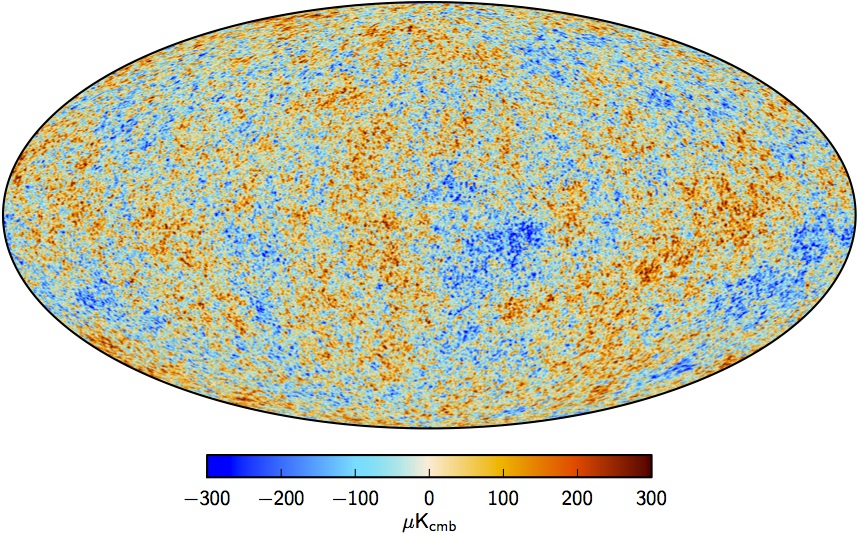}
\caption{Maximum posterior CMB intensity map at 5\arcm\ resolution derived
from the joint baseline analysis of \Planck, \WMAP, and 408\,MHz observations.
A small strip of the Galactic plane, 1.6\,\% of the sky, is filled in by a
constrained realization that has the same statistical properties as the rest
of the sky.}
\label{fig:CMBT}
\end{center}
\end{figure*}

\begin{figure*}
\begin{center}
\includegraphics[width=88mm]{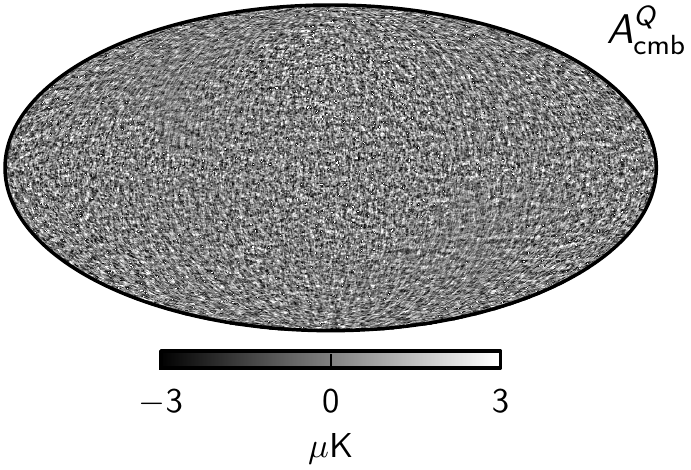}
\includegraphics[width=88mm]{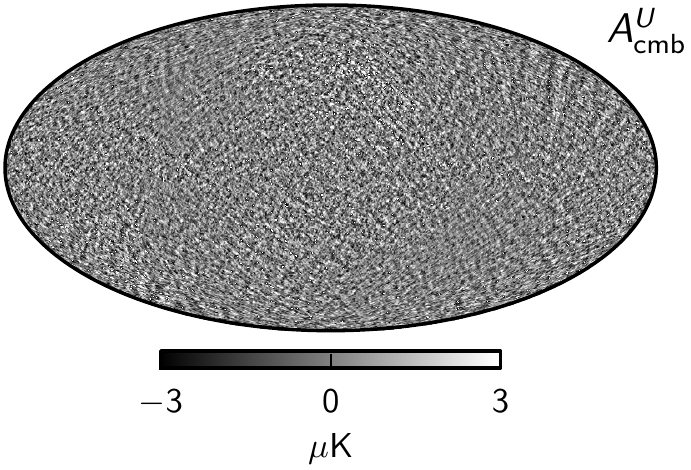}  
\caption{Maximum posterior amplitude Stokes $Q$ ({\it left\/}) and $U$
({\it right\/}) maps derived from \Planck\ observations between 30 and
353\,GHz.  These mapS have been highpass-filtered with a cosine-apodized
filter between $\ell=20$ and 40, and the a 17\,\% region of the Galactic plane
has been replaced with a constrained Gaussian realization
\citep{planck2014-a11}. From \citet{planck2014-a12}.}
\label{fig:CMBQU}
\end{center}
\end{figure*}

As for the \Planck\ 2013 release, we use four different methods to separate
the \Planck\ 2015 frequency maps into physical components
\citep{planck2014-a11}.  The four methods are: {\tt SMICA}
(independent component analysis of power spectra, \citealt{delabrouille2003};
\citealt{cardoso2008}); {\tt NILC} (needlet-based internal linear combination,
\citealt{delabrouille2009}); {\tt Commander} (pixel-based parameter and
template fitting with Gibbs sampling, \citealt{eriksen2006};
\citealt{eriksen2008}); and {\tt SEVEM} (template fitting,
\citealt{fernandez2012}). 
The methods used are conceptually the same as in 2013, but we apply them now
independently to the temperature and polarization maps. Similarly to what was
done in 2013, simulations \citep[in this case FFP8,][]{planck2014-a14} are used
to test the methods and estimate uncertainties in the recovery of components.

All four methods produce CMB maps in Stokes $I$, $Q$, and $U$. In addition,
{\tt Commander} and {\tt SMICA} also separate diffuse astrophysical
``foregrounds'' characterized by their different spectral signatures.
{\tt Commander} does so by fitting physical models of these foregrounds and the
CMB to the sky, whereas {\tt SMICA} extracts a fixed set of independent
components representing CMB, foregrounds and noise; typically, {\tt SMICA}
assumes two ``foregrounds'' are present at low and high frequencies,
respectively. An important change in the implementation of {\tt Commander} in
2015 is the number of input maps used: firstly, the number of \Planck\ maps is
expanded to use detector-level maps rather than maps which combine all
detectors at each frequency; secondly, the inputs include a map of 408\,MHz
emission, and the 9-year \WMAP\ maps. The significant increase in the number of
input maps allows {\tt Commander} to: (a) control much better factors such as
relative calibration and frequency response of individual channels; and
(b) extract a larger number of foreground temperature components, now
matching those that are expected to be present in the sky. 

The 2015 \Planck\ CMB temperature maps produced by all four methods (see an
example in Fig.~\ref{fig:CMBT}) are significantly more sensitive than those
produced in 2013 (by a factor of 1.3). They are used mainly for non-Gaussianity
analysis \citep{planck2014-a19, planck2014-a18} and for the extraction of
lensing deflection maps \citep{planck2014-a17}.  For these analyses, 
all four methods are considered to give equivalently robust results,
and the dispersion between the four gives a reasonable estimate of the
uncertainty of the CMB recovery. Although the statistical properties of these
maps give good results when used to fit cosmological models, the best \Planck\
2015 cosmological parameters are derived from a likelihood code which allows
much more detailed tuning of the contribution of individual frequencies and
$\ell$-by-$\ell$ removal of foregrounds \citep{planck2014-a15}. A
low-resolution version of the {\tt Commander} map is also used in the
pixel-based low-$\ell$ likelihood used to extract our best-fit 2015 cosmology
\citep{planck2014-a13}. 

In polarization, the CMB maps resulting from the 2015 \Planck\ component separation methods represent a dramatic advance in terms of coverage, angular resolution, and sensitivity.  Nonetheless, they suffer from a significantly high level of anomalous features at large angular scales, arising from corresponding systematic effects in the input frequency maps between 100 and
217\,GHz. The characterization of these systematic effects is still ongoing, and it is currently suspected that low-level spurious signals are also present at intermediate angular scales \citep{planck2014-a08, planck2014-a09}.  For this reason, the CMB polarization maps presented here have had their large angular scales $\ell\la30$) filtered out. Filtered Stokes $Q$ and $U$ maps resulting from one of the methods are shown as an example in Fig.~\ref{fig:CMBQU}. They are used only for a very limited number of cosmological analyses, which have been shown to be immune to the undesired features still present: estimation of primordial non-Gaussianity levels \citep{planck2014-a19}; stacking analysis \citep{planck2014-a18}; estimation of primordial magnetic field levels \citep{planck2014-a22}; and estimation of lensing potential \citep{planck2014-a17}.  We note here that
the low-$\ell$ polarization likelihood used in \cite{planck2014-a15} is based exclusively on the 70\,GHz polarization map, cleaned of foregrounds by use of the 30 and 353\,GHz polarization maps.  Both of these frequencies have been shown to be free of the kind of systematic errors that still affect intermediate frequencies \citep{planck2014-a03, planck2014-a09}.

\subsection{CMB power spectra}

\begin{figure*}
\begin{center}
\includegraphics[width=13cm]{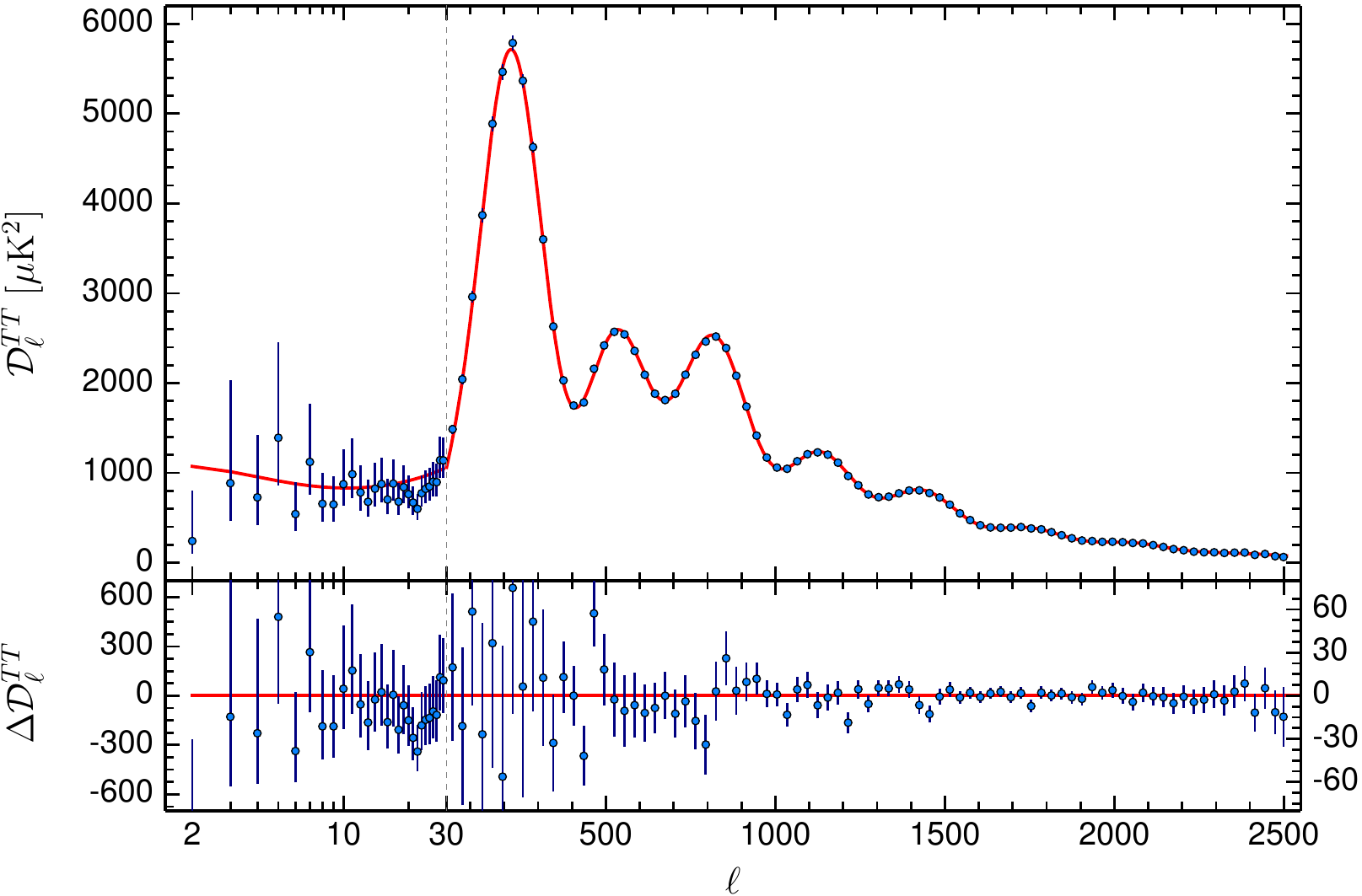}  
\caption{The \Planck\ 2015 temperature power spectrum. At multipoles $\ell\ge30$ we show the maximum likelihood frequency-averaged temperature spectrum computed from the cross-half-mission likelihood with foreground and other nuisance parameters determined from the MCMC analysis of the base \LCDM\ cosmology. In the multipole range $2 \le \ell \le 29$, we plot the power spectrum estimates from the {\tt Commander} component-separation algorithm computed over 94\,\% of the sky.  The best-fit base $\Lambda$CDM theoretical spectrum fitted to the PlanckTT+lowP likelihood is plotted in the upper panel. Residuals with respect to this model are shown in the lower panel.  The error bars show $\pm 1\,\sigma$ uncertainties. From \citet{planck2014-a15}.}
\label{fig:TT}
\end{center}
\end{figure*}

The foreground-subtracted, frequency-averaged, cross-half-mission $TT$
spectrum is plotted in Fig.~\ref{fig:TT}, together with the {\tt Commander}
power spectrum at multipoles $\ell < 29$. The figure also shows the best-fit
base $\Lambda$CDM theoretical spectrum fitted to the PlanckTT+lowP
likelihood, together with residuals (bottom panel) and $\pm 1\,\sigma$
uncertainties. 

\subsubsection{Polarization power spectra}

\begin{figure*}
\begin{center}
\includegraphics[width=88mm]{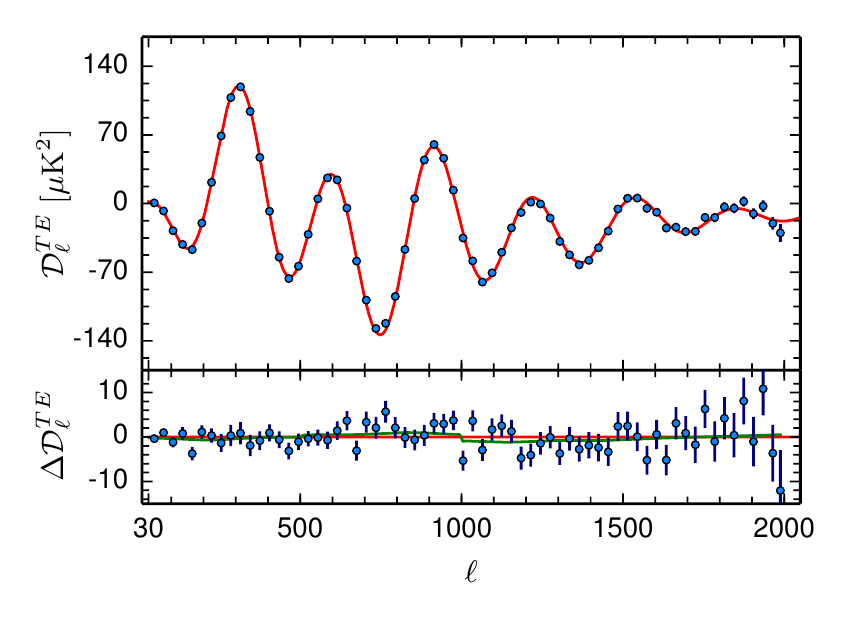}
\includegraphics[width=88mm]{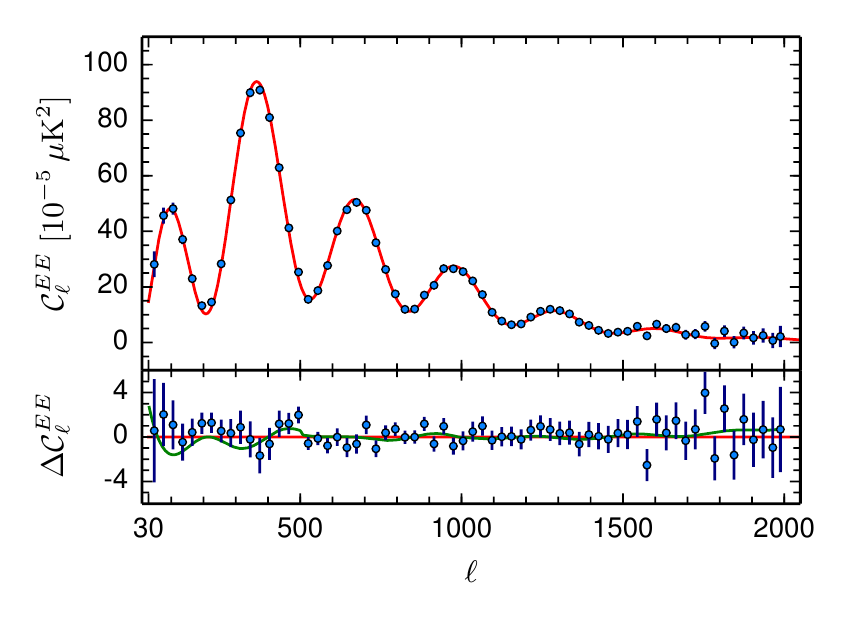}  
\caption{Frequency-averaged $TE$ ({\it left\/}) and $EE$ ({\it right\/})
spectra (without fitting for $T$--$P$ leakage). The theoretical $TE$ and $EE$
spectra plotted in the upper panel of each plot are computed from the best-fit
model of Fig.~\ref{fig:TT}. Residuals with respect to this theoretical model
are shown in the lower panel in each plot.  The error bars show $\pm 1\,\sigma$
errors. The green lines in the lower panels show the best-fit
temperature-to-polarization leakage model, fitted separately to the $TE$ and
$EE$ spectra.  From \citet{planck2014-a15}.}
\label{fig:TEEE}
\end{center}
\end{figure*}

In addition to the $TT$ spectra, the 2015 \Planck\ likelihood includes the
$TE$ and $EE$ spectra.  Figure~\ref{fig:TEEE} shows the $TE$ and $EE$ power
spectra calculated from the 2015 data and including all frequency combinations.
The theory curve shown in the figure is the best-fit base $\Lambda$CDM model
fitted to the temperature spectra using the PlanckTT+lowP likelihood. The
residuals shown in Fig.~\ref{fig:TEEE} are higher than expected and provide
evidence of residual instrumental systematics in the $TE$ and $EE$ spectra.
It is currently believed that the dominant source of errors is beam mismatch
generating leakage from temperature to polarization at low levels of a few
$\mu$K$^2$ in ${\cal D}_\ell$. We urge caution in the interpretation of any
features in these spectra, which should be viewed as work in progress.
Nonetheless, we find a high level of consistency in results between the TT
and the full TT+TE+EE likelihoods. Furthermore, the cosmological parameters
(which do not depend strongly on $\tau$) derived from the $TE$ spectra have
comparable errors to the $TT$-derived parameters, and they are consistent to
within typically 0.5$\,\sigma$ or better.

\subsubsection{Number of modes}

One way of assessing the constraining power contained in a particular
measurement of CMB anisotropies is to determine the effective number of
$a_{\ell m}$ modes that have been measured.  This is equivalent to
estimating 2 times the square of the total S/N in the power spectra, a measure
that contains {\it all\/} the available cosmological information if we
assume that the anisotropies are purely Gaussian (and hence ignore all
non-Gaussian information coming from lensing, the CIB,
cross-correlations with other probes, etc.).  Carrying out this procedure
for the \Planck\ 2013 $TT$ power spectrum data provided in
\cite{planck2013-p08} and \cite{planck2013-p11}, yields the number
826\,000 (which includes the effects of instrumental noise, cosmic
variance and masking).  The 2015 $TT$ data have increased this value to
1\,114\,000, with $TE$ and $EE$ adding a further 60\,000 and 96\,000 modes,
respectively.\footnote{Here we have used the basic (and conservative)
likelihood; more modes are effectively probed by \Planck\ if one includes
larger sky fractions.}  From this perspective the 2015 \Planck\ data constrain
approximately 55\,\% more modes than in the 2013 release.  Of course this
is not the whole story, since some pieces of information are more valuable
than others, and in fact \Planck\ is able to place considerably tighter
constraints on particular parameters (e.g., reionization optical depth
or certain extensions to the base $\Lambda$CDM model) by including new
polarization data.

\subsubsection{Peaks in the power spectra}

\def\cltt{$C_\ell^{TT}$}
\def\clte{$C_\ell^{TE}$}
\def\cltb{$C_\ell^{TB}$}
\def\clee{$C_\ell^{EE}$}
\def\clbb{$C_\ell^{BB}$}
\def\cleb{$C_\ell^{EB}$}
\def\dlee{${\cal D}_\ell^{EE}$}
\def\dlbb{${\cal D}_\ell^{BB}$}
\def\dlte{${\cal D}_\ell^{TE}$}
\def\dltb{${\cal D}_\ell^{TB}$}
\def\dleb{${\cal D}_\ell^{EB}$}

The fidelity with which \Planck\ has measured the \cltt, \clte,
and \clee\ power spectra enables us to precisely estimate the underlying
cosmological parameters (see Sect.~\ref{sec:CMBcosmology}), but the $C_\ell$s
are themselves a set of cosmological observables, whose properties can be
described independently of any model.  The acoustic peaks in the $C_\ell$s
reveal the underlying physics of oscillating sound waves in the coupled
photon-baryon fluid, driven by dark matter potential perturbations, and one
can talk about the fundamental mode, the first harmonic, and so on.  Hence
it is natural to ask about the positions of the individual peaks in the power
spectra as empirical information that becomes part of the canon of facts now
known about our Universe.

Here we use the \Planck\ data directly to fit for the multipoles of individual
features in the measured $TT$, $TE$, and $EE$ power spectra.  We specifically
use the CMB-only bandpowers given in \cite{planck2014-a13}, adopting the same
weighting scheme within each bin.  Fitting for the positions and amplitudes of
features in the bandpowers is a topic with a long history, with approaches
becoming more sophisticated as the fidelity of the data improved
(e.g., \citealt{Scott94}, 
\citealt{Hancock97}, \citealt{Knox00}, \citealt{deBernardis02}, 
\citealt{bond03}, \citealt{page2003b}, \citealt{Durrer03}, 
\citealt{Readhead04}, \citealt{Jones06}, \citealt{hinshaw2007}, 
\citealt{Corasaniti08}, \citealt{Pryke09}).
Following earlier approaches, we fit Gaussians to the peaks in \cltt\ and
\clee, but parabolas to the \clte\ peaks.  We have to remove a featureless
damping tail to fit the higher \cltt\ region and care has to be taken to treat
the lowest-$\ell$ ``recombination'' peak in \clee. We explicitly focus on peaks
(ignoring the troughs) in the conventional quantity
${\cal D}_\ell\equiv\ell(\ell+1)C_\ell/2\pi$; note that other quantities
(e.g., $C_\ell$) will have maxima at slightly different multipoles, and that
the choice of bandpowers to use for fitting each peak is somewhat subjective.
Our numerical values, presented in in Table~\ref{table_peaks}, are consistent
with previous estimates, but with a dramatically increased number of peaks
measured.  \Planck\ detects 19 peaks (with the eighth \cltt\ peak detection
being marginal), and an essentially equivalent number of troughs.

\begin{table}[htbp]
\newdimen\tblskip \tblskip=5pt
\caption{\Planck\ peak positions and amplitudes.}
\label{table_peaks}
\vskip -9mm
\footnotesize
\setbox\tablebox=\vbox{
 \newdimen\digitwidth
 \setbox0=\hbox{\rm 0}
 \digitwidth=\wd0
 \catcode`*=\active
 \def*{\kern\digitwidth}
 \newdimen\signwidth
 \setbox0=\hbox{+}
 \signwidth=\wd0
 \catcode`!=\active
 \def!{\kern\signwidth}
 \newdimen\pointwidth
 \setbox0=\hbox{\rm .}
 \pointwidth=\wd0
 \catcode`?=\active
 \def?{\kern\pointwidth}
 \halign{\tabskip=0pt\hbox to 1.0in{#\leaderfil}\tabskip=2em&
       \hfil#\hfil&
       \hfil#\hfil\tabskip=0pt\cr
\noalign{\doubleline}
\multispan3\hfil\sc Peak\hfil\cr
\noalign{\vskip -3pt}
\multispan3\hrulefill\cr
\noalign{\vskip 3pt}
\omit\hfil Number\hfil&Position [$\ell]$&Amplitude [$\muK^2$]\cr
\noalign{\vskip 5pt\hrule\vskip 8pt}
\omit $TT$ power spectrum\hfil\cr
\noalign{\vskip 10pt}
\hglue 1.5em First&   $*220.0\pm*0.5$& $5717?*\pm35?*$\cr
\hglue 1.5em Second&  $*537.5\pm*0.7$& $2582?*\pm11?*$\cr
\hglue 1.5em Third&   $*810.8\pm*0.7$& $2523?*\pm10?*$\cr
\hglue 1.5em Fourth&  $1120.9\pm*1.0$& $1237?*\pm*4?*$\cr
\hglue 1.5em Fifth&   $1444.2\pm*1.1$& $*797.1\pm*3.1$\cr
\hglue 1.5em Sixth&   $1776?*\pm*5?*$& $*377.4\pm*2.9$\cr
\hglue 1.5em Seventh& $2081?*\pm25?*$& $*214?*\pm*4?*$\cr
\hglue 1.5em Eighth&  $2395?*\pm24?*$& $*105?*\pm*4?*$\cr
\noalign{\vskip 10pt}
\omit $TE$ power spectrum\hfil\cr
\noalign{\vskip 5pt}
\hglue 1.5em First&   $*308.5\pm0.4$& $115.9\pm1.1$\cr
\hglue 1.5em Second&  $*595.3\pm0.7$& $*28.6\pm1.1$\cr
\hglue 1.5em Third&   $*916.9\pm0.5$& $*58.4\pm1.0$\cr
\hglue 1.5em Fourth&  $1224?*\pm1.0$& $**0.7\pm0.5$\cr
\hglue 1.5em Fifth&   $1536?*\pm2.8$& $**5.6\pm1.3$\cr
\hglue 1.5em Sixth&   $1861?*\pm4?*$& $**1.2\pm1.0$\cr
\noalign{\vskip 10pt}
\omit $EE$ power spectrum\hfil\cr
\noalign{\vskip 5pt}
\hglue 1.5em First&   $*137?*\pm6?*$& $*1.15\pm0.07$\cr
\hglue 1.5em Second&  $*397.2\pm0.5$& $22.04\pm0.14$\cr
\hglue 1.5em Third&   $*690.8\pm0.6$& $37.35\pm0.25$\cr
\hglue 1.5em Fourth&  $*992.1\pm1.3$& $41.8*\pm0.5*$\cr
\hglue 1.5em Fifth&   $1296?*\pm4?*$& $31.6*\pm1.0*$\cr
\noalign{\vskip 4pt\hrule\vskip 6pt}
}}
\endPlancktable
\end{table}

\subsection{CMB lensing products}
\label{sec:CMBLensing}

\begin{figure*}[!htpb]
\includegraphics[width=18cm]{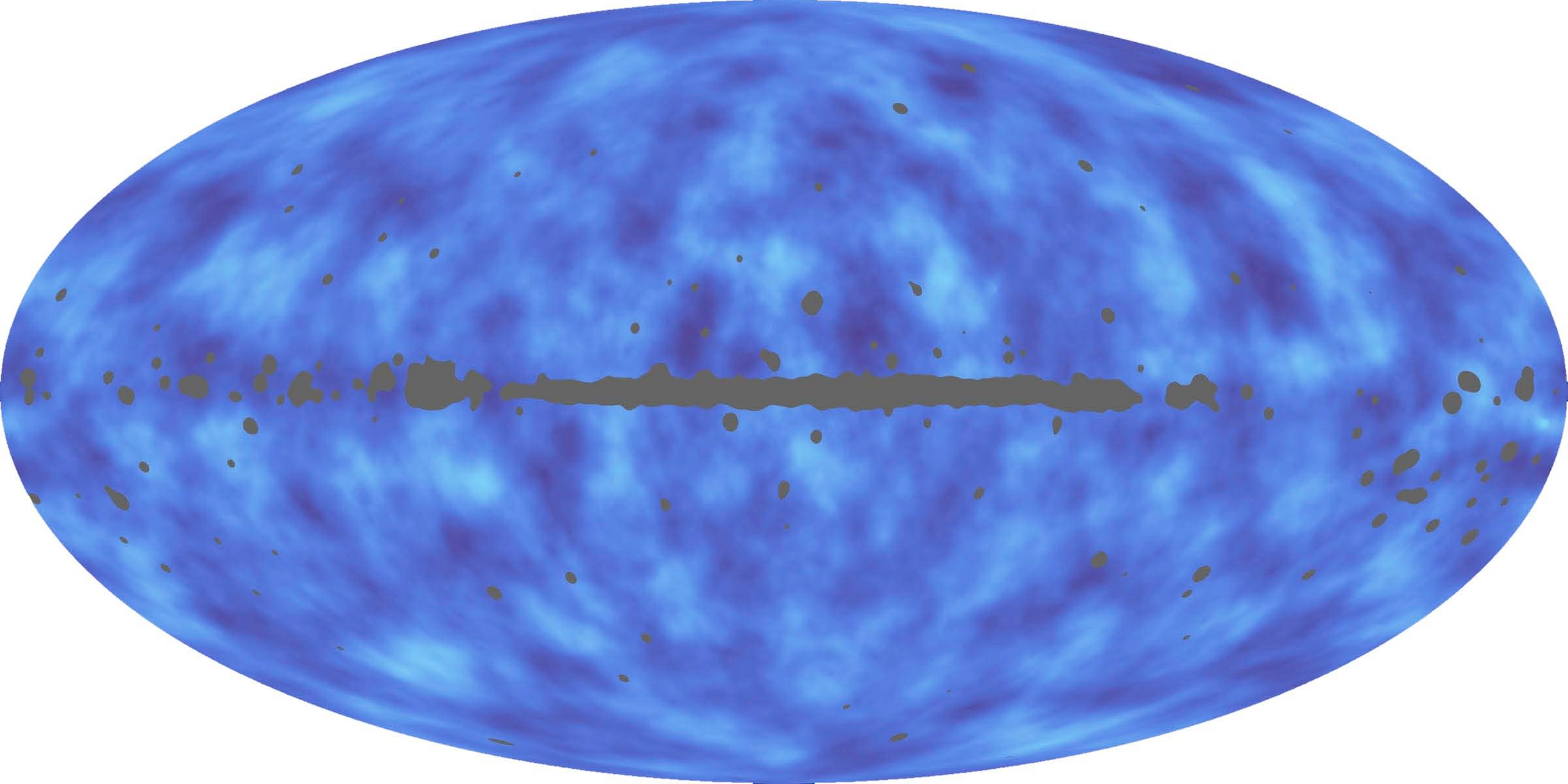}
\caption{Wiener-filtered lensing potential estimate with minimal masking
(using the {\tt NILC} component separated map), in Galactic coordinates with
a Mollweide projection \citep{planck2014-a17}. The reconstruction has been
bandlimited to $8 \le L \le 2048$ (where, following convention, $L$ is used as
the multipole index in the lensing power spectrum).}
\label{fig:xphimap}
\end{figure*}

\Planck\ is the first experiment with the sky coverage, resolution and
sensitivity to form a full-sky reconstruction of the projected mass, along
every line of sight back to the surface of last scattering.
Figure~\ref{fig:xphimap} shows the 2015 \Planck\ lensing map
\citep{planck2014-a17} using as input the
CMB maps produced by the {\tt SMICA} code. The map combines five possible
quadratic estimators based on the various correlations of the CMB temperature
($T$) and polarization ($E$ and $B$).

\subsection{Likelihood code}
\label{sec:Like}

\subsubsection{CMB likelihood}
\label{sec:CMBLike}	

We adopt the same general methodology for the 2015 likelihood as in 2013, extended to include Planck polarization data. The likelihood is a hybrid combination of a low-multipole pixelbased likelihood with a high-multipole likelihood constructed from cross-spectra -- see \citet{planck2014-a13} for details. Note that we use the notation ``Planck TT'' when we are referring to the likelihood deriving from the $TT$ spectrum.

At low multipoles we now use \planck\ instead of WMAP for polarization information.  The 70\,GHz LFI polarization maps are cleaned with the LFI 30\,GHz and HFI 353\,GHz maps to mitigate foreground contamination. Based on null tests, the LFI-cleaned polarization maps are used over 46\,\% of the sky in the low multipole likelihood (referred to as ``lowP'').  The {\tt Commander} temperature solution, constructed from all \Planck\ frequency maps, together with the Haslam 408\,MHz and WMAP maps, is used over 93\,\% of the sky. The temperature and polarization data are then treated in a unified low-resolution pixel-based
manner for the multipole range $\ell=2$ to $29$.

The high-$\ell$ likelihood uses pseudo-$C_\ell$ cross-spectra from HFI 100, 143, and 217\,GHz maps in a ``fiducial Gaussian'' approximation, employing analytic covariance matrices calculated for a fiducial cosmological model.  Unresolved foregrounds are modelled parametrically using power spectrum templates, with only minor changes to the model adopted in the 2013 analysis.  The baseline high-multipole likelihood uses cross-spectra between frequency maps constructed from the first and second halves of the full mission data, to reduce any possible biases from co-temporal systematics.  We also make more aggressive use of sky at all frequencies in the 2015 analysis.  The most significant change is the addition of the option to include the $TE$ and $EE$ power
spectra and the associated covariance matrices into the scheme, to form a combined $TT$, $TE$, $EE$ likelihood at high multipoles (referred to as PlanckTT,TE,EE).  Although we find firm evidence for systematics associated with temperature-to-polarization leakage in the $TE$ and $EE$ spectra, these systematics are at low levels.  We find a high level of consistency between the $TT$, $TE$, and $EE$ spectra for the cosmological models analyzed in the 2015 \Planck\ papers. However, in this data release, we regard the combined $TT$, $TE$, and $EE$ \Planck\ results as preliminary and hence recommend the $TT$ likelihood as the baseline.

\subsubsection{Lensing likelihood}
\label{sec:LensLike}

Our power spectrum measurement constrains the lensing potential power spectrum to a precision of $\pm2.5\,\%$, corresponding to a $1.2\,\%$ constraint on the overall amplitude of matter fluctuations ($\sigma_8$), a measurement with considerable power for constraining cosmology.  We have constructed two Gaussian bandpower likelihoods based on the lensing power spectrum measurement
described in Sect.~\ref{sec:CMBLike} and plotted in Fig.~\ref{fig:phispec}.  The first likelihood uses a conservative bandpower range, $40 \le L \le 400$, with linear binning, following the temperature-only likelihood released in 2013.  The second likelihood uses a more aggressive range with $8 \le L \le 2048$, and bins that are linear in $L^{0.6}$.  Both likelihoods
incorporate temperature and polarization data.  We incorporate uncertainties in the estimator normalization and bias corrections directly into the likelihood, using precalculated derivatives of these terms with respect to the CMB temperature and polarization power spectra.  The construction of the lensing likelihood is described in \cite{planck2014-a17}, and its cosmological implications are discussed in detail in \cite{planck2014-a15}.

\section{Astrophysics products}
\label{sec:AstroProds}

\subsection{The Second Planck Catalogue of Compact Sources}
\label{sec:PCCS2}

The Second Planck Catalogue of Compact Sources
\citep[PCCS2;][]{planck2014-a35} is the catalogue of sources detected from
the full duration of \Planck\ operations, referred to as the ``extended''
mission. It consists of compact sources, both Galactic and extragalactic,
detected over the entire sky. Compact sources are detected in the
single-frequency maps and assigned to one of two sub-catalogues, the PCCS2 or
PCCS2E. The first of these allows the user to produce additional sub-catalogues
at higher reliabilities than the target 80\,\%\ reliability of the full
catalogue. The second list contains sources whose reliability cannot be
estimated, because they are embedded in a bright and complex (e.g. filamentary) 
background of emission.

The total number of sources in the catalogue ranges from 1560 at
30\,GHz up to 48\,181 sources at 857\,GHz. Both sub-catalogues include
polarization measurements, in the form of polarized flux densities and
orientation angles, or upper-limits, for all seven polarization-sensitive
Planck channels. The number of sources with polarization information (other
than upper-limits) in the catalogue ranges from 113 in the lowest polarized
frequency channel (30\,GHz) up to 666 in the highest polarized frequency
channel (353\,GHz). The improved data processing of the full-mission maps and
their reduced instrumental noise levels allow us to increase the number of
objects in the catalogue, improving its completeness for the target 80\,\%\
reliability as compared with the previous versions, the PCCS
\citep{planck2013-p05} and the Early Release Compact Source Catalogue
\citep[ERCSC;][]{planck2011-6.1}.
The improvements are most pronounced for the LFI channels, due to the much
larger increase in the data available. The completeness of the 857\,GHz
channel, however, has not improved; this is due to a more refined
reliability assessment, which resulted in a higher S/N threshold being applied
in the selection function of this catalogue. The reliability of the PCCS2 catalogue at
857\,GHz, however, is higher than that of the PCCS.

\subsection{The Second Planck Catalogue of Clusters}
\label{sec:PSZ2}

\newcommand{\yfive}{$Y_{500}$}
\newcommand{\yfiver}{$Y_{5\text{R}500}$}
\newcommand{\pszone}{PSZ1}
\newcommand{\psztwo}{PSZ2}

\begin{figure}
\begin{center}
\includegraphics[angle=0,width=0.48\textwidth]{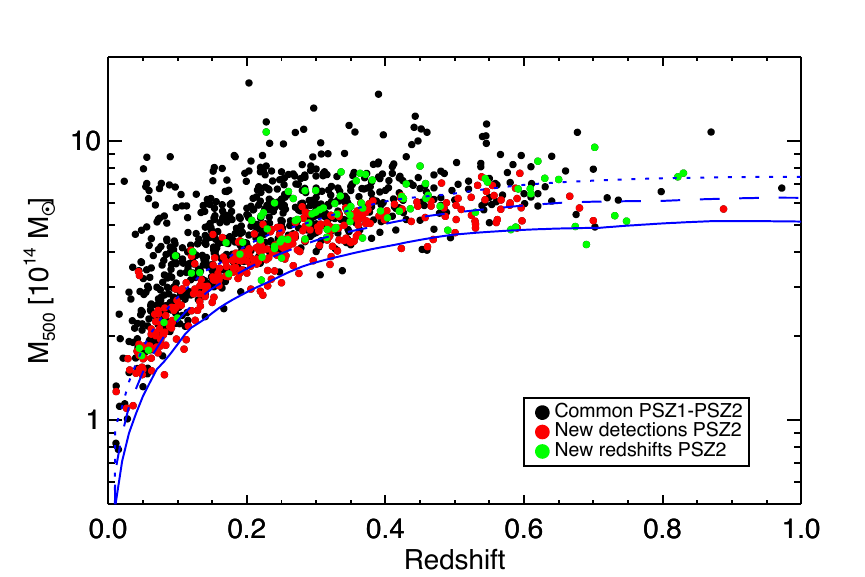}  
\caption{Distribution of the 1093 \psztwo\ clusters with counterparts with
known redshift in the $M_{500}$--$z$ plane.  New \psztwo-detected clusters are
indicated with red dots, while commmon \pszone\ and \psztwo\ clusters are
indicated by black dots. Green dots mark the common \psztwo--\pszone\
detections where we have updated the redshift value for the \psztwo.
The solid, dashed, and dotted lines indicate the limiting mass at 20\,\%,
50\,\%, and 80\,\% survey completeness, respectively.}
\label{fig:msz_z_distr}
\end{center}
\end{figure}

The Second \Planck\ Catalogue of SZ Sources
\citep[\psztwo;][]{planck2014-a36}, based on the full mission data, is the
largest SZ-selected sample of galaxy clusters yet produced and the deepest
all-sky catalogue of galaxy clusters.  It contains 1653 detections, of which
1203 are confirmed clusters with identified counterparts in external data sets,
and is the first SZ-selected cluster survey containing $>10^3$ confirmed
clusters.  A total of 937 sources from the half-mission catalogue (PSZ1)
released in 2013 are included, as well as 716 new detections.  The
completeness, which is provided as a product with the catalogue, is determined
using simulated signal injection, validated through comparison to external
data, and is shown to be consistent with semi-analytic expectations.  The
reliability is characterized using high-fidelity simulated observations and a
machine-learning-based quality assessment, which together place a robust lower
limit of 83\,\% on the purity.  Using simulations, we find that the
\yfiver\ estimates are robust to pressure-profile variations and beam
systematics; however, accurate conversion to \yfive\ requires the use of prior
information on the cluster extent. Results of a multi-wavelength search for
counterparts in ancillary data (which makes use of radio, microwave, infra-red,
optical, and X-ray data-sets, and which places emphasis on the robustness of
the counterpart match) are included in the catalogue.  We discuss the physical
properties of the new sample and identify a population of low-redshift X-ray
under-luminous clusters revealed by SZ selection.
Figure~\ref{fig:msz_z_distr} shows the masses and redshifts for the 1093
\psztwo\ clusters with known redshifts.

\subsection{The Planck Catalogue of Galactic Cold Clumps}
\label{sec:PCCC}

The {\Planck} catalogue of Galactic Cold Clumps \citep[PGCC,][]{planck2014-a37}
contains Galactic sources that have been identified as cold sources in
\Planck\ data.  We ran the {\tt CoCoCoDeT} \citep{Montier2010} multi-frequency
point source detection algorithm on the \Planck\ 857, 545, and 353\,GHz data
and the IRIS 3\,THz data \citep{Miville2005}, at a resolution of 5\arcm.
This selects point sources exhibiting submillimetre excess in the 857, 545,
and 353\,GHz {\Planck} bands simultaneously, compared to the average colour of
the background, which is typical of sources appearing colder than their
environment.

The PGCC catalogue is the full version of the Early Cold Core (ECC) catalogue
released in 2011, which was part of the ERCSC \citep{planck2011-1.10}.  The
ECC catalogue was built on the first 295 days of \Planck\ data, and contains
915 sources selected to ensure $T < 14$\,K and  S/N\,$>$\,15.  A
statistical description of the ECC and the extended catalogue (including
sources at all temperatures and with S/N\,$>$\,4) is given in
\cite{planck2011-7.7a}, while a detailed description of a subsample of 10
sources was presented in \cite{planck2011-7.7b}.  The PGCC catalogue, included
in the 2015 \Planck\ release, has now been built on the full \Planck\ mission
data, and contains 13\,188 Galactic sources, plus 54 sources located in the
Large and Small Magellanic Clouds.

The morphology of each source is obtained using a Gaussian elliptical fit,
which is then used to estimate flux densities in all bands through aperture
photometry. Depending on the S/N of the flux density estimates, three
categories of sources are identified: 6993 sources with reliable flux
densities in all bands ({\tt FLUX\_QUALITY}=1); 3755 sources with flux density
estimates in all bands except 3\,THz ({\tt FLUX\_QUALITY}=2), which are
considered very cold candidates; and 2440 sources without reliable flux density
estimates ({\tt FLUX\_QUALITY}=3), usually due to a complex environment, which
are considered poor candidates.

Distance estimates have been obtained for 5574 PGCC sources by combining seven
different methods.  While PGCC sources are mainly located in the solar
neighbourhood, with 88\,\% of sources with reliable distance estimates lying
within 2\,kpc of the Sun, distance estimates range from a few hundred parsecs
towards local molecular clouds to 10.5\,kpc towards the Galactic centre.

The temperature of each source is obtained by fitting a modified blackbody to
the spectral energy density from 3\,THz to 353\,GHz, considering the spectral
index $\beta$ as a free parameter when possible.  PGCC sources have an average
temperature of 13--14.5\,K, depending on flux quality category, and range from
5.8 to 20\,K.  Other physical parameters have been derived, such as the
$\mathrm{H}_2$ column density, the physical size, the mass, the density, and
the luminosity.  It appears that the PGCC contains a large variety of objects
with very different properties, from compact and dense cores to large and
massive molecular clouds, located all over the sky.  While a large
{\it Herschel\/} program (HKP-GCC) already followed-up 315 PGCC sources with
the PACS and SPIRE instruments, the PGCC catalogue is the first all-sky
sample of Galactic cold sources obtained with a homogeneous method, and hence
represents a goldmine for investigations of the early phases of star formation
in various environments.

\begin{figure}[th!]
\center
\includegraphics[width=88mm]{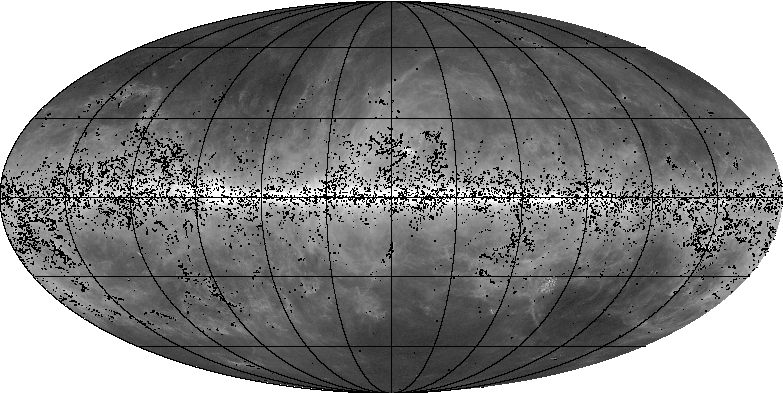} 
\caption{All-sky distribution of the 13\,188 PGCC Galactic cold clumps (black
dots) and the 54 cold sources (grey dots) located in the Large and Small
Magellanic Clouds. The background map is the 857\,GHz \Planck\ band, shown in
logarithmic scale from $10^{-2}$ to $10^2$\,$\rm{MJy\,sr^{-1}}$.}
\label{fig:allsky}
\end{figure}

\subsection{Diffuse Galactic foregrounds from CMB component separation}
\label{sec:CommFGs}

As was done in 2013, in \citet{planck2014-a12} we establish a single parametric
model of the microwave sky, accounting simultaneously for all significant
diffuse astrophysical components and relevant instrumental effects using the
Bayesian \texttt{Commander} analysis framework
\citep{eriksen2004,eriksen2006,eriksen2008}. The 2015 analysis is extended
in multiple directions. First, instead of 15.5~months of temperature data,
the new analysis includes the full \Planck\ mission data---50\,months of LFI
and 29\,months of HFI data---in both temperature and
polarization. Second, we now also include the 9-year
\WMAP\ observations between 23 and 94\,GHz \citep{bennett2012} and a
408\,MHz survey map \citep{haslam1982}, providing enough frequency
constraints to decompose the low-frequency foregrounds into separate
synchrotron, free-free, and spinning dust components.  Third, we now
include the \Planck\ 545 and 857\,GHz frequency bands, allowing us to
constrain the thermal dust temperature and emissivity index with
greater precision, thereby reducing degeneracies between CMB, CO, and
free-free emission.  
Fourth, the present analysis implements a multi-resolution strategy to
provide component maps at high angular resolution. Specifically, the
CMB is recovered with angular resolution 5\arcm\ FWHM \citep{planck2014-a11},
thermal dust emission and CO $J\,{=}\,2\,{\rightarrow}\,1$
lines are recovered at 7\parcm5 FWHM, and synchrotron, free-free, and
spinning dust are recovered at 1\deg\ FWHM. 

An important difference with respect to 2013 is that we employ individual
detector and detector set maps as inputs, instead of fully combined frequency
maps. The increase in the number of input maps allows us to make many null
tests that are used to reject individual maps exhibiting significant levels of
systematic effects. In addition, in our analysis we allow our model to fit for
two important instrumental effects: relative calibration between detectors;
and bandpass uncertainties. 

The sum of these improvements allows us to reconstruct a total of six primary
emission mechanisms in temperature: CMB; synchrotron; free-free;
spinning dust; CO; and thermal dust emission---in addition to two secondary
components, namely thermal SZ emission around the Coma and Virgo regions, and
molecular line emission between 90 and 100\,GHz. For polarization, we
reconstruct three primary emission mechanisms: CMB; synchrotron; and thermal
dust. All of these components are delivered as part of the 2015 \Planck\
release.

\begin{figure*}[th!]
\center
\includegraphics[width=18cm]{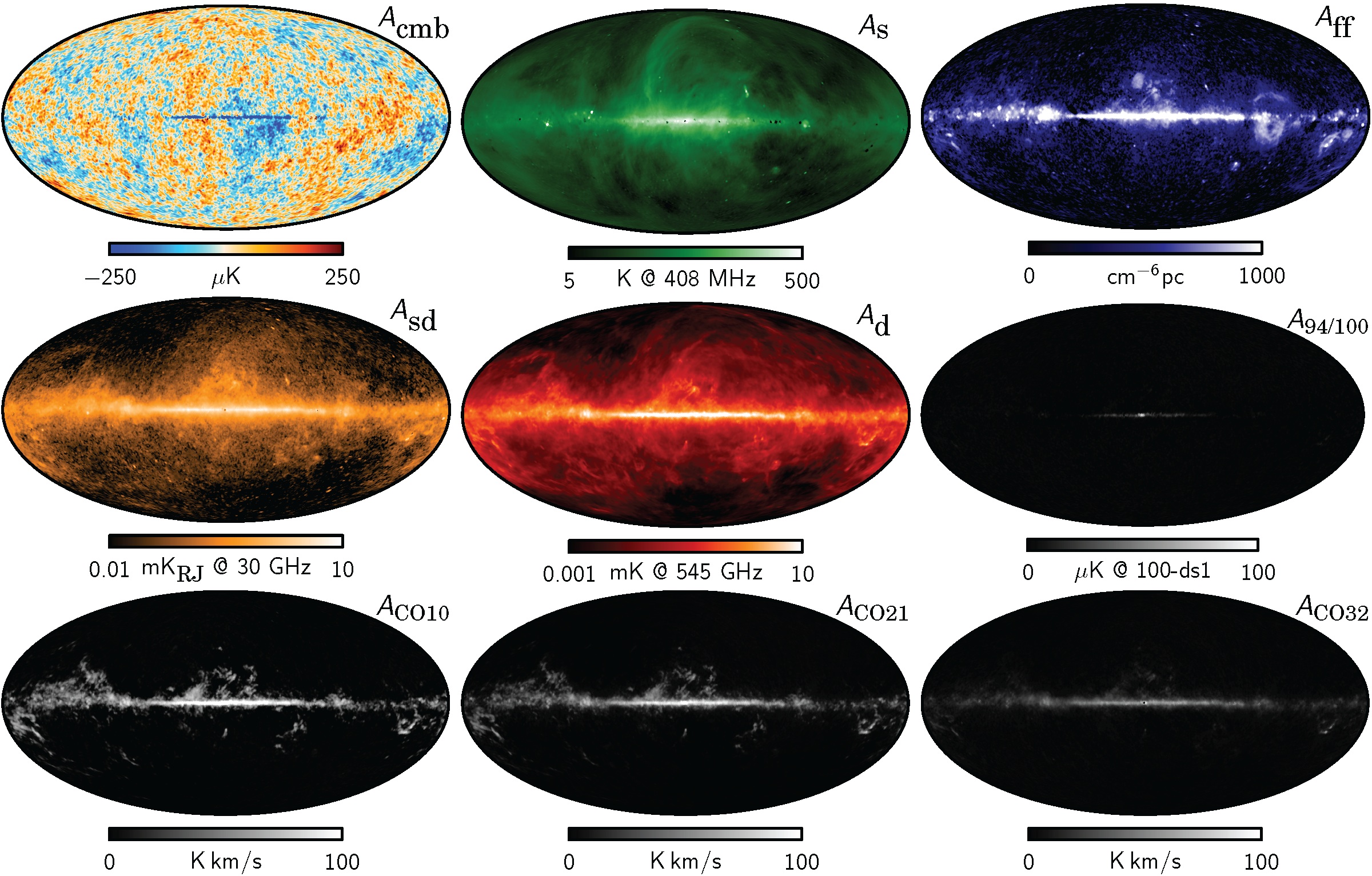} 
\caption{Maximum posterior intensity maps derived from the joint analysis of \Planck, WMAP, and 408\,MHz observations \citep{planck2014-a12}.  From left to right, top to bottom: CMB; synchrotron; free-free; spinning dust; thermal dust; line emission around 90\,GHz; CO\,$J\,{=}\,1\,{\rightarrow}\,0$; CO\,$J\,{=}\,2\,{\rightarrow}\,1$, and CO\,$J\,{=}\,3\,{\rightarrow}\,2$.}
\label{fig:Tcomps}
\end{figure*}

\begin{figure*}[th!]
\center
\includegraphics[width=15cm]{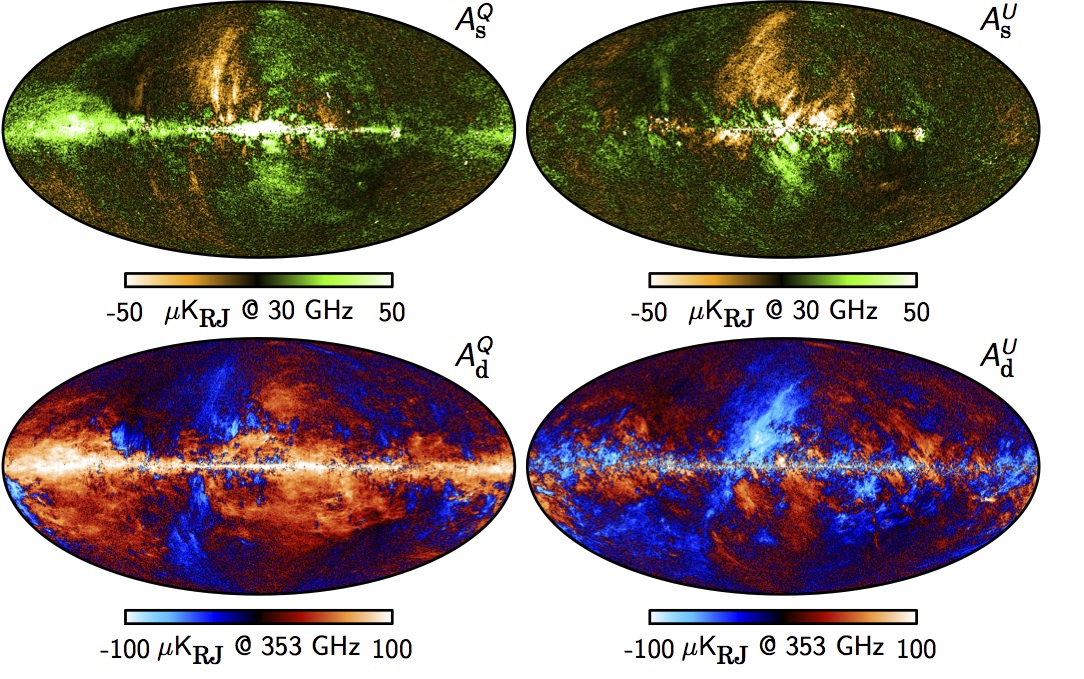} 
\caption{Maximum posterior amplitude polarization maps derived from the \Planck\ observations between 30 and 353\,GHz \citep{planck2014-a12}. The left and right columns show the Stokes $Q$ and $U$ parameters, respectively. Rows show, from top to bottom: CMB; synchrotron polarization at 30\,GHz; and thermal dust polarization at 353\,GHz. The CMB map has been highpass-filtered with a cosine-apodized filter between $\ell=20$ and 40, and the Galactic plane (defined by the 17\,\% CPM83 mask) has been replaced with a constrained Gaussian realization \citep{planck2014-a11}.}
\label{fig:QUcomps}
\end{figure*}

\begin{figure*}[th!]
\center
\includegraphics[width=9cm]{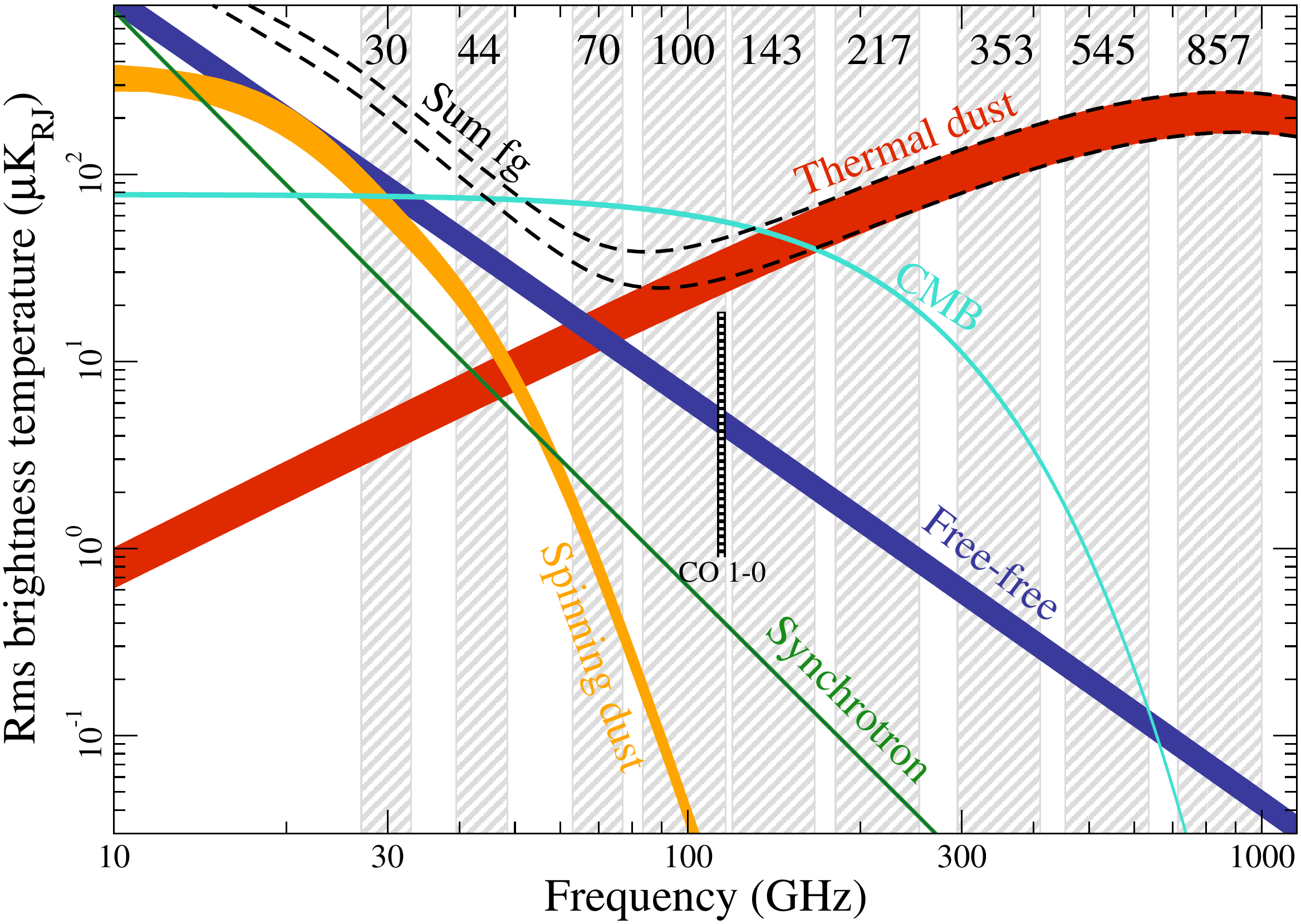}
\includegraphics[width=9cm]{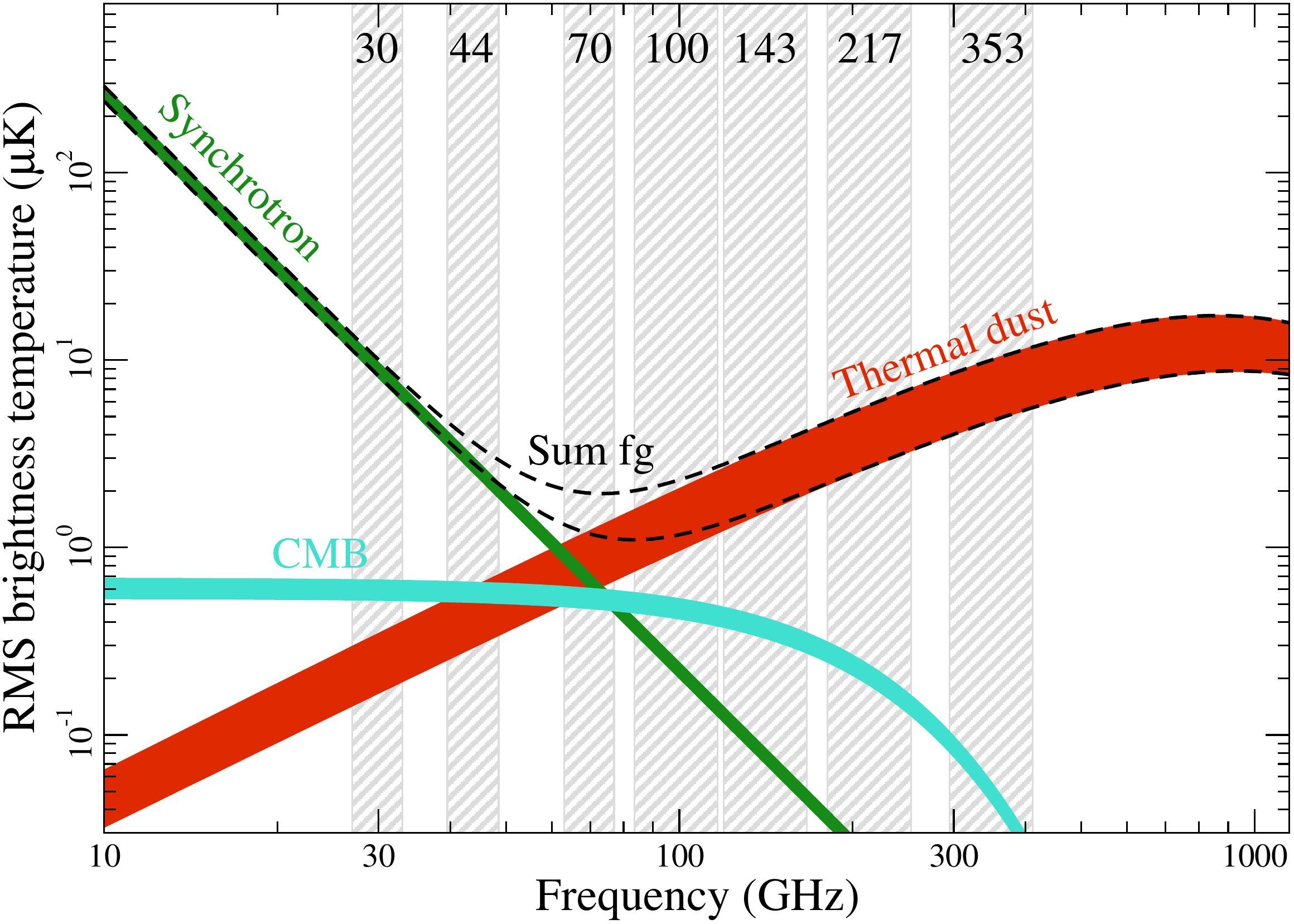} 
\caption{Brightness temperature rms of the high-latitude sky as a function of frequency and astrophysical component for temperature ({\it left\/}) and polarization ({\it right\/}). For temperature, each component is smoothed to an angular resolution of 1\deg\ FWHM, and the lower and upper edges of each line are defined by masks covering 81 and 93\,\% of the sky, respectively. For polarization, the corresponding smoothing scale is 40\arcm, and the sky fractions are 73 and 93\,\%. }
\label{fig:foregroundspectra}
\end{figure*}

Figures~\ref{fig:Tcomps} and \ref{fig:QUcomps} \citep{planck2014-a12} show the
diffuse high-latitude Galactic foreground components determined from component separation in
temperature and polarization.  Figure~\ref{fig:foregroundspectra} shows the
spectra of fluctuations of diffuse foreground components in temperature and
polarization, compared to the CMB.  The sky model presented in this paper
provides an impressive fit to the current data, with temperature residuals at
the few microkelvin level at high latitudes across the CMB-dominated
frequencies, and with median fractional errors below 1\,\% in the
Galactic plane across the \Planck\ frequencies. For polarization, the
residuals are statistically consistent with instrumental noise at high
latitudes, but limited by significant temperature-to-polarization
leakage in the Galactic plane. Overall, this model represents the most
accurate and complete description currently available of the
astrophysical sky between 20 and 857\,GHz.

\subsection{Carbon monoxide emission}
\label{sec:CO}

Carbon monoxide emission lines are present in all HFI frequency bands, except
143\,GHz. Using component separation techniques, the three lowest rotational
transitions can be extracted from \Planck\ data, providing full sky maps of
the CO\,$J\,{=}\,1\,{\rightarrow}\,0$, $J\,{=}\,2\,{\rightarrow}\,1$, and
$J\,{=}\,3\,{\rightarrow}\,2$ transitions \citep{planck2013-p03a}. For the 2015
release, data from the full mission and better control of systematic errors
lead to better maps.  Table~\ref{tab:co_summary} summarizes the products.
Figure~\ref{fig:Tcomps} shows the {\tt Commander} maps of all three
transitions.

\begin{table*}[tb]                                                                                                                                                  
\begingroup                                                                                                                                      
\newdimen\tblskip \tblskip=5pt
\caption{Summary of main CO product characteristics.}
\label{tab:co_summary}
\vskip -8mm
\footnotesize
\setbox\tablebox=\vbox{
\newdimen\digitwidth
\setbox0=\hbox{\rm 0}
\digitwidth=\wd0
\catcode`*=\active
\def*{\kern\digitwidth}
\newdimen\signwidth
\setbox0=\hbox{+}
\signwidth=\wd0
\catcode`!=\active
\def!{\kern\signwidth}
\newdimen\decimalwidth
\setbox0=\hbox{.}
\decimalwidth=\wd0
\catcode`@=\active
\def@{\kern\signwidth}
\halign{ \hbox to 0.7in{#\leaderfil}\tabskip=1em&
  \hfil#\hfil\tabskip=1em&
  \hfil#\hfil\tabskip=1em&
  \hfil#\hfil\tabskip=1em&
  \hfil#\hfil\tabskip=0.5em&
  \hfil#\hfil\tabskip=1em&
  \hfil#\hfil\tabskip=1.0em&
  \hfil#\hfil\tabskip=0pt\cr
\noalign{\doubleline}
\omit& \omit&\omit&\omit& \multispan2\hfil Noise rms [$\textrm{K}_{\textrm{RJ}}\,\textrm{km}\,\textrm{s}^{-1}$] \hfil& \multispan2\hfil Analysis details\hfil\cr
\noalign{\vskip -3pt}
\omit&\omit&\omit&\lower3pt\hbox{Resolution}&\multispan2\hrulefill&\multispan2\hrulefill\cr
\noalign{\vskip 2pt}
\omit\hfil Map\hfil& Algorithm& CO line& [arcmin]& $15\arcm$ FWHM& $60\arcm$ FWHM& Frequencies [GHz]& Model\cr
\noalign{\doubleline}
\sc Type 1& \texttt{MILCA}&$J\,{=}\,1\,{\rightarrow}\,0$& 9.6& 1.4*& 0.34*& 100 (bol maps)& CO, CMB\cr
\omit& \texttt{MILCA}&$J\,{=}\,2\,{\rightarrow}\,1$& 5.0& 0.53& 0.16*& 217 (bol maps)& CO, CMB, dust\cr
\omit& \texttt{MILCA}&$J\,{=}\,3\,{\rightarrow}\,2$& 4.8& 0.55& 0.18*& 353 (bol maps)& CO, dust\cr
\noalign{\vskip 5pt}
\sc Type 2& \texttt{MILCA}& $J\,{=}\,1\,{\rightarrow}\,0$& 15& 0.39& 0.085& 70, 100, 143, 353& CO, CMB, dust, free-free\cr
\omit& \texttt{MILCA}&$J\,{=}\,2\,{\rightarrow}\,1$& 15& 0.11& 0.042& 70, 143, 217, 353& CO, CMB, dust, free-free\cr
\noalign{\vskip 5pt}
\omit& \texttt{Commander}&$J\,{=}\,1\,{\rightarrow}\,0$& 60& $\cdots$& 0.084& 0.408--857& Full\cr
\omit& \texttt{Commander}&$J\,{=}\,2\,{\rightarrow}\,1$& 60& $\cdots$& 0.037& 0.408--857& Full\cr
\omit& \texttt{Commander}&$J\,{=}\,3\,{\rightarrow}\,2$& 60& $\cdots$& 0.060& 0.408--857& Full\cr
\noalign{\vskip 5pt}
\sc Type 3& \texttt{Commander}&$J\,{=}\,2\,{\rightarrow}\,1$\rlap{$^{\rm a}$}& 7.5& *0.090& 0.031& 143--857& CO, CMB, dust\cr
\noalign{\vskip 5pt}
\omit& \texttt{Commander-Ruler}&$J\,{=}\,1\,{\rightarrow}\,0$\rlap{$^{\rm b,c}$}&5.5&0.19&0.082&30--353&CO, CMB, dust, low-freq\cr
\noalign{\vskip 3pt\hrule\vskip 3pt}}}
\endPlancktablewide
\tablenote {{\rm a}} Formally a weighted average of CO\,$J\,{=}\,2\,{\rightarrow}\,1$
and $J\,{=}\,3\,{\rightarrow}\,2$, but strongly dominated by
CO\,$J\,{=}\,2\,{\rightarrow}\,1$.\par
\tablenote {{\rm b}} Formally a weighted average of CO\,$J\,{=}\,1\,{\rightarrow}\,0$,
 $J\,{=}\,2\,{\rightarrow}\,1$ and $J\,{=}\,3\,{\rightarrow}\,2$, but strongly dominated
 by CO\,$J\,{=}\,1\,{\rightarrow}\,0$.\par
\tablenote {{\rm c}} Only published in 2013.\par
\endgroup
\end{table*}

\begin{itemize}
\item \typeone\ maps are produced by a single-channel analysis, where
individual bolometer maps are linearly combined to produce maps of the \cooz,
\coto, and \cott\ emission lines at the native resolution of the \Planck\
maps. Although noisier than the other approaches, using information from a
single channel strongly limits contamination from other Galactic components,
such as dust or free-free emission. This makes \typeone\ maps suitable for
studying emission in the Galactic disk and CO-rich regions, but not for the
high-Galactic latitudes where the CO emission is below the noise level.

\vskip 3pt

\item \typetwo\ maps of \cooz\ and \coto\ have been obtained using
multi-channel information (i.e., using linear combination of \Planck\ channel
maps smoothed to 15\arcm). Using frequency maps, this type of map has a higher
signal-to-noise ratio, allowing for their use in fainter high-Galactic
latitude regions. They are, however, more susceptible to dust contamination,
especially for \coto, which makes them less suitable in the Galactic plane
than \typeone\ maps.

\vskip 3pt

\item A high-resolution \typethree\ map, as defined in \citet{planck2013-p03a},
is not being delivered in the 2015 data release. Alternatively, another set of
CO maps has been produced as part of the full {\tt Commander} baseline
multi-component model, which is described in \citet{planck2014-a12}.

\end{itemize}

\typeone\ and \typetwo\ maps are released with associated standard
deviation maps, error maps, and masks. The suite of tests detailed in
\citet{planck2013-p03a} has been repeated on the new \typeone\ and \typetwo\
maps, which have been found to perform as well as their 2013 counterparts,
even though small variations ($\la 2$ to 5\,K\,km\,s$^{-1}$) exist in the
Galactic plane.

\subsubsection{All-sky Sunyaev-Zeldovich emission}
\label{sec:SZDiff}

\begin{figure*}[htpb!]
\begin{center}
\includegraphics[width=18cm]{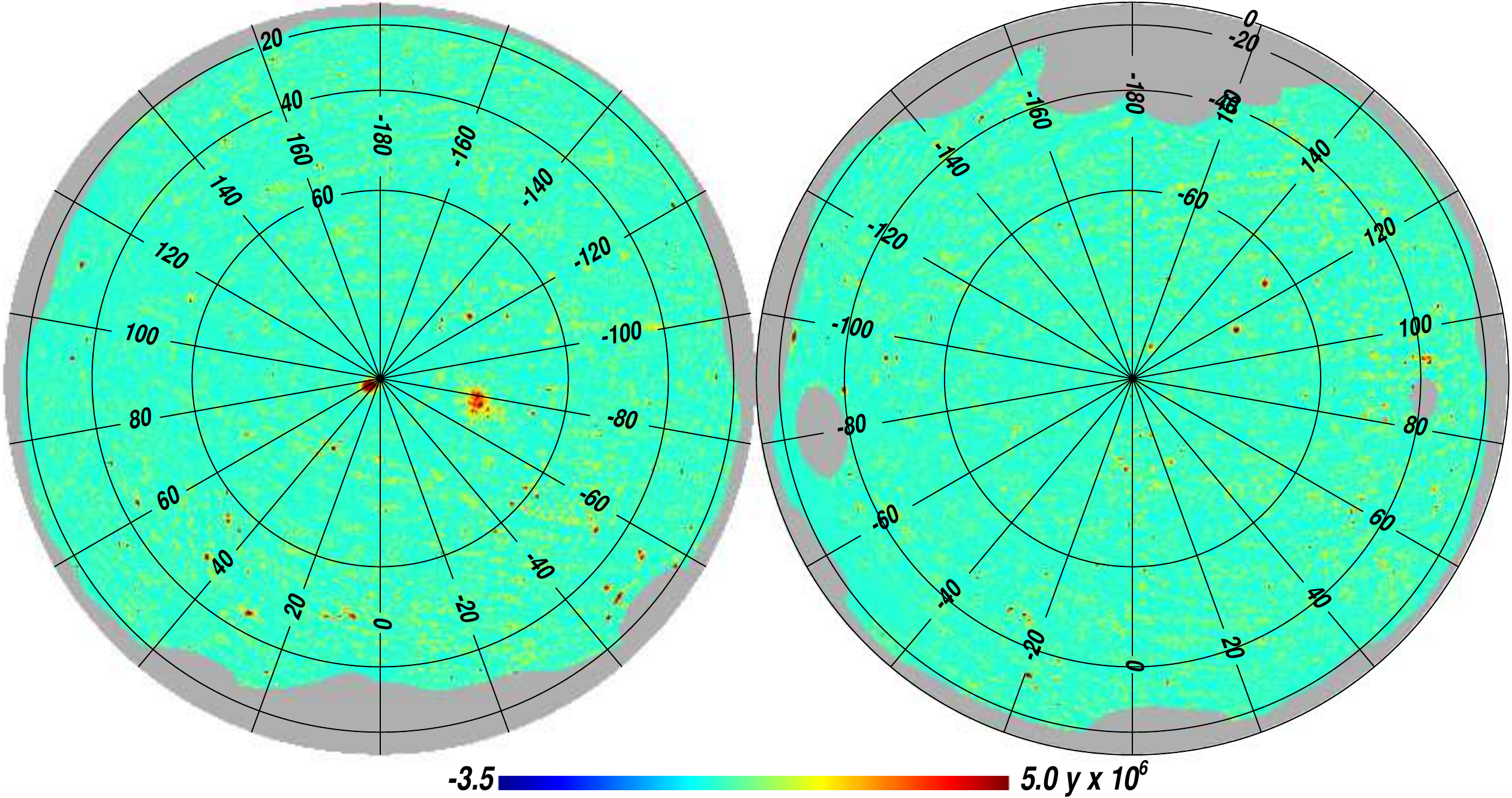}
\end{center}
\caption{Orthographic projection of the reconstructed \Planck\ all-sky y-map in Compton parameter units \citep{planck2014-a28}. For illustration purposes and to enhance the tSZ signal-to-noise ratio, the y-map has been Wiener filtered. Positive sources in the map correspond to clusters and super-clusters of galaxies with strong tSZ emission.  In particular, the Coma and Virgo clusters are clearly visible near the north Galactic pole. The region of strongest contamination from Galactic foreground emission in the Galactic plane has been partially masked.
\label{fig:planck_y_map}}
\end{figure*}

The 30 to 857\,GHz frequency channel maps from the \Planck\ satellite survey
were used to construct an all-sky map of the thermal Sunyaev-Zeldovich (tSZ)
effect \cite{planck2014-a28}.  As discussed in \cite{planck2013-p05b}, we
apply to those maps specifically tailored component separation algorithms,
{\tt MILCA} \cite{milca} and {\tt NILC} \cite{nilc},
that allow us to separate the tSZ
emission from foreground contamination, including the CMB.  An orthographic
view of the reconstructed Compton $y$-map in {\tt Healpix} pixelization is
presented in Fig.~\ref{fig:planck_y_map}.  This $y$-map has been characterized
in terms of noise properties and residual foreground contamination, mainly
thermal dust emission at large angular scales and CIB and extragalactic point
sources at small angular scales. Blindly-detected clusters in this map are
consistent with those from the PSZ2 catalogue \cite{planck2014-a36}, both in
terms of cluster numbers and integrated flux. Furthermore, by stacking
individually undetected groups and clusters of galaxies we find that the
$y$-map is consistent with tSZ emission even for low S/N regions.  Using
foreground models derived in \cite{planck2014-a29} we are able to measure the
tSZ angular power spectrum over 50\,\% of the sky. We conclude that the
$y$-map is dominated by tSZ signal in the multipole range, $20<\ell<800$.
Similar results are obtained from a high-order statistic analysis.
The reconstructed $y$-map is delivered as part of the \Planck\ 2015 release.
We also deliver a foreground mask (with known point sources and regions
with strong contamination from Galactic emission masked out), a noise variance
map, the estimated power spectrum, and the weights for the {\tt NILC} algorithm.

\section{\Planck\ 2015 cosmology results}
\label{sec:CMBcosmology}

Since their discovery, anisotropies in the CMB have contributed significantly
to defining our cosmological model and measuring its key parameters.  The
standard model of cosmology is based upon a spatially flat, expanding Universe
whose dynamics are governed by General Relativity and dominated by cold dark
matter and a cosmological constant ($\Lambda$).
The seeds of structure have Gaussian statistics and form an almost
scale-invariant spectrum of adiabatic fluctuations.
The 2015 \Planck\ data remain in excellent agreement with this paradigm, and
continue to tighten the constraints on deviations and reduce the uncertainty
on the key cosmological parameters.

The major methodological changes in the steps going from sky maps to
cosmological parameters are discussed in \citet{planck2014-a14,planck2014-a15}.
These include the use of \Planck\ polarization data instead of WMAP, changes
to the foreground masks to include more sky and dramatically reduce the number
of point source ``holes,'' minor changes to the foreground models,
improvements to the data processing and use of cross-half-mission likelihoods
\citep{planck2014-a13,planck2014-a14}.  We find good agreement with our
earlier results, with increased precision.

\providecommand{\Omm}{\Omega_{\mathrm{m}}}
\providecommand{\omb}{\omega_{\mathrm{b}}}
\providecommand{\omc}{\omega_{\mathrm{c}}}
\providecommand{\omm}{\omega_{\mathrm{m}}}

\begin{table}[tmb]
\begingroup
\newdimen\tblskip \tblskip=5pt
\caption{Parameter 68\,\% confidence levels for the base $\Lambda$CDM
cosmology computed from the \Planck\ CMB power spectra, in combination with
the CMB lensing likelihood (``lensing'').}
\label{tab:params}
\nointerlineskip
\vskip -3mm
\footnotesize
\setbox\tablebox=\vbox{
   \newdimen\digitwidth
   \setbox0=\hbox{\rm 0}
   \digitwidth=\wd0
   \catcode`*=\active
   \def*{\kern\digitwidth}
   \newdimen\signwidth
   \setbox0=\hbox{+}
   \signwidth=\wd0
   \catcode`!=\active
   \def!{\kern\signwidth}
\halign{\hbox to 1.0in{$#$\leaderfil}\tabskip 1.2em&
     \hfil$#$\hfil\tabskip 0pt\cr
\noalign{\doubleline}
\omit\hfil Parameter\hfil&\omit\hfil \Planck\ TT+lowP+lensing\hfil\cr
\noalign{\vskip 3pt\hrule\vskip 5pt}
\Omega_{\rm b}h^2&  0.02226\pm 0.00023\cr
\Omega_{\rm c}h^2&  0.1186\pm0.0020\cr
100\theta_{\rm MC}& 1.04103\pm0.00046\cr
\tau&               0.066\pm0.016\cr
\ln(10^{10}A_{\rm s})&    3.062\pm 0.029\cr
n_s&                0.9677\pm0.0060\cr
\noalign{\vskip 8pt}
H_0&                67.8\pm0.9*\cr
\Omega_{\rm m}&     0.308\pm0.012\cr
\Omega_{\rm m}h^2&  0.1415\pm0.0019\cr
\Omega_{\rm m}h^3&  0.09591\pm 0.00045\cr
\sigma_8&           0.815\pm0.009\cr
\sigma_8\Omega_{\rm m}^{0.5}&0.4521\pm0.0088\cr
{\rm Age/Gyr}& 13.799\pm 0.038*\cr
r_{\rm drag}& 147.60\pm 0.43**\cr
k_{\rm eq}& 0.01027\pm 0.00014\cr
\noalign{\vskip 5pt\hrule\vskip 3pt}}}
\endPlancktable
\endgroup
\end{table}

\subsection{Cosmological parameters}
\label{sec:CosmoPar}

\Planck's measurements of the cosmological parameters derived from the full
mission are presented and discussed in \citet{planck2014-a15}.  As in our
previous release, the data are in excellent agreement with the predictions of
the 6-parameter $\Lambda$CDM model (see Table~\ref{tab:params}), with
parameters tightly constrained by the angular power spectrum.  The best-fit
model parameters from the full mission are typically within a fraction of a
standard deviation of their results from \citet{planck2013-p11}, with no
outliers. The constraints on the parameters of the base $\Lambda$CDM model have
improved by up to a factor of 3.  The largest shifts are in the scalar spectral
index, $n_{\rm s}$, which has increased by $0.7\,\sigma$, and the baryon
density, $\omega_{\rm b}$, which has increased by $0.6\,\sigma$.  Both of these
shifts are partly due to correction of a systematic error that contributed to
a loss of power near $\ell=1800$ in the 2013 results \citet{planck2014-a15}.
This systematic also biased the inferences on $H_0$ slightly low (by less than
$0.5\,\sigma$).  In addition, the overall amplitude of the observed spectrum
has shifted upwards by $2\,\%$ (in power), due to a calibration change, and
the optical depth to Thomson scattering, $\tau$, has shifted down by nearly
$1\,\sigma$.  These shifts approximately cancel in the derived normalization
of the matter power spectrum.  The remaining shifts are consistent with the
known changes in noise level, time-stream filtering, absolute calibration,
beams, and other aspects of the data processing.

Both the angular size of the sound horizon, $\theta_\ast$, and the cold dark
matter density, $\omega_{\rm c}$, have become significantly better determined.
The data at high-$\ell$ are now so precise, and the polarization data so
constraining, that we not only see very strong evidence for three species of
light neutrinos, but can measure the effective viscosity of the neutrino
``fluid'' to be non-zero at the $9\,\sigma$ level.  The constraint on the
baryon density, $\omega_{\rm b}$, is now comparable with the best quoted
errors from big bang nucleosynthesis and suggests the possibility of
calibrating nuclear capture cross-sections from CMB observations.  The
addition of polarization data has improved by an order of magnitude our upper
limit on the annihilation rate of dark matter.

Despite trying a wide range of extensions to the basic, 6-parameter
$\Lambda$CDM model, we find no significant evidence for a failure of the
model.  Within each extension of the parameter space, the default parameter
values for the $\Lambda$CDM model remain a good fit to the data.  This
continues to hold when we combine the \Planck\ data with other measurements,
such as the distance scale measured by baryon acoustic oscillations (BAO) in
galaxy surveys or Type~Ia supernovae, or the growth of structure determined by
redshift-space distortions.  Since our best-fit cosmology has shifted by very
little since our 2013 release, we continue to see tensions with some analyses
of other astrophysical data sets (e.g., the abundance of clusters of galaxies
and weak gravitational lensing of galaxies or cosmic shear, and distances
measured by BAO in the Ly$\alpha$ forest at high-$z$).  \citet{planck2014-a15}
shows that these tensions cannot be resolved with standard single parameter
extensions of the base $\Lambda$CDM model.  Resolving these discrepancies
remains an area of active research.

\subsection{Constraints from large angular scales}
\label{sec:Low-ellScience}

The anisotropy at large angular scales, particularly the polarization, allows
us to place tight constraints on the optical depth to Thomson scattering,
$\tau$, and the tensor-to-scalar ratio, $r$.  The \Planck\ temperature data,
in combination with CMB lensing and low-$\ell$ polarization measured at
$70\,$GHz, prefer a lower optical depth, $\tau=0.066\pm 0.016$, than the
earlier inference from WMAP9 ($\tau\,{\approx}\,0.09$, which was used in our 2013
analysis), which implies a lower redshift of reionization
($z_{\rm re}=8.8^{+1.7}_{-1.4}$).  When cleaned of foregrounds using our
$353\,$GHz channel, the WMAP polarization data are in good agreement with a
lower optical depth.  With the dramatic improvement in our CMB lensing
detection, we are able to independently constrain $\tau$, finding comparably
tight and consistent results ($\tau=0.071\pm 0.016$) without the use of
low-$\ell$ polarization.  This provides additional confidence in the results.

While improved constraints on polarization at low-$\ell$ will eventually allow
us to study the reionization epoch in more detail, at present the largest
impact of the change in $\tau$ comes from the implied downward shift in the
inferred matter power spectrum normalization, $\sigma_8$.  As it happens, much
of the downward shift in this parameter is largely cancelled by the upward
shift in the CMB spectrum arising from the improved calibration in the current
data release.

Gravitational waves entering the horizon between recombination and today give
a ``tensor'' contribution to the large-scale temperature and polarization
anisotropies.  Our strongest \Planck-only constraint still comes from
temperature anisotropies at $\ell<10^2$ (or $k\la0.01\,{\rm Mpc}^{-1}$),
and is thus limited by cosmic variance and model-dependent.  Tensor modes also
generate a $B$-mode signal, which peaks at $\ell\approx 10^2$, slightly smaller
scales than the bulk of the temperature signal.  The cosmological landscape
became more complicated earlier this year with the detection of $B$-mode
polarization anisotropy by the BICEP2 team \citep{Bicep:2014}.  Analysis
of \Planck\ polarization data at high Galactic latitudes demonstrated that no
region of the sky can be considered dust-free when searching for primordial
$B$-modes \citep{planck2014-XXX}, and a joint
analysis of BICEP2/Keck Array observations and \Planck\ polarization
data \citep{pb2015} shows that polarized dust emission contributes a
significant part of the BICEP2 signal.  Combining the \Planck\ and revised
BICEP2/Keck Array likelihoods leads to a 95\,\% upper limit of
$r_{0.002} < 0.09$.  This eliminates any tension between the BICEP2 and
\Planck\ results, and in combination with our other constraints disfavours
inflationary models with a $\phi^2$ potential.  This and other implications
for inflationary models in the early Universe are discussed more fully in
\citet{planck2014-a15,planck2014-a24}.

\subsection{Dark energy and modified gravity}
\label{sec:DE&MG}

Even though much of the weight in the \Planck\ data lies at high redshift,
\Planck\ can still provide tight constraints on dark energy and modified
gravity, especially when used in combination with other probes.  This is
explored in \citet{planck2014-a16}, which focuses on tests of dark energy and
modified gravity on the scales where linear theory is most applicable, since
these are the most theoretically robust.  As for \citet{planck2014-a15}, the
results are consistent with the simplest scenario, $\Lambda$CDM, though all
constraints on dark energy models (including minimally-coupled scalar field
models or evolving equation of state models) and modified gravity models
(including effective field theory, phenomenological, $f(R)$, and coupled dark
energy models) are considerably improved with respect to past analyses.
In particular, we improve significantly the constraint on the density of dark
energy at early times, finding that it has to be below $2\,\%$
(95\,\% confidence) of the critical density, even if it only plays a role below
$z=50$.  Constraints are tighter if early dark energy is present since
recombination, with $\Omega_{\rm e} < 0.0071$ (for the data combination
PlanckTT+lensing+BAO+SNe+$H_0$), and an even tighter bound if high-$\ell$
polarization is included. In models where perturbations are modified, even if
the background is \lcdm, a few tensions appear, mainly driven by external
data sets.

\subsection{Lensing of the CMB}
\label{sec:CMBLens}

The CMB fluctuations measured by \Planck\ provide a slightly perturbed image
of the last-scattering surface, due to the effects of gravitational lensing by
large-scale structure.  Lensing slightly washes out the acoustic peaks of the
CMB power spectrum, an effect we see in the \Planck\ data at high significance.
Lensing also introduces distinctive non-Gaussian features into the CMB maps,
which allow us to map and make statistical measurements of the gravitational
potentials, and the associated matter.  These are studied in detail in
\cite{planck2014-a17}.  The lensing signal is consistent with the basic,
6-parameter, $\Lambda$CDM model that best fits the temperature data. This
gives us a very strong consistency check on the gravitational instability
paradigm and the growth of structure over more than two decades in expansion
factor.

\begin{figure*}[!htpb]
\begin{center}
\begin{overpic}[width=\textwidth]{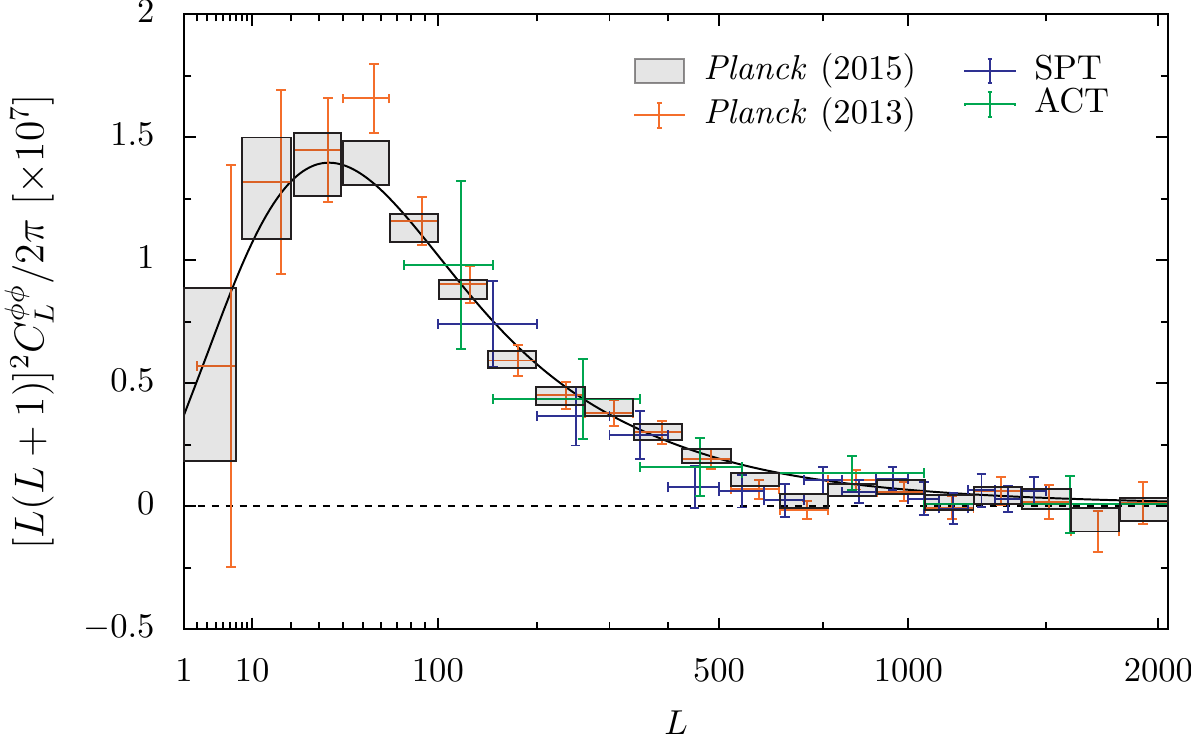}
\end{overpic}
\end{center}
\caption{Lensing potential power spectrum estimate from the 2015 data release
\citep{planck2014-a17}, based on the {\tt SMICA} CMB map, as well as previous
reconstructions from \Planck\ as well as other experiments for comparison.}
\label{fig:phispec}
\end{figure*}

Since it provides sensitivity to the growth of structure between the surface
of last scattering and the present epoch, the lensing signal allows us to
measure a number of important parameters by breaking parameter degeneracies.
Figure~\ref{fig:phispec} shows the lensing power spectrum, which for the first
time is measured with higher accuracy than it is predicted by the base
$\Lambda$CDM model that fits the temperature data. With the temperature-only
nominal mission data from the 2013 \Planck\ data release, we were able to make
the most powerful measurement of lensing to that date (at a level of
$25\,\sigma$).  In the current release, incorporating additional temperature
data, as well as entirely new polarization information, we have nearly doubled
the power of this measurement to $40\,\sigma$.  This is the most significant
detection to date, allowing lensing to be used as part of our precision
cosmology suite.

\subsection{Inflation}
\label{sec:CMBInflation}

The release of the 2013 \Planck\ data and findings had an enormous impact upon
the inflationary community, and the \Planck\ 2015 results continue to stress
the importance of this window into the early Universe.  \citet{planck2014-a24}
presents our constraints on inflationary models.  The \Planck\ data are
consistent with a purely adiabatic, power-law spectrum
of initial fluctuations, whose spectral index ($n_{\rm s}=0.9677\pm0.006$)
is significantly different from unity.  The addition of polarization data has
significantly improved the constraints on any isocurvature modes, which are
now constrained at the percent level.  Despite a detailed search, and study of
several models, we see no statistically significant evidence for departures
from a power law.  The combination of \Planck\ data with the
BICEP2/Keck Array data provide a strong upper limit on the tensor-to-scalar
ratio, and disfavour all monomial models ($V(\phi)\propto\phi^{2p}$) with
$p\ge 1$.  This is an important milestone, since these form the simplest class
of inflationary models.

\subsection{Primordial non-Gaussianity}
\label{sec:PNG}
\citep{planck2014-a19} for the first time uses polarization information to
constrain non-Gaussian signals left by primordial physics. The results
significantly reduce the allowed model space spanned by local, equilateral,
and orthogonal non-Gaussianity, tighenting constraints by up to 45\,\%. In
particular, $f_{\rm NL}^{\rm local} = 0.8 \pm 5.0$,
$f_{\rm NL}^{\rm equil} = -4 \pm 43$,
and $f_{\rm NL}^{\rm ortho} = - 26 \pm 21$. In addition, the \Planck\ 2015
analysis covers a greatly extended range of primordial 3-point and 4-point
signals, constraining inflationary model space as well as some proposed
alternatives to inflation. The global picture that emerges is one of
consistency with the premises of $\Lambda$CDM cosmology, namely that the
structure we observe today is the consequence of the passive evolution of
adiabatic, Gaussian, nearly scale-invariant, primordial seed perturbations.

\subsection{Isotropy and statistics}

The Planck 2013 results determined the presence of statistically anisotropic signals in the CMB, confirming previous studies made using WMAP data. Such anomalies therefore constitute real features of the microwave sky, and potentially challenge fundamental assumptions of the standard cosmological model. \citet{planck2014-a18} extends these studies based mainly on the full \Planck\ mission for temperature, but
also including some polarization measurements. A large number of statistical tests indicate consistency with Gaussianity, while a power deficit at large angular scales is manifested in several ways, for example low map variance. The well-known `Cold Spot' is identified through various methods. Tests of directionality suggest the presence of angular clustering from large to small scales, but at a significance that is dependent on the details of the approach. On large-angular scales, a dipolar power asymmetry is investigated through several approaches, and we address the subject of a posteriori correction. Our ability to include results based on polarization data is limited by two factors. First, CMB polarization maps have been high-pass filtered to mitigate residual large-scale systematic errors in the HFI channels, thus eliminating structure in the maps on angular scales larger than about 10\deg. Second, an observed noise mismatch between the simulations and the data prevents robust conclusions from being reached based on the nulll-hypothesis approach adopted throughout the paper. Nevertheless, we perform the first examination of polarization data via a stacking analysis, in which the stacking of the data themselves necessarily acts to lower the effect of the noise mismatch.  We find that the morphology of the stacked peaks is consistent with the expectations of statistically isotropic simulations.  Further studies of the large angular scale structure of the CMB polarization anisotropy will be conducted with data of improved quality expected to be released in 2016.

\subsection{Cosmology from clusters}
\label{sec:SZsci}

In 2013 we found an apparent tension between our primary CMB constraints and
those from the \Planck\ cluster counts, with the clusters preferring a lower
normalization of the matter power spectrum, $\sigma_8$.  The comparison is
interesting because the cluster counts directly measure $\sigma_8$ at low
redshift and hence any tension could signal the need for extensions of the
base model, such as non-minimal neutrino mass.  However, limited knowledge of
the normalization of the scaling relation between SZ signal and mass
(usually called ``mass bias'') continues to hamper the interpretation of this
result.

Our 2015 cluster analysis benefits from a larger catalogue (438 objects versus
the 189 in 2013), greater control of the selection function, and recent
gravitational lensing determinations of the mass bias for \Planck\ clusters.
With the larger sample, we now fit the counts in the two-dimensional plane of
redshift and S/N, allowing us to simultaneously constrain the slope of the
scaling relation and the cosmological parameters.  We examine three new
empirical determinations of the mass bias from gravitational lensing:
Weighing the Giants \citep[WtG]{wtgplanck}; the Canadian Cluster Comparison
Project (CCCP; Hoekstra et al., private communication); and results from a new
method based on CMB lensing \citep{mb14}. We use these three results as priors
because they measure the mass scale directly on samples of \Planck\ clusters.

The cluster constraints on $\sigma_8$ and $\Omega_{\rm m}$ are statistically
identical to those of 2013 when adopting the same scaling relation and mass
bias; in this sense, we confirm the 2013 results with the larger 2015
catalogue.  Applying the three new mass bias priors, we find that the WtG
calibration reduces the tension with the primary CMB constraints to slightly
more than $1\,\sigma$ in the base model, and CCCP results in tension at just
over $2\,\sigma$, similar to the case for the CMB lensing calibration.  More
detailed discussion of constraints from \Planck\ cluster counts can be found in
\citet{planck2014-a30}.

\section{\Planck\ 2015 astrophysics results}
\label{sec:astroresults}

\subsection{Low frequency foregrounds}
\label{sec:lowfreqfore}

\citet{planck2014-a31} discusses Galactic foreground emission between 20 and 100\,GHz, based primarily on the {\tt Commander} component
separation of \citep{planck2014-a12}.  The total intensity in this part of the spectrum is dominated by free-free and spinning dust emission, while polarization is dominated by synchrotron emission. 

Comparison with radio recombination line templates verifies the recovery of the free-free emission along the Galactic plane. Comparison of the high-latitude H$\alpha$ emission with our free-free map shows residuals that correlate with dust optical depth, consistent with a fraction ($\sim$30\,\%) of H$\alpha$ having been scattered by high-latitude dust. We highlight a number of diffuse spinning dust morphological features at high latitude. There is substantial spatial variation in the spinning dust spectrum, with the emission peak (in $I_\nu$) ranging from below 20\GHz\ to more than 50\GHz. There is a strong tendency for the spinning dust component near many prominent \ion{H}{ii} regions to have a higher peak frequency, suggesting that this increase in peak frequency is associated with dust in the photo-dissociation regions around the nebulae. The emissivity of spinning dust in these diffuse regions is of the same order as previous detections in the literature. Over the entire sky, the {\tt Commander} solution finds more anomalous microwave emission (AME) than the WMAP component maps, at the expense of synchrotron and free-free emission. Although the {\tt Commander} model fits the data exceptionally well, as noted in Sect~\ref{sec:CommFGs}, the discrepancy is largely driven by differences in the assumed synchrotron spectrum and the more elaborate model of spinning dust designed to allow for the variation in peak frequency noted above. Future surveys, particularly at 5--20\GHz, will greatly improve the separation, since the predicted brightness between the two models disagrees substantially in that range.

\begin{figure*}
\begin{center}
\includegraphics[width=17cm]{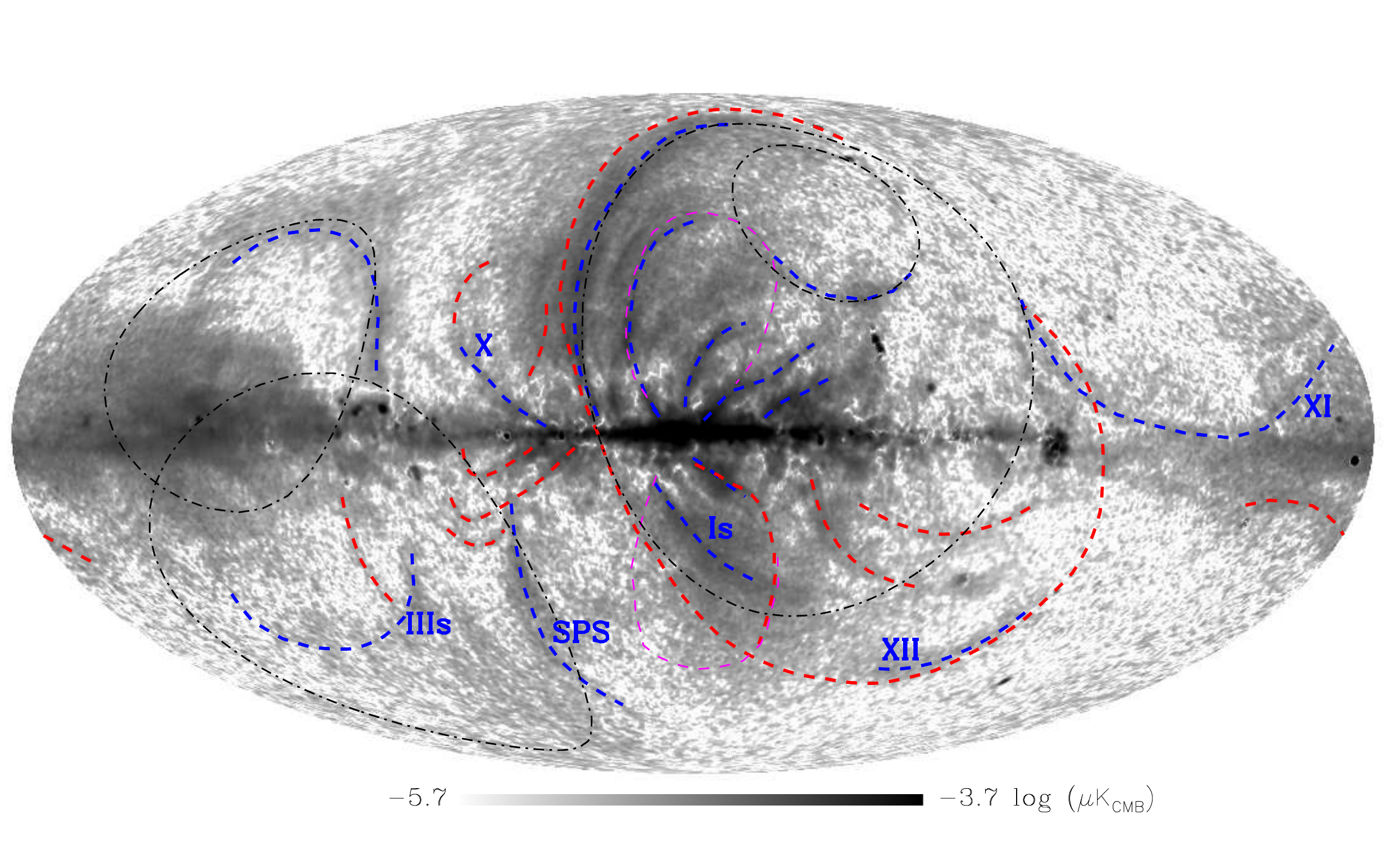}
\end{center}
\caption{Synchrotron polarization amplitude map, $P=\sqrt{Q^2+U^2}$, at 30\,GHz, smoothed to an angular resolution of $60\arcm$, produced by a weighted sum of \Planck\ and WMAP data as described in \citep{planck2014-a31}.  The traditional locii of radio loops I--IV are marked in black, a selection of the spurs identified by \citet{Vidal14} in blue, the outline of the {\it Fermi} bubbles in magenta, and
features discussed for the first time in \citep{planck2014-a31} in red.  Our measured outline for Loop~I departs substantially from the traditional small circle.}
\label{fig:Psynch}
\end{figure*}

\begin{figure*}[th!]
\centering
\includegraphics[width=17cm]{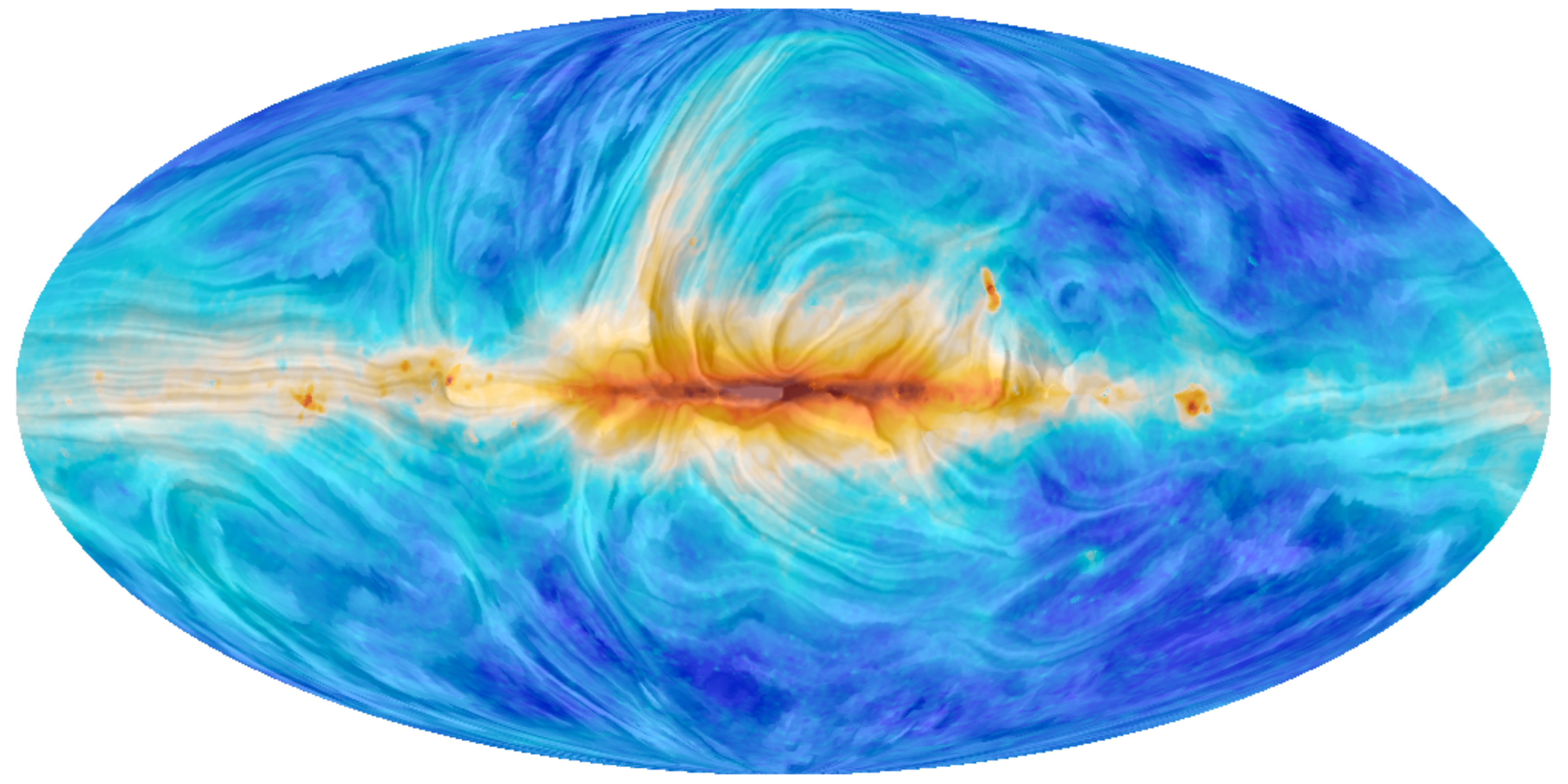}
\caption{All-sky view of the angle of polarization at 30\,GHz, rotated by 90\deg\ to indicate the direction of the Galactic magnetic field projected on the plane of the sky. The colours represent intensity, dominated at this frequency by synchrotron emission.   The ``drapery'' pattern was obtained by applying the line integral convolution \citep[LIC;][]{cabral1993} using an IDL implementation provided by Diego Falceta-Goncalves (\url{http://each.uspnet.usp.br/fgoncalves/pros/lic.pro}).  Where the field varies significantly along the line of sight, the orientation pattern is irregular and difficult to interpret.}
\label{fig:LIC30_all_sky_map}
\end{figure*}

In polarization, synchrotron emission completely dominates on angular scales larger than 1\deg\ and frequencies up to 44\,GHz.  We combine \Planck\ and \WMAP\ data to make the highest signal-to-noise ratio map yet of the intensity of the all-sky polarized synchrotron emission at frequencies above a few gigahertz, where Faraday rotation and depolarization are negligible (Figs.~\ref{fig:Psynch} and Fig.~\ref{fig:LIC30_all_sky_map}).  Most of the high-latitude polarized emission is associated with distinct large-scale loops and spurs, and we re-discuss their structure following the earlier study of \citet{Vidal14} based on \WMAP\ observations. We argue that nearly all the emission at $-90\deg < l < 40\deg$ is part of the Loop~I structure, and show that the emission extends much further in to the southern Galactic hemisphere than previously recognized, giving \mbox{Loop I} an ovoid rather than circular outline. However, it does not continue as far as the ``{\it Fermi} bubble/microwave haze'', which probably rules out an association between the two structures. The South
Polar Spur (SPS, see Fig.~\ref{fig:Psynch}) is bordered by a polarized dust filament and associated low-velocity \ion{H}{i}, analogous to
the cold features long known to border Loop~I around the North Polar Spur.  We find two structures that could correspond to distant analogues of the radio loops, as predicted by \citet{Mertsch2013}, including one surrounding the Cygnus~X star forming region, both of which are again associated with dust polarization.

We identify a number of other faint features in the polarized sky, including a dearth of polarized synchrotron emission directly  correlated with a narrow, roughly $20\deg$ long filament seen in H$\alpha$ at high Galactic latitude, and also visible in the Faraday rotation map of \citet{Oppermann2012}. Finally, we look for evidence of polarized AME; however, many AME regions are significantly contaminated by polarized synchrotron emission, and we find a 2$\sigma$ upper limit of 1.6\,\% in the Perseus region.

\subsection{Polarized thermal dust emission}
\label{sec:Dust}

\begin{figure*}
\begin{center}
\includegraphics[width=18cm]{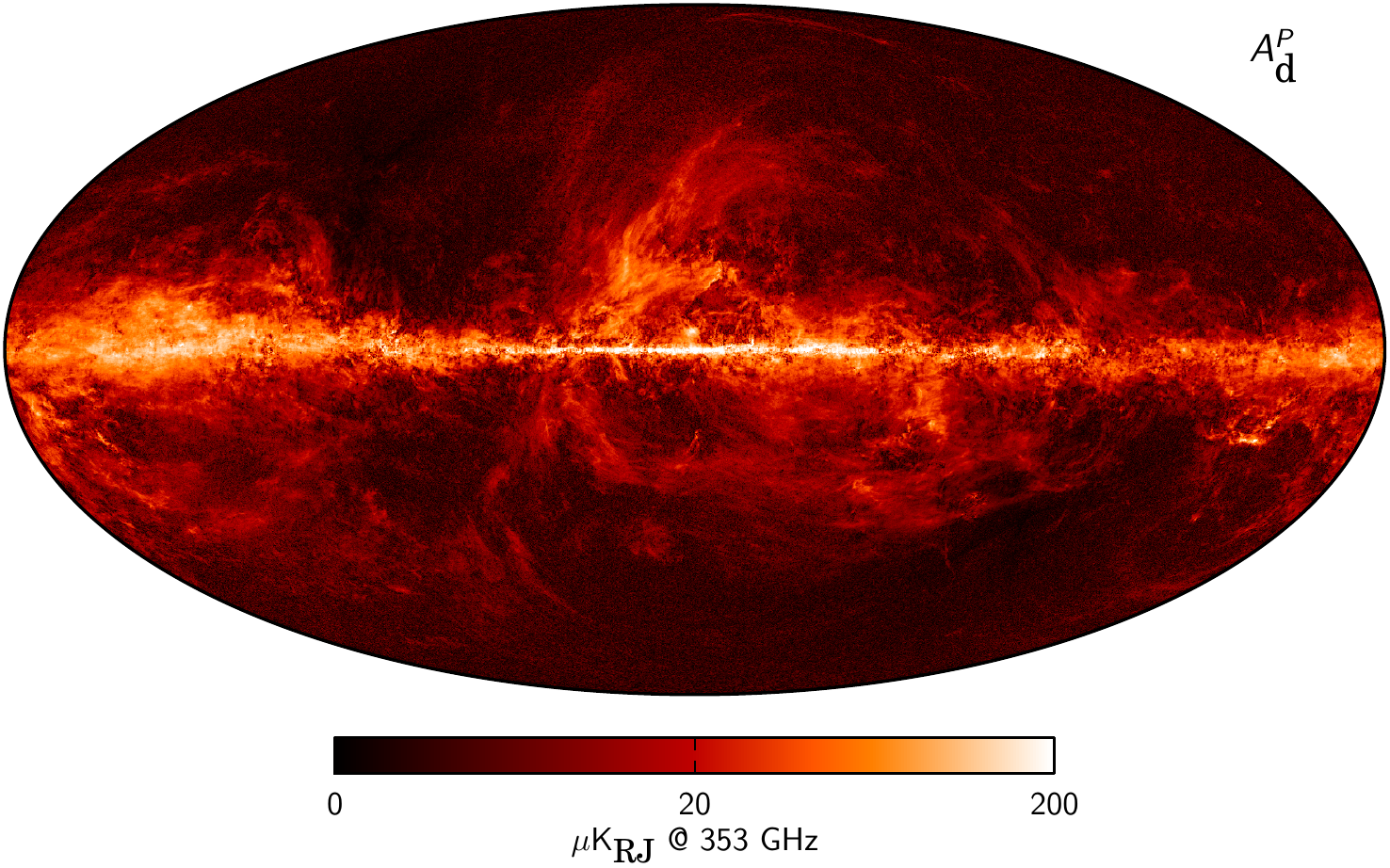}
\end{center}
\caption{Dust polarization amplitude map, $P=\sqrt{Q^2+U^2}$, at 353\,GHz, smoothed to an angular resolution of $10\arcm$, produced by the diffuse component separation process described in \citep{planck2014-a12} using \Planck\ and WMAP data.}
\label{fig:Pdust}
\end{figure*}

\begin{figure*}[th!]
\centering
\includegraphics[width=18cm]{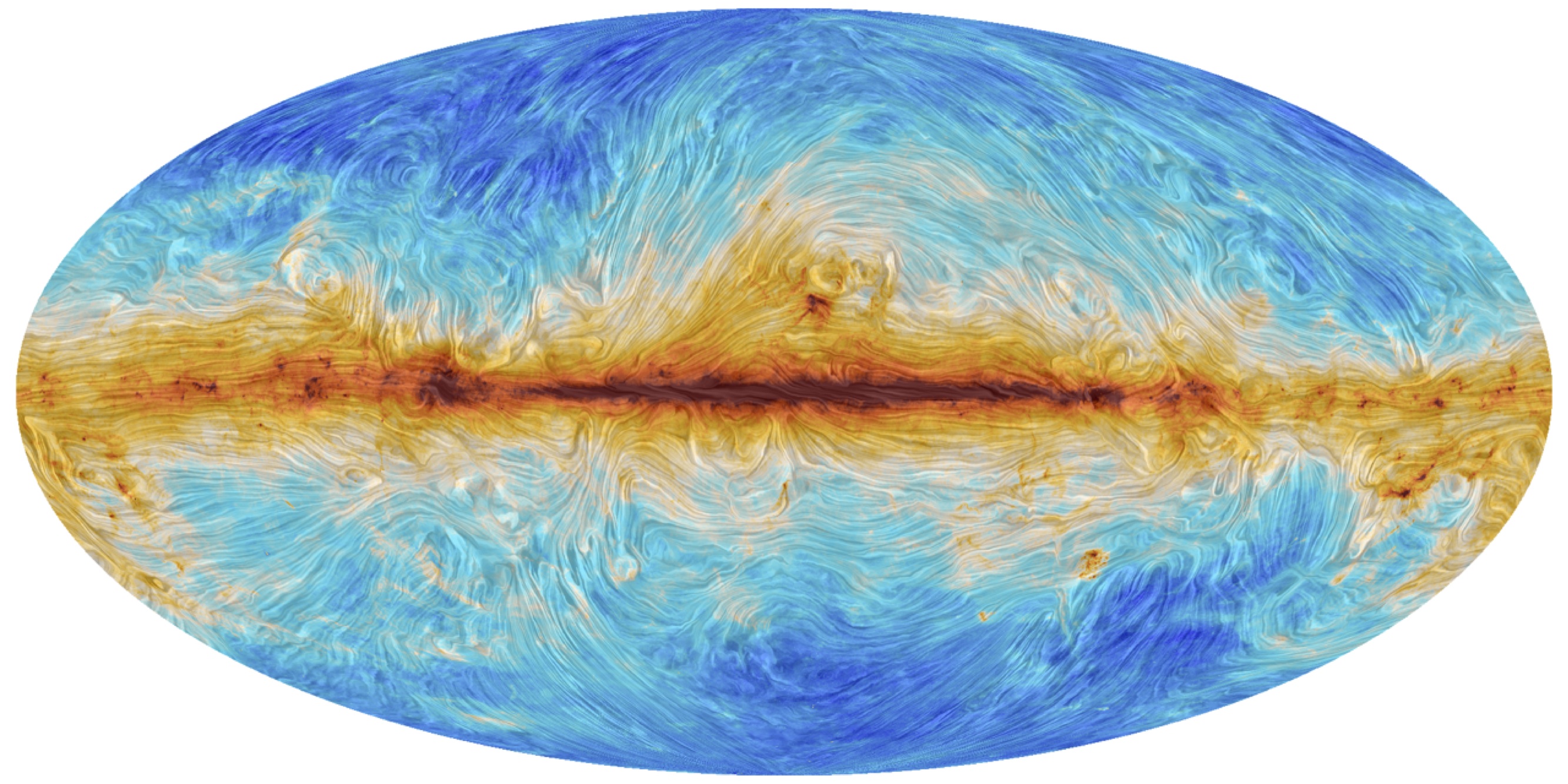}
\caption{All-sky view of the angle of polarization at 353 GHz, rotated by 90\deg\ to indicate the direction of the Galactic magnetic field projected on the plane of the sky. The colours represent intensity, dominated at this frequency by thermal dust emission.   The ``drapery'' pattern was obtained by applying the line integral convolution \citep[LIC;][]{cabral1993} using an IDL implementation provided by Diego Falceta-Goncalves (\url{http://each.uspnet.usp.br/fgoncalves/pros/lic.pro}).  Where the field varies significantly along the line of sight, the orientation pattern is irregular and difficult to interpret.}
\label{fig:LIC_all_sky_map}
\end{figure*}

\Planck\ has produced the first all-sky map of the polarized emission from
dust at submm wavelengths (Figs.~\ref{fig:QUcomps} and \ref{fig:Pdust}).
Compared with earlier ground-based and balloon-borne observations
\citep[e.g.,][]{Benoit04,Ward09,Matthews09, Koch10, Matthews14} this survey is
an immense step forward in sensitivity, coverage, and statistics.  It provides
new insight into the structure of the Galactic magnetic field and the
properties of dust, as well as the first statistical characterization of one
of the main foregrounds to CMB polarization. The wealth of information encoded
in the all-sky maps of polarized intensity, $P$, polarization fraction, $p$,
and polarization angle, $\psi$, presented in \citet{planck2014-a12} is
illustrated in Fig.~\ref{fig:LIC_all_sky_map}. Here we summarize the main
results from the data analysis by the Planck Consortium.  The release of the
data to the science community at large will trigger many more studies.

\subsubsection{The dust polarization sky}

\citet{planck2014-XIX} presents an overview of the polarized sky as seen by
\Planck\ at 353\,GHz, the most sensitive \Planck\ channel for polarized
thermal dust emission, focusing on the statistics of $p$ and $\psi$.  At all
$\NH$ below  $10^{22}\NHUNIT$, $p$ displays a large scatter.  The maximum $p$,
observed in regions of moderate hydrogen column density
($\NH < 2\times10^{21}\NHUNIT$), is high ($p_{\rm max} \approx 20\,\%$).
There is a general decrease in $p$ with increasing column density above
$\NH\approx 1\times10^{21}\NHUNIT$ and in particular a sharp drop above
$\NH\approx 10^{22}\NHUNIT$.   

The spatial structure of $\psi$ is characterized using the angle dispersion
function $\mathcal{S}$, the local dispersion of $\psi$ introduced by
\citet{Hildebrand09}.  The polarization fraction is found to be anti-correlated
with $\mathcal{S}$.  The polarization angle is ordered over extended areas of
several square degrees.  The ordered areas are separated by long, narrow
structures of high $\mathcal{S}$ that highlight interfaces where the sky
polarization changes abruptly. These structures have no clear counterpart in
the map of the total intensity, $I$.  They bear a morphological resemblance to
features detected in gradient maps of radio polarized emission
\citep{Iacobelli14}.

\subsubsection{The Galactic magnetic field}

The \Planck\ maps of $p$ and $\psi$ contain information on the magnetic field
structure.  The data have been compared to synthetic polarized emission maps
computed from simulations of anisotropic magnetohydrodynamical turbulence,
assuming simply a uniform intrinsic polarization fraction of dust grains
\citep{planck2014-XX}.  The turbulent structure of the magnetic field is able
to reproduce the main statistical properties of $p$ and $\psi$ that are
observed directly in a variety of nearby clouds (dense cores excluded). The
large-scale field orientation with respect to the line of sight plays a major
role in the quantitative analysis of these statistical properties.  This study
suggests that the large scatter of $p$ at $\NH$ smaller than about
$10^{22}\NHUNIT$ is due mainly to fluctuations in the magnetic field
orientation along the line of sight, rather than to changes in grain shape
and/or the efficiency of grain alignment. 

The formation of density structures in the interstellar medium involves
turbulence, gas cooling, magnetic fields, and gravity.  Polarization of
thermal dust emission is well suited to studying the role of the magnetic
field, because it images structure through an emission process that traces the
mass of interstellar matter \citep{planck2013-p06b}.  The \Planck\ $I$ map
shows elongated structures (filaments or ridges) that have counterparts in
either the Stokes $Q$ or $U$ map, or in both, depending on the mean
orientation.  The correlation between Stokes maps characterizes the relative
orientation between the ridges and the magnetic field.  In the diffuse
interstellar medium, the ridges are preferentially aligned with the magnetic
field measured on the structures. This statistical trend becomes more striking
for decreasing column density and, as expected from the potential effects of
projection, for increasing polarization fraction \citep{planck2014-XXXII}.
Towards nearby molecular clouds the relative orientation changes progressively
from preferentially parallel in areas with the lowest $\NH$ to preferentially
perpendicular in the areas with the highest $\NH$ \citep{planck2015-XXXV}.
This change in relative orientation might be a signature of the formation of
gravitationally-bound structures in the presence of a dynamically-important
magnetic field.

The relation between the structure of matter and the magnetic field is also
investigated in \citet{planck2014-XXXIII}, modelling the variations of the
Stokes parameters across three filaments for different hypotheses on $p$.
For these representative structures in molecular clouds the magnetic fields in
the filaments and their background have an ordered component with a mean
orientation inferred from \Planck\ polarization data.  However, the mean
magnetic field in the filaments does not have the same orientation as in the
background, with a different configuration in all three cases examined.
\citet{planck2015-XXXIV} analyzes the magnetic field in a massive star forming
region, the Rosette Nebula and parent molecular cloud, combining Faraday
rotation measures from the ionized gas with dust polarized emission from the
swept-up shell. This same methodology and modelling framework could be used to
study the field structure in a sample of massive star forming regions. 

\subsubsection{Dust polarization properties}

Galactic interstellar dust consists of components with different sizes and
compositions and consequently different polarization properties.  The relatively
large grains that are in thermal equilibrium and emit the radiation seen by
\Planck\ in the submm also extinguish and polarize starlight in the
visible \citep{Martin07}.  Comparison of polarized emission and starlight
polarization on lines of sight probed by stars is therefore a unique
opportunity to characterize the properties of polarizing grains.  For this
comparison, \citet{planck2014-XXI} use $P$ and $I$ in the \Planck\ $353\,$GHz
channel and stellar polarization observations in the $V$ band, the degree of
polarization, $p_V$, and the optical depth to the star, $\tau_V$.  Lines of
sight through the diffuse interstellar medium are selected with comparable
values of the column density as estimated in the submm and visible and
with polarization directions in emission and extinction that are close to
orthogonal.  Through correlations involving many lines of sight two ratios are
determined, $R_{S/V} = (P/I)/(p_V/\tau_V)$ and $R_{P/p} = P/p_V$, the latter
focusing directly on the polarization properties of the grains contributing to
polarization. The first ratio, $R_{S/V}$, is compatible with predictions based
on a range of dust models that have been developed for the diffuse interstellar
medium \citep[e.g.,][]{Martin07,Draine09}. This estimate provides new empirical
validation of many of the common underlying assumptions of the models, but is
not very discriminating among them. The second ratio, $R_{P/p}$, is higher than
model predictions by a factor of about 2.5. A comparable difference between
data and model is observed for $I/\tau_V$ \citep{planck2014-XXIX}. To address
this, changes will be needed in the optical properties of the large dust grains
contributing to the submm emission and polarization. 

The spectral dependence in the submm is also important for
constraining dust models.  In \citet{planck2014-XXII} the \Planck\ and WMAP
data are combined to characterize the frequency dependence of emission that is
spatially correlated with dust emission at $353\,$GHz, for both intensity and
polarization, in a consistent manner.  At $\nu \ge 100\,$GHz, the mean spectral
energy distribution (SED) of the correlated emission is well fit by a modified
blackbody spectrum for which the mean dust temperature of 19.6\,K (derived from
an SED fit of the dust total intensity up to $3000\,$GHz, i.e., $100\,\mu$m) 
is adopted.  It is found that the opacity has a spectral index $1.59\pm0.02$
for polarization and $1.51\pm0.01$ for intensity. The difference between the
two spectral indices is small but significant. It might result from differences
in polarization efficiency among different components of interstellar dust.
\citet{planck2014-XXII} also finds that the spectral energy distribution
increases with decreasing frequency at $\nu < 60$\,GHz, for both intensity and
polarization. The rise of the polarization SED towards low frequency might be
accounted for by a synchrotron component correlated with dust, with no need
for any polarization  of the anomalous microwave emission.

\subsubsection{Polarized dust and the CMB}

The polarized thermal emission from diffuse Galactic dust is the main
foreground present in measurements of the polarization of the CMB at
frequencies above 100\,GHz.  The \Planck\ sky coverage, spectral coverage from
100 to 353\,GHz for HFI, and sensitivity are all important for component
separation of the polarization data.  \citet{planck2014-XXX} measures the
polarized dust angular power spectra \clee\ and \clbb\ over the multipole
range $40 <\ell<600$ well away from the Galactic plane, providing cosmologists
with a precise characterization of the dust foreground to CMB polarization. 

The polarization power spectra of the dust are well described by power laws in
multipole, $C_\ell\propto \ell^{\,\alpha}$, with exponents
$\alpha = -2.42 \pm 0.02$ for both the $EE$ and $BB$ spectra. The amplitudes
of the polarization power spectra are observed to scale with the average dust
brightness as $\langle I\rangle^{1.9}$,  similar to the scaling found earlier
for power spectra of $I$ \citep{Miville07}. 
The frequency dependence of the power spectra for polarized thermal dust
emission is consistent with that found for the modified blackbody emission in
\citet{planck2014-XXII}.  A systematic difference is discovered between the
amplitudes of the Galactic $B$- and $E$-modes, such that
$C_\ell^{BB} /C_\ell^{EE} = 0.5$. There is additional information coming from
the dust $TE$ and $TB$ spectra.  These general properties apply at
intermediate and high Galactic latitude in regions with low dust column
density. The data show that there are no windows in the sky where primordial
CMB $B$-mode polarization can be measured without subtraction of polarized
dust emission.

\section{Summary and Conclusions}
\label{sec:summary}

This paper is an overview of the \Planck\ 2015 release, summarizing the main features of the products being released and the main scientific conclusions that we draw from them at this time.  Some of the highlights of this release are listed below.
\begin{itemize}

\item Data from the entire mission are now used, including both temperature and polarization, and significant improvements have been made in the understanding of beams, pointing, calibration, and systematic errors. As a result, the new products are less noisy, but even more importantly they are much better understood and the overall level of confidence is significantly increased.

\item The residual systematics in the \Planck\ 2015 polarization maps have been dramatically reduced compared to 2013, by as much as two orders of magnitude in some cases. Nevertheless, on angular scales greater than 10\deg, systematic errors in the polarization maps between 100 and 217\,GHz are still non-negligible compared to the expected cosmological signal.  It was not possible, for this data release, to fully characterize the large-scale residuals due to these systematic errors from the data or from simulations. Therefore all results published by the \Planck\ Collaboration in 2015 have used CMB polarization maps that have been high-pass filtered to remove the large angular scales. Users of the \Planck\ CMB maps are warned that they are not useable for cosmological analysis at $\ell$<30. 
 
\item A large set of simulations accompanies the release, including up to  10\,000 realizations of signal and noise; this has been used to test and verify methods of analysis, and also to estimate uncertainties.

\item We measure the amplitude and direction of the Solar dipole to the best precision so far. 

\item One of the most notable improvements in this data set is that now LFI, HFI, and WMAP agree to within a few tenths of a percent on angular  scales from the dipole through the first acoustic peak.

\item Polarization is a new product in this release, and especially on large angular scales systematic effects are not yet fully controlled. This is an area where we expect to make significant improvements in the coming months.  Nevertheless, our polarization data are already making important contributions in a variety of analyses.

\item More specifically, we are able to use the $TE$ and $EE$ angular power spectra at small scales (and to a more limited extent, at large angular scales) over the full sky, reaching the expected sensitivity. This allows us to estimate cosmological parameters independently of $TT$, and in combination with $TT$.

\item At large angular scales, we are now able to use \Planck-only products to carry out cosmological analysis.  Specifically, we can estimate the optical depth of reionization, $\tau$, independently of other experiments. The value of $\tau$ is smaller than found in previous determination, implying later reionization.

\item Foregrounds can be separated effectively over larger areas of the sky, allowing more sky to be used for cosmology, and producing high-quality maps of synchrotron, free-free, spinning dust, thermal dust, and CO emission.

\item Our 2015 results for cosmology are very consistent with our 2013 results, but with smaller uncertainties, and covering a greater range of science. 

\item Our best-fit 2015 cosmological parameters comfirm the basic 6-parameter $\Lambda$CDM scenario that we determined in 2013.  There is no compelling evidence for any extensions to the 6-parameter model, or any need for new physics. Depending somewhat on what data combinations are used, five of the six parameters are now measured to better than 1\,\% precision. Areas that were in ``tension'' in 2013 ($\sigma_8$ and weak galaxy lensing), have been confirmed to remain in tension today, although the disagreement is lessened when only particular subsets of the external data are considered.

\item Using only \Planck\ data, we find that the Universe is flat to 0.7\,\% ($1\,\sigma$).  Including BAO data, the constraint tightens to a remarkable 0.25\,\%.

\item Using the \Planck\ temperature data over the whole sky, together with our recent work combining \Planck\ and BICEP2/Keck data, we have obtained the best current upper limits on the tensor-to-scalar ratio obtained to date.

\item We have obtained improved limits on primordial non-gaussianity ($f_{\rm NL}$), which are about 30\,\% tighter before, reaching the expected sensitivity of \Planck\ when including polarization. 

\item Models of inflation are more tightly constrained than ever before, with the simplest $\phi^n$ models being ruled out for $n\ge2$.

\item We have obtained the tightest limits yet on the amplitude of primordial magnetic fields.

\item \Planck's measurement of lensing of the CMB has the highest signal-to-noise ratio yet achieved, 40$\,\sigma$.

\item The second \Planck\ catalogues of compact sources, Sunyaev-Zeldovich clusters, and Galactic cold clumps, are larger than the previous ones and better-characterized in terms of completeness and reliability.

\end{itemize}

\Planck\ continues to provide a rich harvest of data for cosmology and astrophysics.


\begin{acknowledgements}
Planck is a project of the European Space Agency in cooperation with the scientific community, which started in 1993.  ESA led the project, developed the satellite, integrated the payload into it, and launched and operated the satellite.  Two Consortia, comprising around 100~scientific institutes within Europe, the USA, and Canada, and funded by agencies from the participating countries, developed and operated the scientific instruments LFI and HFI.  The Consortia are also responsible for scientific processing of the acquired data. The Consortia are led by the Principal Investigators: J.-L. Puget in France for HFI (funded principally by CNES and CNRS/INSU-IN2P3) and N. Mandolesi in Italy for LFI (funded principally via ASI).  NASA's US \Planck\ Project, based at JPL and involving scientists at many US institutions, contributes significantly to the efforts of these two Consortia.  A third Consortium, led by H.U. Norgaard-Nielsen and supported by the Danish Natural Research Council, contributed to the reflector programme.  These three Consortia, together with ESA's Planck Science Office, form the \Planck\ Collaboration.  A description of the \Planck\ Collaboration and a list of its members, indicating which technical or scientific activities they have been involved in, can be found at
\url{http://www.cosmos.esa.int/web/planck/planck-collaboration}.  The \Planck\ Collaboration acknowledges the support of: ESA; CNES and
CNRS/INSU-IN2P3-INP (France); ASI, CNR, and INAF (Italy); NASA and DoE (USA); STFC and UKSA (UK); CSIC, MINECO, JA, and RES (Spain); Tekes, AoF, and CSC (Finland); DLR and MPG (Germany); CSA (Canada); DTU Space (Denmark); SER/SSO (Switzerland); RCN (Norway); SFI (Ireland); FCT/MCTES (Portugal); ERC and PRACE (EU).  We thank Diego Falceta-Goncalves for the line-integral-convolution maps in Figs.~\ref{fig:LIC30_all_sky_map} and \ref{fig:LIC_all_sky_map} (see \url{http://each.uspnet.usp.br/fgoncalves/pros/lic.pro}).
\end{acknowledgements}

\bibliographystyle{aat}

\bibliography{Planck_bib,Overview_bib}
\end{document}